\documentclass[article,twocolumn,usenames,dvipsnames]{aastex62}
\usepackage{lineno}
\usepackage{textcomp}
\usepackage{seqsplit}
\usepackage{color}
\usepackage{xcolor}
\usepackage{colortbl}
\usepackage{multirow}
\usepackage{array}
\definecolor{LightRed}{rgb}{1,.5,.5}
\definecolor{LightBlue}{rgb}{.33,.66,1.}
\newcolumntype{H}{>{\setbox0=\hbox\bgroup}c<{\egroup}@{}}

\newcommand{\Fermi}{\emph{Fermi}}
\newcommand{\FermiGBM}{\emph{Fermi}-GBM}
\newcommand{\Swift}{\emph{Swift}}
\newcommand{\SwiftBAT}{\emph{Swift}-BAT}

\newcommand{\pastro}{\ensuremath{p_{\mathrm{astro}}}}
\newcommand{\Msun}{M$_\odot$}

\newcommand{\jw}[1]{{\color{purple} #1}}

\begin{document}

\title{A Joint \FermiGBM\ and \SwiftBAT\ Analysis of Gravitational-Wave Candidates from the Third Gravitational-wave Observing Run}

%
%
%

\author[0000-0002-0186-3313]{C.~Fletcher}
\affiliation{Science and Technology Institute, Universities Space Research Association, Huntsville, AL 35805, USA}
\author[0000-0001-9012-2463]{J.~Wood}
\affiliation{NASA Marshall Space Flight Center, Huntsville, AL 35812, USA}

\author[0000-0003-0761-6388]{R.~Hamburg}
\affiliation{Department of Space Science, University of Alabama in Huntsville, Huntsville, AL 35899, USA}
\affiliation{Center for Space Plasma and Aeronomic Research, University of Alabama in Huntsville, Huntsville, AL 35899, USA}
\affiliation{Universit\'e Paris-Saclay, CNRS/IN2P3, IJCLab, 91405 Orsay, France}

\author[0000-0002-2149-9846]{P.~Veres}
\affiliation{Department of Space Science, University of Alabama in Huntsville, Huntsville, AL 35899, USA}
\affiliation{Center for Space Plasma and Aeronomic Research, University of Alabama in Huntsville, Huntsville, AL 35899, USA}

\author[0000-0002-0468-6025]{C.~M.~Hui}
\affiliation{NASA Marshall Space Flight Center, Huntsville, AL 35812, USA}

\author[0000-0001-9935-8106]{E.~Bissaldi}
\affiliation{Dipartimento Interateneo di Fisica, Politecnico di Bari, Via E. Orabona 4, 70125, Bari, Italy}
\affiliation{INFN - Sezione di Bari, Via E. Orabona 4, 70125, Bari, Italy}

\author{M.~S.~Briggs}
\affiliation{Department of Space Science, University of Alabama in Huntsville, Huntsville, AL 35899, USA}
\affiliation{Center for Space Plasma and Aeronomic Research, University of Alabama in Huntsville, Huntsville, AL 35899, USA}

\author[0000-0002-2942-3379]{E. Burns}
\affiliation{Department of Physics \& Astronomy, Louisiana State University, Baton Rouge, LA 70803, USA}

\author[0009-0003-3480-8251]{W.~H.~Cleveland}
\affiliation{Science and Technology Institute, Universities Space Research Association, Huntsville, AL 35805, USA}

\author{M.~M.~Giles}
\affiliation{Jacobs Space Exploration Group, Huntsville, AL 35806, USA}

\author[0000-0002-0587-7042]{A.~Goldstein}
\affiliation{Science and Technology Institute, Universities Space Research Association, Huntsville, AL 35805, USA}

\author[0000-0001-9556-7576]{B.~A.~Hristov}
\affiliation{Center for Space Plasma and Aeronomic Research, The University of Alabama in Huntsville, Huntsville, AL 35899}

\author{D.~Kocevski}
\affiliation{NASA Marshall Space Flight Center, Huntsville, AL 35812, USA}

\author[0000-0001-8058-9684]{S.~Lesage}
\affiliation{Department of Space Science, University of Alabama in Huntsville, 320 Sparkman Drive, Huntsville, AL 35899, USA}
\affiliation{Center for Space Plasma and Aeronomic Research, University of Alabama in Huntsville, Huntsville, AL 35899, USA}

\author[0000-0002-2531-3703]{B.~Mailyan}
\affiliation{Department of Aerospace, Physics and Space Sciences, Florida Institute of Technology, Melbourne, FL 32901, USA}

\author[0000-0002-0380-0041]{C.~Malacaria}
\affiliation{International Space Science Institute (ISSI), Hallerstrasse
6, 3012 Bern, Switzerland}

\author[0000-0002-6269-0452]{S.~Poolakkil}
\affiliation{Department of Space Science, University of Alabama in Huntsville, Huntsville, AL 35899, USA}
\affiliation{Center for Space Plasma and Aeronomic Research, University of Alabama in Huntsville, Huntsville, AL 35899, USA}

\author[0000-0002-0221-5916]{A.~von Kienlin} \affil{Max-Planck-Institut f\"{u}r extraterrestrische Physik, Giessenbachstrasse 1, D-85748 Garching, Germany}

\author[0000-0002-8585-0084]{C.~A.~Wilson-Hodge}
\affiliation{NASA Marshall Space Flight Center, Huntsville, AL 35812, USA}

 \collaboration{\Fermi\ Gamma-Ray Burst Monitor Team}

\collaboration{}
%

\author[0000-0002-7604-1779]{M.~Crnogor\v{c}evi\'{c}}
\affiliation{Department of Astronomy, University of Maryland, College Park, MD 20742, USA}
\affiliation{Center for Research and Exploration in Space Science and Technology, NASA Goddard Space Flight Center, Greenbelt, MD 20771, USA}
\affiliation{Stockholm University and The Oskar Klein Centre for Cosmoparticle Physics, Alba Nova, 10691 Stockholm, Sweden}
\author[0000-0001-5229-1995]{J.~DeLaunay}
\affiliation{Department of Physics and Astronomy, University of Alabama, Tuscaloosa, AL 35487, USA}
\affiliation{Department of Physics, Pennsylvania State University, University Park, PA 16802, USA}
\affiliation{Center for Multimessenger Astrophysics, Institute for Gravitation and the Cosmos, Pennsylvania State University, University Park, PA 16802, USA}
\author[0000-0002-2810-8764]{A.~Tohuvavohu}
\affiliation{Department of Astronomy \& Astrophysics, University of Toronto, Toronto, Ontario M5S 1A1, Canada}
\author[0000-0002-9280-836X]{R.~Caputo}
\affiliation{Astrophysics Science Division, NASA Goddard Space Flight Center, MC 661, Greenbelt, MD 20771, USA}

\author{S.~B.~Cenko}
\affiliation{Astrophysics Science Division, NASA Goddard Space Flight Center, MC 661, Greenbelt, MD 20771, USA}
\affiliation{Joint Space-Science Institute, University of Maryland, College Park, MD 20742, USA}

\author[0000-0003-2714-0487]{S.~Laha}
\affiliation{Center for Research and Exploration in Space Science and Technology, NASA Goddard Space Flight Center, Greenbelt, MD 20771, USA}
\affiliation{Astrophysics Science Division, NASA Goddard Space Flight Center, Greenbelt, MD 20771, USA}
\affiliation{Center for Space Science and Technology, University of Maryland Baltimore County, 1000 Hilltop Circle, Baltimore, MD 21250, USA}

\author[0000-0002-4299-2517]{T.~Parsotan}
\affiliation{Center for Research and Exploration in Space Science and Technology, NASA Goddard Space Flight Center, Greenbelt, MD 20771, USA}
\affiliation{Astrophysics Science Division, NASA Goddard Space Flight Center, Greenbelt, MD 20771, USA}

\affiliation{Center for Space Science and Technology, University of Maryland Baltimore County, 1000 Hilltop Circle, Baltimore, MD 21250, USA}

\collaboration{}
%

\author{R.~Abbott}
\affiliation{LIGO Laboratory, California Institute of Technology, Pasadena, CA 91125, USA}
\author{H.~Abe}
\affiliation{Graduate School of Science, Tokyo Institute of Technology, Meguro-ku, Tokyo 152-8551, Japan  }
\author{F.~Acernese}
\affiliation{Dipartimento di Farmacia, Universit\`a di Salerno, I-84084 Fisciano, Salerno, Italy  }
\affiliation{INFN, Sezione di Napoli, Complesso Universitario di Monte S. Angelo, I-80126 Napoli, Italy  }
\author[0000-0002-8648-0767]{K.~Ackley}
\affiliation{OzGrav, School of Physics \& Astronomy, Monash University, Clayton 3800, Victoria, Australia}
\author[0000-0002-4559-8427]{N.~Adhikari}
\affiliation{University of Wisconsin-Milwaukee, Milwaukee, WI 53201, USA}
\author[0000-0002-5731-5076]{R.~X.~Adhikari}
\affiliation{LIGO Laboratory, California Institute of Technology, Pasadena, CA 91125, USA}
\author{V.~K.~Adkins}
\affiliation{Louisiana State University, Baton Rouge, LA 70803, USA}
\author{V.~B.~Adya}
\affiliation{OzGrav, Australian National University, Canberra, Australian Capital Territory 0200, Australia}
\author{C.~Affeldt}
\affiliation{Max Planck Institute for Gravitational Physics (Albert Einstein Institute), D-30167 Hannover, Germany}
\affiliation{Leibniz Universit\"at Hannover, D-30167 Hannover, Germany}
\author{D.~Agarwal}
\affiliation{Inter-University Centre for Astronomy and Astrophysics, Pune 411007, India}
% ...

\collaboration{The LIGO Scientific Collaboration, the Virgo Collaboration, and the KAGRA Collaboration}

\date{\today}

\begin{abstract}
%
%
We present \textit{Fermi} Gamma-ray Burst Monitor (\FermiGBM) and \textit{Swift} Burst Alert Telescope (\SwiftBAT) searches for gamma-ray/X-ray counterparts to gravitational wave (GW) candidate events 
identified during the third observing run of the Advanced LIGO and Advanced Virgo detectors.
Using \FermiGBM\ on-board triggers and sub-threshold gamma-ray burst (GRB) candidates found in the \FermiGBM\ ground analyses, the Targeted Search and the Untargeted Search, we investigate whether there are any coincident GRBs associated with the GWs. We also search the \SwiftBAT\ rate data around the GW times to determine whether a GRB counterpart is present. No counterparts are found. Using both the \FermiGBM\ Targeted Search and the \SwiftBAT\ search, we calculate flux upper limits and present joint upper limits on the gamma-ray luminosity of each GW. Given these limits, we constrain theoretical models for the emission of gamma-rays from binary black hole mergers.

\end{abstract}

\section{Introduction}\label{sec:intro}

The detection of GW170817 \citep{GW170817} coincident with  the short gamma-ray burst GRB 170817A \citep{Goldstein+17170817a, savchenko_2017} was a ground-breaking discovery for the multimessenger era. Not only was it the first binary neutron star (BNS) merger detected by the gravitational-wave (GW) instruments Advanced LIGO \citep{2015CQGra..32g4001L} and Advanced Virgo \citep{Acernese_2015}, it was also the first, and to date only, GW detection with a confirmed electromagnetic (EM) counterpart. Since then, the search for EM emission from more of these extreme events has been at the forefront of multimessenger astronomy, particularly in the gamma-ray energy band since GRB 170817A demonstrated that BNS mergers are a progenitor of short gamma-ray bursts (GRBs) \citep{Abbott+17170817GWGRB}. GWs have also been observed from the mergers of other compact objects, such as binary black hole (BBH) and  neutron star--black hole (NSBH) systems \citep{GWTC-1,GWTC-2,GWTC2.1, GWTC-3}; however, no additional EM counterparts have been confirmed as they have been inconclusive \citep{Connaughton+16BBH,2019GCN.25406....1L, 2019GCN.25465....1L} or are still under debate \citep{PhysRevLett.124.251102, Ashton2021, Bustillo:2021tga, 2020MNRAS.499L..87D, 2021ApJ...914L..34P}.

GRB 170817A was first reported by the \textit{Fermi} Gamma-ray Burst Monitor (GBM; \citealt{Meegan_2009}), a space-based gamma-ray instrument sensitive from 8 keV to 40 MeV. 
This wide energy range of \FermiGBM\ combined with its large field-of-view (FoV) and rapid alert abilities make it an ideal platform to search for gamma-ray counterparts to GWs in real time. \FermiGBM\ also provides continuous time tagged event (CTTE) data with a 6-hour latency that enables sensitive searches for short GRBs on the ground. Two of these searches are the Untargeted Search, a blind search of \FermiGBM\ data for short GRBs, and the Targeted Search, which uses an external time to search for a short GRB \citep{Blackburn2015,Goldstein2019}. Both were previously used to look for sub-threshold GRBs coincident with GWs from the first two LIGO-Virgo observing runs. 

Additionally, the Burst Alert Telescope (BAT; \citealt{Barthelmy_2005}) on-board the \textit{Neil Gehrels Swift Observatory} (hereafter referred to as \emph{Swift}) provides excellent sensitivity to detecting hard X-ray and gamma-ray transients \citep{Gehrels_2004}. 
\textit{Swift}-BAT primarily runs in a survey mode that continuously evaluates photon rate increases and potential GRB triggers. An increase in the observed photon rate can trigger the on-board image-processing algorithms which can yield $\sim$arcminute GRB localizations.

Ideally, \SwiftBAT\ would detect and localize a GRB produced by a binary merger independently of the GW detection. If a GRB does not trigger an on-board detection, continuous count rate lightcurves are still available for offline ground searches. Although \SwiftBAT\ has been used to search for public and sub-threshold GWs during the LIGO-Virgo observing runs, this work presents the first systematic search of \SwiftBAT\ data from a LIGO-Virgo observing run.

The first observing run (O1) operated from September 2015 to January 2016, producing the first detection of GWs from a BBH merger (GW150914; \citealt{Abbott+16GW150914}). \cite{Burns2019} used the \FermiGBM\ Targeted Search to identify both triggered and sub-threshold GRB candidates in coincidence with GW candidates from O1. The most significant gamma-ray candidate found by the search was within 0.4 s of GW150914; however, it could not be confirmed as a counterpart due to its weak signal and poor localization \citep{Connaughton+16BBH, Greiner_2016, Connaughton_2018}.

The second observing run (O2) took place from November 2016 to August 2017, resulting in the detections of GW170817, GRB 170817A, and the kilonova AT2017gfo  \citep{Chornock_2017,Cowperthwaite_2017,Nicholl_2017,Soares-Santos_2017,Tanvir_2017,GW170817_review}.
Following O2, the LIGO Scientific and Virgo Collaboration published its first catalog of GW signals called the Gravitational-Wave Transient Catalog 1 (GWTC-1; \citealt{GWTC-1}) using a re-analysis of data from both O1 and O2. 
\cite{Hamburg+2020paper} searched for GRBs coincident to the GWs reported in GWTC-1, using \FermiGBM\ triggers as well as sub-threshold GRB candidates from the Untargeted and Targeted Searches, but found no additional counterparts beyond GRB 170817A.

The third observing run (O3) occurred from April 2019 to March 2020 with a month-long commissioning break during October 2019. It benefited from improvements to the sensitivity and duty cycle of the GW detectors made after O2~\citep{PhysRevD.102.062003,GWTC-2,VirgoO3}. This observing run provided 56 public GW candidates in real-time with information from their preliminary analysis. More detailed analyses were published by the LIGO, Virgo, and KAGRA (LVK) Collaboration in a series of  GWTCs (GWTC-2; \citealt{GWTC-2}, GWTC-2.1; \citealt{GWTC2.1}, GWTC-3; \citealt{GWTC-3}) with GWTC-3 providing a cumulative list of 79 GW signals from O3 with a probability of astrophysical origin ($\pastro$) $>$ 0.5 -- an 8-fold increase relative to O2.
Among these candidates was the detection of a second confident signal classified as a BNS merger, GW190425, whose total mass is larger than that known from Galactic neutron star binaries \citep{Abbott_2020_190425}.

Additionally, GW191219\_163120 and GW200115\_042309 provided the first detections of NSBH systems with \pastro\ $>$ 0.5.
Another possible NSBH, GW200105\_162426, fell just outside the \pastro\ $>$ 0.5 criterion in the GWTC-3 analysis \citep{NSBH_discovery_paper}. There were also two confident detections with ambiguous classifications, GW190814 \citep{190814} and GW200210\_092254, that represent a black hole merging with either a light black hole or a heavy neutron star. 
An overwhelming majority of the remaining candidates are most likely BBH in origin.

In this paper, we search \FermiGBM\ and \SwiftBAT\ data for short GRB counterparts to GW candidates from O3, discussed in Section \ref{ssec:method_cbc}.
Section \ref{ssec:method_gbm} provides an overview of the \FermiGBM\ Untargeted Search as well as improvements made to the \FermiGBM\ Targeted Search. Section~\ref{ssec:method_swift} describes the \SwiftBAT\ sub-threshold search. We present the results of the search with \FermiGBM\ triggers and the Untargeted Search in Section \ref{ssec:results_gbm}; with the \FermiGBM\ Targeted Search, including a new joint ranking statistic that takes the spatial coincidence into account, in Section \ref{ssec:results_targeted}; and with the \SwiftBAT\ sub-threshold search in \ref{ssec:Swift_results}. 
Section \ref{ssec:marginal_results} presents the results from both \FermiGBM\ and \SwiftBAT\ for the marginal GWs identified in Section~\ref{ssec:method_cbc}. Furthermore, Section \ref{sec:discussion} divides the discussion of GWs with \pastro\ $>$ 0.5 into two groups depending on their estimated secondary component mass $m_2$. For mergers with a possible neutron star component, 
we present the flux and isotropic equivalent luminosity upper limits from both \FermiGBM\ and \SwiftBAT\ (Section \ref{sec:upperlimits_region1}). 
For the BBH mergers,
we compare the lack of observed gamma-ray emission to that predicted by theoretical models (Section \ref{sec:bbh_region3}).
We discuss upper limits to the marginal GWs in Section~\ref{ssec:marginal_discussion}. Finally, in Section \ref{sec:conclusion} we summarize our results and discuss future plans for using the sub-threshold searches for GWs.

\section{Method}\label{sec:method}
In this section, we summarize the set of GW signals that we analyze from O3. We also present the search methods used to find coincident gamma-ray and hard X-ray emission with \FermiGBM\ and \SwiftBAT.

\subsection{GW Trigger Selection}\label{ssec:method_cbc}

The analysis reported here focuses on GW candidates identified during O3.
These were selected by four separate analysis pipelines (i.e., GstLAL, Multi-Band Template Analysis (MBTA), PyCBC, and cWB) and published in GWTC-3 \citep{GWTC-3}. Each pipeline calculates both a false alarm rate (FAR) from a background noise hypothesis and a \pastro\ for each candidate assuming a compact binary coalescence source. Candidate signals with $\pastro > 0.5$ in any pipeline are selected for detailed analysis with a full estimation of the potential astrophysical source parameters. The one exception is GW candidates identified by the minimally modeled cWB pipeline, which requires a time-matched confirmation with $\pastro>0.1$ in one of the other pipelines in order to ensure they originated from a compact binary coalescence. In total, there were 79 GWs identified with \pastro~$>$~0.5 during O3. Table \ref{tab:GWevents} shows the candidate identifier, date, time, and $\pastro$ for these GWs.

The remaining subset of GW signals with a FAR below 2 yr$^{-1}$ and $\pastro \leq 0.5$ in a given pipeline are considered marginal GW candidates. As of GWTC-3, there are 6 marginal candidates which cannot be attributed to instrumental or environmental causes (Table~\ref{tab:margGWevents}). 
We exclude these candidates from our main analysis; 
however, since the existence of a gamma-ray counterpart could potentially prove an astrophysical origin, we perform separate searches around each marginal candidate.

\begin{table*}[ht]
    \centering
    \caption{GW candidates from O3 with $\pastro > 0.5$ \citep{GWTC-3}.}
    \begin{tabular}{Hlccc|lHlcccr}
    \hline
    & Event Name & Date & Time (UTC) & $\pastro$ && & Event Name & Date & Time (UTC) & $\pastro$\\ \hline\hline 
 1 & GW190403\_051519 & 04-03-2019 & 05:15:19 &   0.60 &&  1 & GW191103\_012549 & 11-03-2019 & 01:25:49 &  0.94 \\
 2 & GW190408\_181802 & 04-08-2019 & 18:18:02 & $>$0.99 &&  2 & GW191105\_143521 & 11-05-2019 & 14:35:21 & $>$0.99 \\
 3 & GW190412 & 04-12-2019 & 05:30:44 & $>$0.99 &&  3 & GW191109\_010717 & 11-09-2019 & 01:07:17 & $>$0.99 \\
 4 & GW190413\_052954 & 04-13-2019 & 05:29:54 &  0.92 &&  4 & GW191113\_071753 & 11-13-2019 & 07:17:53 &  0.68 \\
 5 & GW190413\_134308 & 04-13-2019 & 13:43:08 &  0.99 &&  5 & GW191126\_115259 & 11-26-2019 & 11:52:59 &   0.70 \\
 6 & GW190421\_213856 & 04-21-2019 & 21:38:56 & $>$0.99 &&  6 & GW191127\_050227 & 11-27-2019 & 05:02:27 &  0.74 \\
 7 & GW190425 & 04-25-2019 & 08:18:05 &  0.69 &&  7 & GW191129\_134029 & 11-29-2019 & 13:40:29 & $>$0.99 \\
 8 & GW190426\_190642 & 04-26-2019 & 19:06:42 &  0.73 &&  8 & GW191204\_110529 & 12-04-2019 & 11:05:29 &  0.74 \\
 9 & GW190503\_185404 & 05-03-2019 & 18:54:04 & $>$0.99 &&  9 & GW191204\_171526 & 12-04-2019 & 17:15:26 & $>$0.99 \\
10 & GW190512\_180714 & 05-12-2019 & 18:07:14 & $>$0.99 && 10 & GW191215\_223052 & 12-15-2019 & 22:30:52 & $>$0.99 \\
11 & GW190513\_205428 & 05-13-2019 & 20:54:28 & $>$0.99 && 11 & GW191216\_213338 & 12-16-2019 & 21:33:38 & $>$0.99 \\
12 & GW190514\_065416 & 05-14-2019 & 06:54:16 &  0.75 && 12 & GW191219\_163120 & 12-19-2019 & 16:31:20 &  0.82 \\
13 & GW190517\_055101 & 05-17-2019 & 05:51:01 & $>$0.99 && 13 & GW191222\_033537 & 12-22-2019 & 03:35:37 & $>$0.99 \\
14 & GW190519\_153544 & 05-19-2019 & 15:35:44 & $>$0.99 && 14 & GW191230\_180458 & 12-30-2019 & 18:04:58 &  0.96 \\
15 & GW190521 & 05-21-2019 & 03:02:29 & $>$0.99 && 15 & GW200112\_155838 & 01-12-2020 & 15:58:38 & $>$0.99 \\
16 & GW190521\_074359 & 05-21-2019 & 07:43:59 & $>$0.99 && 16 & GW200115\_042309 & 01-15-2020 & 04:23:09 & $>$0.99 \\
17 & GW190527\_092055 & 05-27-2019 & 09:20:55 &  0.83 && 17 & GW200128\_022011 & 01-28-2020 & 02:20:11 & $>$0.99 \\
18 & GW190602\_175927 & 06-02-2019 & 17:59:27 & $>$0.99 && 18 & GW200129\_065458 & 01-29-2020 & 06:54:58 & $>$0.99 \\
19 & GW190620\_030421 & 06-20-2019 & 03:04:21 &  0.99 && 19 & GW200202\_154313 & 02-02-2020 & 15:43:13 & $>$0.99 \\
20 & GW190630\_185205 & 06-30-2019 & 18:52:05 & $>$0.99 && 20 & GW200208\_130117 & 02-08-2020 & 13:01:17 & $>$0.99 \\
21 & GW190701\_203306 & 07-01-2019 & 20:33:06 & $>$0.99 && 21 & GW200208\_222617 & 02-08-2020 & 22:26:17 &   0.70 \\
22 & GW190706\_222641 & 07-06-2019 & 22:26:41 & $>$0.99 && 22 & GW200209\_085452 & 02-09-2020 & 08:54:52 &  0.97 \\
23 & GW190707\_093326 & 07-07-2019 & 09:33:26 & $>$0.99 && 23 & GW200210\_092254 & 02-10-2020 & 09:22:54 &  0.54 \\
24 & GW190708\_232457 & 07-08-2019 & 23:24:57 & $>$0.99 && 24 & GW200216\_220804 & 02-16-2020 & 22:08:04 &  0.77 \\
25 & GW190719\_215514 & 07-19-2019 & 21:55:14 &  0.91 && 25 & GW200219\_094415 & 02-19-2020 & 09:44:15 & $>$0.99 \\
26 & GW190720\_000836 & 07-20-2019 & 00:08:36 & $>$0.99 && 26 & GW200220\_061928 & 02-20-2020 & 06:19:28 &  0.62 \\
27 & GW190725\_174728 & 07-25-2019 & 17:47:28 &  0.96 && 27 & GW200220\_124850 & 02-20-2020 & 12:48:50 &  0.83 \\
28 & GW190727\_060333 & 07-27-2019 & 06:03:33 & $>$0.99 && 28 & GW200224\_222234 & 02-24-2020 & 22:22:34 & $>$0.99 \\
29 & GW190728\_064510 & 07-28-2019 & 06:45:10 & $>$0.99 && 29 & GW200225\_060421 & 02-25-2020 & 06:04:21 & $>$0.99 \\
30 & GW190731\_140936 & 07-31-2019 & 14:09:36 &  0.83 && 30 & GW200302\_015811 & 03-02-2020 & 01:58:11 &  0.91 \\
31 & GW190803\_022701 & 08-03-2019 & 02:27:01 &  0.97 && 31 & GW200306\_093714 & 03-06-2020 & 09:37:14 &  0.81 \\
32 & GW190805\_211137 & 08-05-2019 & 21:11:37 &  0.95 && 32 & GW200308\_173609 & 03-08-2020 & 17:36:09 &  0.86 \\
33 & GW190814 & 08-14-2019 & 21:10:39 & $>$0.99 && 33 & GW200311\_115853 & 03-11-2020 & 11:58:53 & $>$0.99 \\
34 & GW190828\_063405 & 08-28-2019 & 06:34:05 & $>$0.99 && 34 & GW200316\_215756 & 03-16-2020 & 21:57:56 & $>$0.99 \\
35 & GW190828\_065509 & 08-28-2019 & 06:55:09 & $>$0.99 && 35 & GW200322\_091133 & 03-22-2020 & 09:11:33 &  0.62 \\
36 & GW190910\_112807 & 09-10-2019 & 11:28:07 & $>$0.99 && &&&& \\
37 & GW190915\_235702 & 09-15-2019 & 23:57:02 & $>$0.99 && &&&& \\
38 & GW190916\_200658 & 09-16-2019 & 20:06:58 &  0.62 && &&&& \\
39 & GW190917\_114630 & 09-17-2019 & 11:46:30 &  0.74 && &&&& \\
40 & GW190924\_021846 & 09-24-2019 & 02:18:46 & $>$0.99 && &&&& \\
41 & GW190925\_232845 & 09-25-2019 & 23:28:45 &  0.99 && &&&& \\
42 & GW190926\_050336 & 09-26-2019 & 05:03:36 &  0.51 && &&&& \\
43 & GW190929\_012149 & 09-29-2019 & 01:21:49 &  0.86 && &&&& \\
44 & GW190930\_133541 & 09-30-2019 & 13:35:41 & $>$0.99 && &&&& \\ 
     \hline\hline
    
    \end{tabular}
    
    \label{tab:GWevents}
\end{table*}

\begin{table*}[ht]
    \centering
    \caption{Marginal GWs from O3 without clear instrumental or environmental causes \citep{GWTC2.1, GWTC-3}}
    \begin{tabular}{Hlcccc}
    \hline
    & Event Name & Date & Time (UTC) & $\pastro$  \\ \hline\hline
1 & GW190426\_152155 & 04-26-2019 & 15:21:55 & 0.14  \\
2 & GW190531\_023648 & 05-31-2019 & 02:36:48 & 0.28  \\
3 & GW191118\_212859 & 11-18-2019 & 21:28:59 & 0.05  \\
4 & GW200105\_162426 & 01-05-2020 & 16:24:26 & 0.36  \\
5 & GW200201\_203549 & 02-01-2020 & 20:35:49 & 0.12  \\
6 & GW200311\_103121 & 03-11-2020 & 10:31:21 & 0.19  \\ \hline\hline
    
    \end{tabular}
    
    \label{tab:margGWevents}
\end{table*}

\subsection{\FermiGBM\ Searches}\label{ssec:method_gbm}
\FermiGBM\ has 12 sodium iodide (NaI) and 2 bismuth germanate (BGO) detectors that  are strategically positioned to cover the full sky, unocculted by the Earth
\citep{Meegan_2009}. 
The flight software on-board \FermiGBM\ triggers on an event when there is an influx of gamma rays at a level greater than $4.5 \sigma$ above the background rate in at least two 
NaI detectors
\citep{Paciesas_2012}.
Additionally, the downlink of CTTE data enables searches for GRBs below \FermiGBM's on-board triggering threshold using ground-based computing resources. With 2~\textmu s timing resolution and full coverage of the unocculted sky over the energy range from 8 keV to 40 MeV, CTTE data has significantly expanded the sensitivity of the \FermiGBM\ instrument and its sub-threshold searches.

\subsubsection{Untargeted Search}\label{ssec:untargeted_search}
The \FermiGBM\ Untargeted Search is a blind search that automatically scans the CTTE data for significant count rate increases in at least two NaI detectors. The algorithm was originally developed for detecting terrestrial gamma-ray flashes \citep{briggs_2013} and has since been adapted to search for short GRBs with fluxes below the on-board triggering threshold.
The Untargeted Search runs through eighteen timescales ranging from 64 ms to 31 s and five energy bins from 27 keV to 985 keV, and short GRB candidates are identified when at least two detectors exceed $2.5\sigma$ and $1.25\sigma$ above the background rate.
Each candidate is given a reliability score based on whether the geometry of the detectors with significant flux is consistent with the observation of a distant astrophysical source.
Currently, short GRB candidates with durations less than $2.8$ s and reliability classifications of low, medium, and high are publicly distributed via GCN.\footnote{\url{https://gcn.gsfc.nasa.gov/fermi_gbm_subthresh_archive.html}}

In this work, we combine short GRB candidates detected by the Untargeted Search with GBM-triggered GRBs and examine their temporal offsets from the  GWs listed in Table \ref{tab:GWevents}. 
Theoretical models predict the temporal offset between merger time and the production of gamma-rays to range from 0.01 s to 10 s depending on the conditions producing the gamma-ray emission \citep{zhang_2019}. 
For GRB 170817A, the only known short GRB associated with a GW, the temporal offset was 1.7~s with a duration ($T_{90}$) over which 5--95\% of the GRB flux (50-300 keV) was detected of 2~s \citep{Abbott+17170817GWGRB}. 
This is consistent with a range of physically viable scenarios (e.g., \citealt{Lin2018_timeoffset, Salafia2018_timeoffset, Zhang2018_timeoffset}) where the temporal offset is correlated with burst duration.
We therefore choose to subtract the burst duration timescale from the temporal offset when performing our analysis. Doing so increases the observed significance of simulated short GRB counterparts and
yields no loss in detection sensitivity at the 3$\sigma$ level in alternative scenarios where the temporal offset is the same for all simulated GRBs.

After calculating the time offsets for each GW--GRB pair minus the burst duration, the smallest resulting time offset for each GW is taken. For GBM-triggered GRBs, we use the $T_{90}$ as a measure of the duration. For GRB candidates from the Untargeted Search, we use the most significant timescale over which the GRB candidate was detected, which scales linearly with $T_{90}$ for on-board triggered GRBs. A background distribution is produced in the same way by replacing the observed GW times with random times during which there are no reported GW signals. This yields a distribution of temporal offsets minus the burst duration between unrelated GWs and the GRB sample. In both the search and background samples, positive and negative time offsets are allowed, with no maxima imposed. GW triggers occurring during Fermi passage through South Atlantic Anomaly (SAA) are also included. See the results presented in Section~\ref{ssec:results_gbm} for a comparison of the cumulative signal and background distributions.

\subsubsection{GBM Targeted Search}\label{sec:TS_method}

The \FermiGBM\ Targeted Search was developed for multimessenger follow-up observations \citep{Blackburn2015}. It uses CTTE data to scan around an external trigger time for gamma-ray emission typical of a short GRB. For follow-up of the GWs in Table~\ref{tab:GWevents}, we search from $-1$~s to +30~s around the GW time to ensure we do not miss unexpectedly delayed gamma-ray emission from a counterpart short GRB, even after accounting for temporal offsets up to 10~s relative to the GW time. Starting $1$~s before the GW time provides a comfortable buffer to account for the fact that the trigger times can vary by a few milliseconds for GW signals that are identified by multiple pipelines. The scan is repeated for eight characteristic emission timescales which increase by factors of 2 from 64~ms to 8.192~s. Each emission timescale begins the search centered at the start of the scan window and then advances until the end using a fixed time step size. Emission timescales greater than 256~ms use a time step equal to one-eighth the total emission duration. The remaining emission timescales use a 64 ms step size to limit both the additional trials and the additional computational time associated with the shorter emission timescales.

The Targeted Search achieves greater sensitivity than the on-board triggering algorithm by processing the data from all 14 detectors coherently rather than focusing on significant signals present in detector pairs. This allows for the detection of weaker signals below the \FermiGBM\ on-board triggering threshold \citep{Kocevski2018}. To do this, three spectral templates representing spectrally hard, normal, and soft GRBs (Table~\ref{tab:targeted_search_templates}) are folded through the GBM detector responses to produce an expected count rate for a given astrophysical source location and flux. This expected count rate is then compared to the observed counts through a log-likelihood ratio,
\begin{equation}
\mathcal{L}_j(d, s) = \sum_i \Big[ \mathrm{ln} \frac{\sigma_{n_i}}{\sigma_{d_i}} + \frac{\tilde{d}_i^2}{2 \sigma_{n_i}^2} - \frac{(\tilde{d}_i - r_{i,j} s)^2}{2 \sigma_{d_i}^2} \Big],
\end{equation}
where $\tilde{d}_i$ represents the background-subtracted measurements in each detector, $\sigma_n$ is the standard deviation of the background measurement, $\sigma_{d_i}$ is the standard deviation of the expected data (background$+$signal), $r_{i,j}$ is the location-dependent instrumental response for the spectrum denoted by index $j$, and $s$ is the intrinsic source photon flux at the Earth. See \cite{Blackburn2015} for a full derivation.

\begin{table*}
    \centering
    \caption{Spectral templates used by the \FermiGBM\ Targeted Search. 
             }
    \begin{tabular}{ccl}
    \hline
    Template & Type & \multicolumn{1}{c}{Parameters} \\ \hline\hline
hard & Cut-off Power-law \citep{goldstein2016updates} & $E_\mathrm{peak} = 1500$ keV, $\alpha = -1.5$ \\
normal & Band \citep{Band1993} & $E_\mathrm{peak} = 230$ keV, $\alpha = -1.0$, $\beta = -2.3$ \\
soft & Band \citep{Band1993} & $E_\mathrm{peak} = 70$ keV, $\alpha = -1.9$, $\beta = -3.7$ \\ \hline\hline
    \end{tabular}
    \label{tab:targeted_search_templates}
\end{table*}

The log-likelihood ratio quantifies the probability that an astrophysical source is present versus a background-only hypothesis. It is first computed separately for each point on the sky and spectral template at a given time and emission duration. During this process we estimate the best-fit photon flux for each spectral template by finding the value $s_\mathrm{best}$ that maximizes the log-likelihood ratio. 
Since the best-fit photon flux maximizes the likelihood, which is effectively a product of Gaussian distributions, the variance on this photon flux equals the variance of the likelihood:
\begin{equation}
    \sigma_{\mathcal{L}_j}^2 = \frac{1}{\sum r_{i,j}^2 / \sigma_{d_i}^2},
\end{equation}
where $\sigma_{d_i}$ includes both background and source contributions, with the latter evaluated at $s_\mathrm{best}$. Signal injection studies using the normal spectral template demonstrated that this formulation yields the expected error coverage levels for true source fluxes near $1\times10^{-7}$ erg~cm$^{-2}$~s$^{-1}$ and below, which is the relevant flux range for this sub-threshold analysis in \FermiGBM.

We marginalize the log-likelihood ratio over all possible source amplitudes using a modified power law prior designed to both avoid divergence and to produce a luminosity distribution for the observed source flux that is invariant with respect to source distance \citep{Blackburn2015}:
\begin{equation}
    P(s) = \Big[1 - e^{-(s/2.5 \sigma_\mathcal{L})^{-1}}\Big] s^{-1} \,.
\end{equation}
The net result is a hypothesis test formulation following Bayes' theorem.

The amplitude-marginalized log-likelihood ratios for individual spectral templates, $\mathcal{L'}_j(d)$, at each time and duration are then averaged over all sky positions and templates using a uniform prior to formulate the full marginal log-likelihood ratio,
\begin{equation}
    \Lambda = \sum_{j=1}^3\frac{1}{3} \int \frac{\mathcal{L'}_j(d)}{4 \pi} \, \mathrm{d}\Omega,
\end{equation}
where the sum over $j$ covers the hard, normal, and soft spectral templates. The marginal results from all scanned times and durations are then sorted according to the largest value of $\Lambda$ after filtering out known detector effects. 

Localization maps estimating the probability of finding the true source location at each point on the sky are produced for the top ranking candidates using the log-likelihood ratio of the best-fitting spectrum for each candidate. This is done by noting that the log-likelihood ratio asymptotically approaches the behavior of a $\chi^2$ distribution according to Wilks' theorem \citep{Wilks} with a statistical probability given by
\begin{equation}
    P \propto \mathrm{exp}[\mathcal{L'}_j(d)].
\end{equation}
The statistical probability is then convolved with Gaussian kernels to account for systematic errors, which are predominantly induced by the difference between the true source spectrum and the three spectral templates, imperfect knowledge of the detector response, and whether atmospheric scattering is taken into account for a given spacecraft rocking angle. As a final step, we set the region blocked by the Earth in \FermiGBM\ to zero and re-normalize the map to account for the fact that gamma-ray sources are not visible through the Earth and an implicit assumption that the signal has a non-terrestrial origin.

The Targeted Search method was previously used to search for sub-threshold counterparts to GWs identified during the O1 and O2 observing runs \citep{Hamburg+2020paper}. A number of improvements were made to it
in preparation for O3 \citep{Goldstein2019}:
\begin{enumerate}
    \item Removal of the lowest 4--12 keV energy channel in the NaI detector data helped remove detector noise as well as Galactic transients.
    \item Better background fitting during approach and exit from the SAA. This reduces local particle background triggers that were present in about 1\% of searches and formed the dominant non-GRB background in the high log-likelihood ratio parameter space.
    \item Better detector response models with a more complete treatment of the effects from the back-scattering of high energy gamma-ray photons off the Earth's atmosphere. The atmospheric scattering effects are currently applied when the zenith of \FermiGBM\ is within $\pm 5^\circ$ of its nominal rocking angle of $130^\circ$ with respect to the Earth's geocenter, which occurs for $\sim$70\% of measurements.
    \item Decreasing the resolution from 1$^\circ$ to 5$^\circ$ for the grid of sky positions analyzed during the search. This provided an order of magnitude improvement in execution time with no notable loss in sensitivity or degradation of localization capability.
\end{enumerate}
These changes necessitated a recalculation of the estimated systematic uncertainty applied to the localization maps generated for the top-ranking search candidates. An initial study of 34 sub-threshold short GRB detections modeled this uncertainty as a 2.7$^\circ$ Gaussian systematic \citep{Goldstein2019}. A more detailed model was developed for this work using a larger sample of 3,000 simulated short GRB detections. It consists of a weighted pair of Gaussian shapes normalized over the sky with the standard deviation $\sigma_1$ of the first Gaussian always smaller than that of the second Gaussian, $\sigma_2$. The parameters of each Gaussian were determined as functions of the most probable zenith angle for each candidate and the spacecraft rocking angle relative to the Earth. They range from $1.6^\circ$--$6.0^\circ$ for $\sigma_1$ and $6.4^\circ$--$60.4^\circ$ for $\sigma_2$, with the fractional contribution of the first Gaussian spanning 0.42--0.77.

\subsubsection{GBM Targeted Search Ranking Statistic}\label{ssec:ranking_statistic}
We use a ranking statistic $R$ to characterize the significance of a coincidence between the GW candidates from the catalog and the short GRB candidates found by the Targeted Search. Following the formulation in 
\cite{Hamburg+2020paper}, the statistic takes into account the probability of astronomical origin of the GW, $\pastro$; the fraction of the GW localization not occulted by the Earth for \FermiGBM, $p_\mathrm{visible}$; the time offset of the GRB candidate from the GW time, $\Delta t$; and the FAR from the best-fitting spectral template of the GRB candidate, $\mathrm{FAR}_\mathrm{GBM}$. We update the formulation to include the spatial association probability $p_\mathrm{assoc}$ and the duration $D$ of the gamma-ray emission:
\begin{equation}\label{eq:rank_stat}
    R = \frac{\pastro \times p_\mathrm{visible} \times p_\mathrm{assoc}}{|\Delta t - D| \times \mathrm{FAR}_\mathrm{GBM}}.
\end{equation}

\noindent The spatial association probability $p_\mathrm{assoc}$ quantifies whether the localizations of a sub-threshold gamma-ray candidate and GW are consistent with being produced by the same source. It is computed according to
\begin{equation}
S = \int \rho_\mathrm{GBM} \, \rho_\mathrm{GW} \, \mathrm{d} \Omega \, , \, B = \int \rho_\mathrm{GBM} \, \rho_\mathrm{uniform} \, \mathrm{d} \Omega,
\end{equation}
\begin{equation}
    p_\mathrm{assoc} = \frac{S}{S + B},
\end{equation}
where $S$ represents a signal hypothesis with both localizations produced by the same source and $B$ denotes a background hypothesis where the localizations are unrelated. Both $S$ and $B$ are constructed from integrals over all sky positions. In this context, $\rho_\mathrm{GBM}$ is the probability density per unit area reported by the localization maps produced for gamma-ray candidates identified by the Targeted Search. Likewise, $\rho_\mathrm{GW}$ is the probability density of the localization maps produced for each GW and $\rho_\mathrm{uniform} = 1/4\pi$ is the unit density of a uniform spatial distribution on the sky.

The duration of gamma-ray emission $D$ is incorporated into the temporal weight, 
\begin{equation}
    \frac{1}{|\Delta t - D|},
\end{equation}
where $\Delta t$ is the temporal offset between the GW time and the start of the candidate gamma-ray emission identified by the Targeted Search and $D$ is the candidate emission timescale. As discussed in Section~\ref{ssec:untargeted_search}, 
this is designed to account for scenarios where the observed temporal offset scales with burst duration, which is expected from a broad range of models describing the observations of GW170817/GRB 170817A.
The best-fit value of $D$, given by the candidate with the largest value of $\Lambda$, is a good proxy for burst duration because it scales proportionally with $T_{90}$ when the Targeted Search is applied to confirmed short GRBs in \FermiGBM.

We enforce a minimum value of $|\Delta t -D| =1$ ms to avoid divergence and account for the millisecond scale uncertainty between the GW merger times of signals identified by multiple pipelines. We also apply a minimum value of $\mathrm{FAR}_\mathrm{GBM}$ = 6.43$\times10^{-6}$~Hz. This is equal to observing a single GRB candidate over the length of the background sample used to compute $\mathrm{FAR}_\mathrm{GBM}$.

We tested the impact of these updates to the ranking statistic by using the \FermiGBM\ response generator\footnote{\url{https://fermi.gsfc.nasa.gov/ssc/data/analysis/rmfit/DOCUMENTATION.html}} to inject short GRBs into CTTE data from the locations of modeled BNS mergers in \cite{2020LRR....23....3A}. The start time of each GRB was offset from the GW time using the duration of each burst, as given by $T_{90}$. 
We then applied the Targeted Search to this dataset and ranked the candidates according to Equation~\ref{eq:rank_stat} as well as the older method from \cite{Hamburg+2020paper}. Doing so resulted in a factor of 1.7 increase in the number of joint detections relative to the ranking statistic formulation from our older method.

Since the true time offset model is not known, we repeated the GRB injection study using the following alternative models for the start time of the injected GRB relative to the GW:
\begin{enumerate}
    \item Offset of half $T_{90}$ to test a scaling factor less than the total burst duration.
    \item No time offset to bound emission scenarios where the GRB occurs a few ms after the GW \citep{zhang_2019}.
    \item Fixed offset of 0.5~s assuming most gamma-ray counterparts have a characteristic time delay which is half the median $T_{90}$ of short GRBs observed in \FermiGBM.
\end{enumerate}
These models were chosen with a bias towards testing time offsets shorter than the typical duration of a short GRB since the inclusion of $D$ in the updated temporal weight naturally performs better at longer time offsets than the $1/|\Delta t|$ weight used in \cite{Hamburg+2020paper}. The updated ranking statistic outperformed the older method in all scenarios, albeit with a smaller increase in the relative number of joint detections compared to the scenario where temporal offset scales with burst duration.

\subsubsection{\FermiGBM\ Flux Upper Limits}\label{ssec:method_gbm_ul}
For GW signals without a significant counterpart detection in \FermiGBM\ we compute the gamma-ray flux upper limits as a function of sky position using the Targeted Search because it is the most sensitive analysis method employed by \FermiGBM. To do this, we use the normal spectral template from Table~\ref{tab:targeted_search_templates} and the 1~s gamma-ray emission duration from the Targeted Search since they are characteristic of typical short GRBs \citep{von_Kienlin_2020,Poolakkil_2021}. This results in a set of upper limits for each sky position at times ranging from $-1$~s  to +30~s around the GW time. We then choose the maximum observed upper limit measurement for each sky position, guaranteeing that the specified confidence level of the upper limit applies over the entire search period.

We construct the upper limits from the best-fit photon flux amplitude $s_\mathrm{best}$ and its Gaussian error $\sigma_\mathcal{L}$ discussed in Section~\ref{sec:TS_method} according to
\begin{equation}
    S_{\mathrm{UL}} = s_\mathrm{best} + N \times \sigma_\mathcal{L},
\end{equation}
where $N$ is the significance level of the upper limit. We use a 3$\sigma$ upper limit level for reporting upper limits over the full 10--1000 keV energy range of standard GRB flux measurements in \FermiGBM\ following the convention established in \cite{Goldstein2019}. We also compute a second 5$\sigma$ upper limit over a 15--350 keV range to match the convention used by \SwiftBAT\ (see Section~\ref{ssec:method_swift_ul}) when combining the upper limits from both instruments.

\subsection{\SwiftBAT\ Searches}\label{ssec:method_swift}
\SwiftBAT\ is a coded-aperture, large FoV (2.2 sr at $\>$10\% coding fraction), hard X-ray instrument on-board \textit{Swift}. Its detector plane contains 32,768 CZT detector elements, positioned under a coded aperture mask and a graded-Z fringe shield that helps lower the background rate \citep{Barthelmy_2005}. The BAT 
covers an energy range from 15~keV to 350~keV and monitors large portions of the sky with the goal of detecting GRBs. Once triggered, the BAT can localize a GRB to 1--3~arcmin accuracy, prompting the \textit{Swift} spacecraft to slew and point its two narrow-field instruments---the X-ray Telescope (XRT; \citealt{Burrows2005}) and the Ultraviolet/Optical Telescope (UVOT; \citealt{Roming2005})---for follow-up observations. The BAT's localization accuracy is quantified by the instrument's partial coding fraction, i.e. the fraction of detectors exposed to an event at a given time and sky position. If the coding fraction for a given trigger is 0\%, then BAT will not be able to localize the event. The BAT averages $\sim$90 GRB on-board triggers per year, among which $\sim$10\% are short in duration \citep{Gehrels_2009}. The on-board GRB triggers are complemented by subsequent on-ground rates and image data processing, in turn allowing for dedicated searches for GRB emission. With no GWs from Table~\ref{tab:GWevents} triggering an on-board BAT detection, we conduct an offline follow-up analysis from the ground to search for the corresponding hard X-ray counterpart emission.

\subsubsection{\SwiftBAT\ Rates Data Search}
The BAT flight software continuously assesses the signal-to-noise ratio (SNR) of the observed count rates.
If an SNR exceeds the given threshold value determined by a number of rate-trigger criteria \citep{Fenimore2004}, the triggering algorithm subsequently checks the corresponding image data for the final confirmation and the localization of the potential burst. The detection is confirmed only if the image SNR threshold is surpassed ($\gtrsim 6.5$) and no other sources have been previously reported at the event localization. 
For every confirmed detection, BAT records event data containing counts’ arrival times, location on the detector plane, and energy. With its large effective area ($\sim$2600~cm$^2$ for 100~keV photon detection at launch), the event data volume collected by BAT is too big to be stored on-board and, due to the limitations of the \textit{Swift} downlink bandwidth, it is not possible to transfer all the event data to the ground. As such, until recently, the only way to conduct an offline, on-ground follow-up analysis of untriggered and sub-threshold events relied upon the rates light curves in four energy channels (15--25~keV, 25--50~keV, 50--100~keV, and 100--350~keV) with three time binnings (64~ms, 1~s, and 1.6~s) and their corresponding 64~s images in a single energy bin (15--50~keV). The recently developed Gamma-ray Urgent Archiver for Novel Opportunities (GUANO) technique circumvents this issue by retrieving BAT event data extending to $\sim$200~s long windows surrounding the trigger times from various astrophysical events (e.g., GWs, GRBs, fast radio bursts, neutrinos, etc.; \citealt{GUANO}).
However, in this paper we do not use the GUANO technique since a significant number of the considered GW triggers were detected prior to the GUANO deployment in December 2019. Instead, we conduct the analysis using the regular rates data from BAT and leave the analogous study using the GUANO data for the future GW observing runs.
\par To conduct the untriggered and sub-threshold search for hard X-ray counterparts coincident with the LVK triggers in Table~\ref{tab:GWevents}, we developed a code analogous to \cite{Lien2014}.
The search process begins by extracting the raw light curves from within the central region of the BAT FoV binned in 64~ms, 1~s, and 1.6~s time intervals. We opt to use the 1~s binned data to calculate the average background rate and standard deviation, $\sigma_{\text{bg}}$, starting at $-1$~s before the GW trigger time, and extending to $+30$~s after. Using the raw light curves, we compute the average background rate, $r_{\text{bg}}$, spanning a time window outside the signal interval, and spanning $\sim 800$~s (excluding the instrument slews or SAA). The signal significance, $S$, is then computed from $\sigma_{\text{bg}}$, using
\begin{equation}
S = (r_{\text{sig}} - r_{\text{bg}}) / \sigma_{\text{bg}},
\end{equation}
where $r_{\text{sig}}$ is the threshold signal rate. The background uncertainty is estimated as $\sigma_{\text{bg}} =  \sqrt{\frac{1}{N} \sum_{i=1}^{N} (r_{\text{bg},i} - \overline{r_{\text{bg}}})^2}$, where $N$ is the number of data points in the considered portion of the lightcurve, $r_{\text{bg},i}$ is the $i^{\text{th}}$ background rate measurement, and $\overline{r_{\text{bg}}}$ is the mean background rate over the considered time interval. 
Furthermore, we visually inspect each light curve to ensure that no peaks originate from detector noise. Once this is done, we check whether there are any potential counterparts to the GW, defined as a $\geq 5\sigma$ detection above background.

\subsubsection{NITRATES Response Functions}\label{ssec:method_swift_nitrates}
To produce the BAT instrument response functions appropriate for converting from photon counts to a source flux in the rate data domain, we use the Non-Imaging Transient Reconstruction And TEmporal Search (NITRATES; \citealt{DeLaunay_2021}). The NITRATES response modeling takes into account both coded and uncoded parts of the detector, and thus includes responses appropriate for all counts recorded in the rates data. In addition, these responses allow for potential GRB detection from outside the BAT’s coded FoV, as well as higher sensitivity across the entire FoV. The instrument responses were created by simulating photon beams onto the \textit{Swift} Mass Model (SwiMM) using \textsc{Geant4}, a particle-interaction simulator software toolkit \citep{Allison_2016}. We produce Detector Response Matrices (DRMs) for 31 different incident directions, covering the $\sim2.2$~sr sky area (corresponding to the 10-\% coding fraction) where the responses are well calibrated. For a complete description of the BAT instrument response modeling, see Sec.~5 and Appendix A in \cite{DeLaunay_2021}.

\subsubsection{\SwiftBAT\ Flux Upper Limits}\label{ssec:method_swift_ul}
For GWs without a 5$\sigma$ detection above background in \SwiftBAT, we estimate the flux upper limit from the observed photon counts.
We compute these limits for each time bin in the search by calculating the necessary number of counts that would result in a 5$\sigma$ detection at that time from the estimated background uncertainty $\sigma_{\text{bg}}$, assuming a $1$~s emission duration. We then select the largest counts value obtained in the search bins for each GW since this is guaranteed to satisfy the 5$\sigma$ criteria over the full search window. We convert the photon counts to flux upper limits within the partially coded BAT FoV, over a 15--350 keV energy range as a function of sky position by applying the NITRATES instrument response functions using the  normal spectral template (Table~\ref{tab:targeted_search_templates}) employed for upper limits computed with the \FermiGBM\ Targeted Search in Section~\ref{ssec:method_gbm_ul}. This is done over 31  locations in a grid covering the $\sim2.2$~sr BAT FoV.

\subsection{Combined \& Marginal Flux Upper Limits}\label{ssec:marg_limits}
For GWs without a detected counterpart in \FermiGBM\ or \SwiftBAT, we combine the 5$\sigma$ confidence level flux upper limits described in Sections~\ref{ssec:method_gbm_ul} \&~\ref{ssec:method_swift_ul} to produce joint flux upper limits as a function of sky position using both instruments. We do this by selecting the more constraining upper limit at each position since the individual limits result from independent measurements. This allows us to provide a single upper limit map for each GW that simultaneously leverages the wide FoV provided by \FermiGBM\ as well as the additional coverage and enhanced sensitivity of \SwiftBAT.

We also provide marginalized flux upper limits ($S_{\mathrm{UL,marg}}$) that we compute by integrating the upper limits over the sky using the probability density of the GW localization as a prior
\begin{equation}
    S_\mathrm{UL,marg} = \int S_\mathrm{UL} \, \tilde{\rho}_\mathrm{GW} \, \mathrm{d} \Omega,
\end{equation}
where $S_\mathrm{UL}$ is the position-dependent upper limit at a given confidence level and $\tilde{\rho}_\mathrm{GW}$ is the probability density of the GW localization normalized over the visible portion of the sky. This reduces the set of upper limits for each GW to a single, characteristic upper limit that accounts for the most likely location of the GW source.

\subsection{Isotropic-Equivalent Luminosity Upper Limits}\label{ssec:liso_limits}
We compute upper limits on the isotropic-equivalent gamma-ray luminosity $L_\mathrm{iso}$ in the cosmological rest-frame energy range of 1 keV--10 MeV for GWs without a detected counterpart in \FermiGBM\ or \SwiftBAT\ according to
\begin{equation}
    L_\mathrm{iso} = 4 \pi \, D_\mathrm{L}^2 \, S_\mathrm{UL,marg,} \, k,
\end{equation}
where $D_\mathrm{L}$ is the median luminosity distance of the GW, $S_\mathrm{UL,marg}$ is the marginalized confidence level flux upper limit described in Section~\ref{ssec:marg_limits}, and $k$ is the standard bolometric correction factor given by
\begin{equation}
    k \equiv 
    \frac{ \int_{1 \, \mathrm{keV} / 1+z}^{10 \, \mathrm{MeV} / 1+z} E \, \frac{\mathrm{d}N}{\mathrm{d}E}(E) \, \mathrm{d}E }{
    \int_{15 \, \mathrm{keV}}^{350 \, \mathrm{keV}} E \, \frac{\mathrm{d}N}{\mathrm{d}E}(E) \, \mathrm{d}E },
\end{equation}
where $z$ is the redshift inferred from $D_\mathrm{L}$. In this case, $\frac{\mathrm{d}N}{\mathrm{d}E}(E)$ represents the normal spectral shape from Table~\ref{tab:targeted_search_templates}, which is used to generate the marginalized flux upper limit. 
We chose to use the median $D_\mathrm{L}$ and the marginalized flux upper limit for each GW rather than marginalizing $L_\mathrm{iso}$ from the values estimated at each sky position in order to exclude the low-likelihood modes of the distance luminosity posteriors for GW200308\_173609 and GW200322\_091133, as discussed in \cite{GWTC-3}.
All other GWs yield similar values for $L_\mathrm{iso}$ regardless of whether we use the individual or median values for $D_\mathrm{L}$ in the calculation.

\section{Results}\label{sec:results}
This section presents the results for the searches from \FermiGBM\ and \SwiftBAT\ for coincident gamma-ray emission to the GW candidates presented in Tables \ref{tab:GWevents} and \ref{tab:margGWevents}.

\subsection{Triggered and Untargeted Search Results}\label{ssec:results_gbm}
\begin{figure}[t!]
    \includegraphics[width=0.45\textwidth]{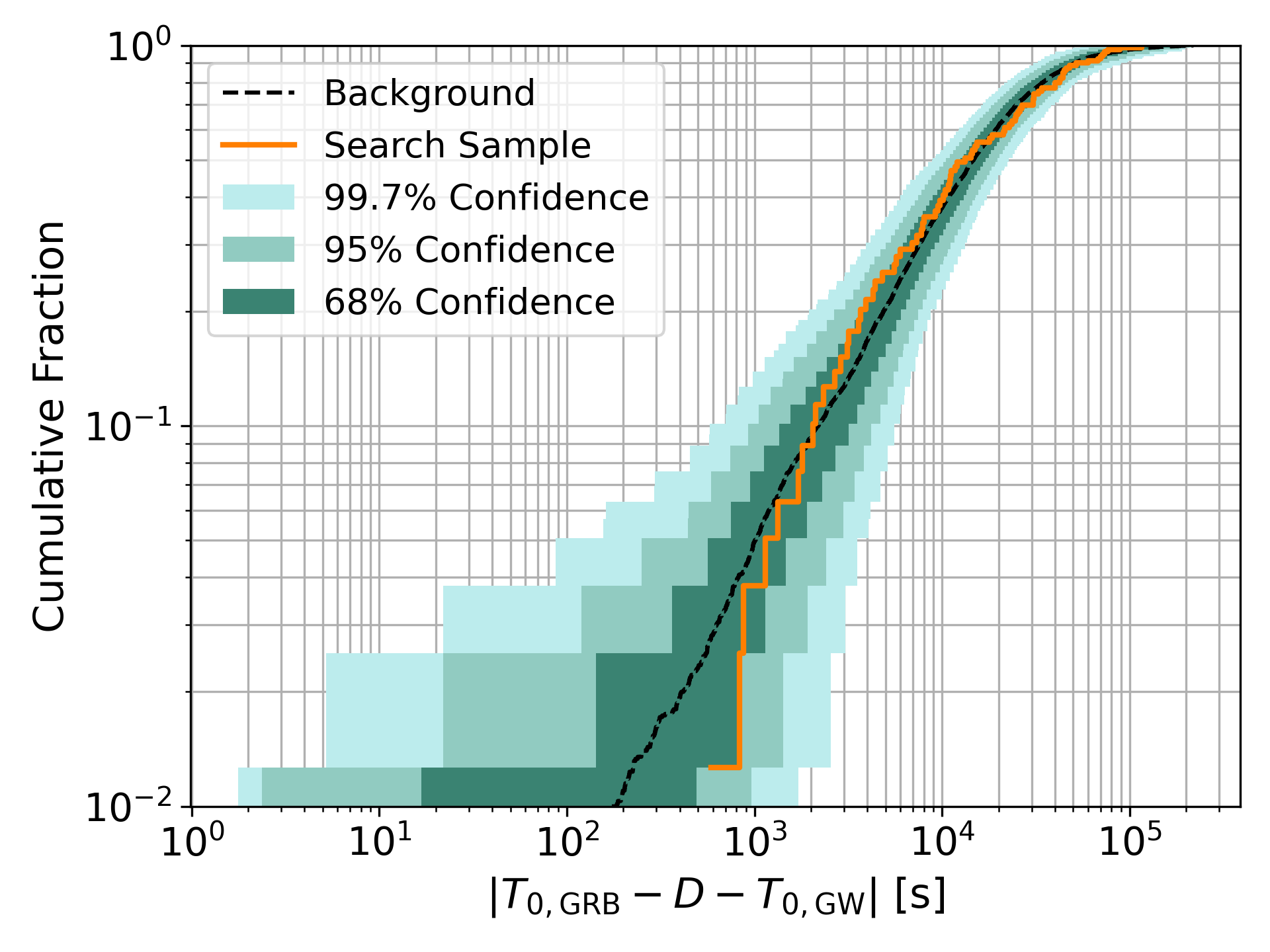}
    \caption{The cumulative distribution for the minimum time offsets between the O3 GW triggers and GRBs found by either the GBM on-board triggering algorithms or the Untargeted Search.
    Confidence intervals for the search sample were determined by Monte Carlo sampling of the background.}
    \label{fig:time_offsets}
\end{figure}

We compare the distribution of temporal offsets between the GWs listed in Table~\ref{tab:GWevents} and the closest gamma-ray signal in \FermiGBM, minus its burst duration, shown in Figure \ref{fig:time_offsets}. The GRB sample here comprises 214 GRBs triggered by \FermiGBM\ during O3 and 
479 short GRB candidates detected by the Untargeted Search. The background distribution was composed by choosing $\sim10^4$ random times in O3 during which there were no reported GWs. Confidence intervals for the search sample were determined empirically by Monte Carlo sampling of the background distribution. 

There are no significant deviations from the background, with the search sample lying largely within the 68\% confidence interval. The shortest temporal offset given by the sample distribution is approximately 10 minutes and is within the 95\% confidence interval. Such large offsets are not expected for on-axis prompt emission from GRBs associated with binary mergers \citep{Vedrenne2009, zhang_2019}; therefore, we find no further evidence for a GW/gamma-ray association.
     
%
\subsection{Targeted Search Results}\label{ssec:results_targeted}
\begin{figure*}
    \centering
    \includegraphics[width = 0.48\textwidth]{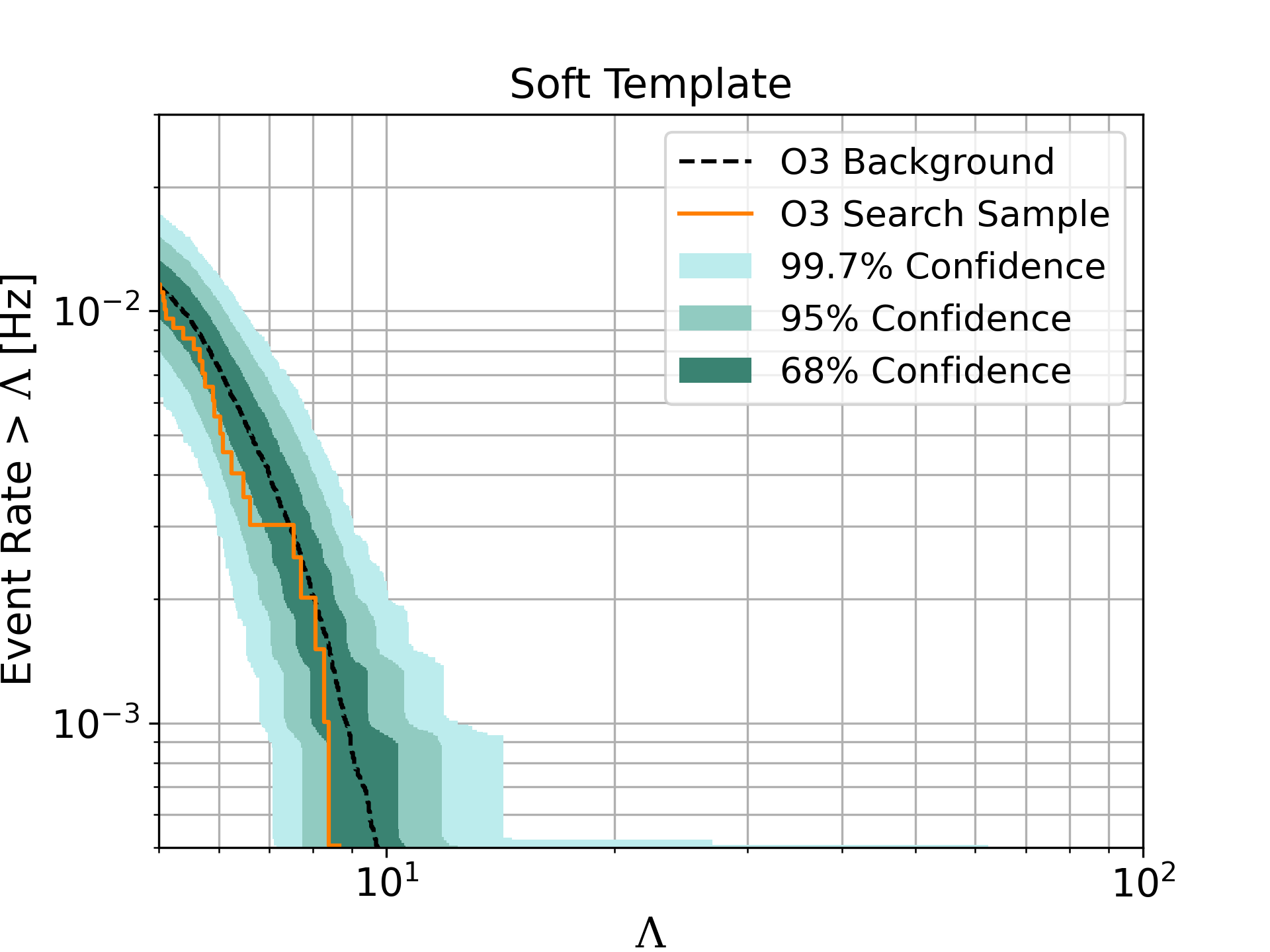}
    \includegraphics[width = 0.48\textwidth]{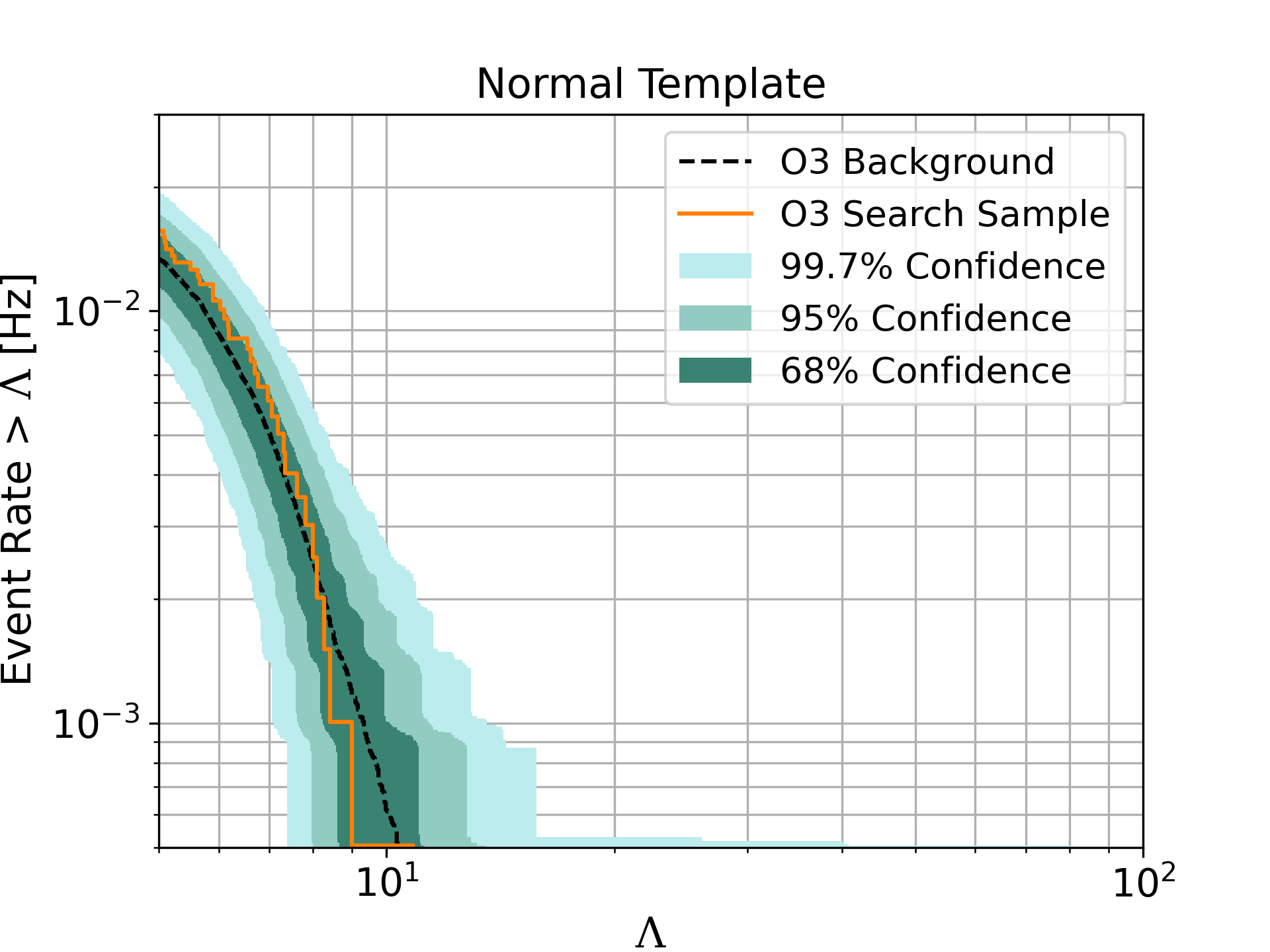}
    \includegraphics[width = 0.48\textwidth]{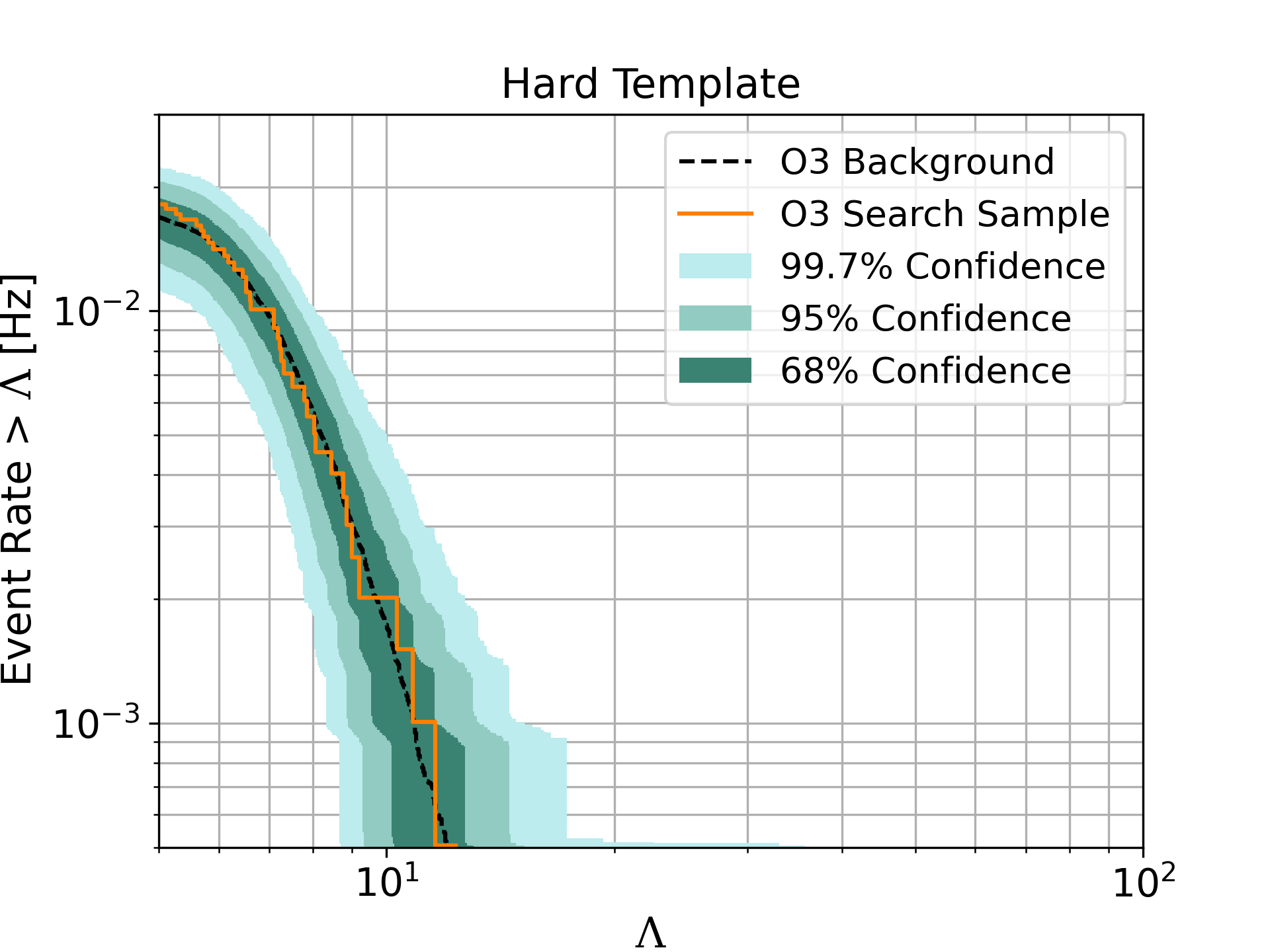}
    \caption{The cumulative rate above a given value of the marginalized log-likelihood ratio $\Lambda$, separated into three plots according to the best fitting spectral template, found in the Targeted Search. The orange line is the foreground distribution of GRB candidates found with the Targeted Search around the given GW merger time. The black dotted line is the randomly selected background sample and the green shading represents the 68\%, 95\% and 99.7\% confidence intervals around it.
    }
    \label{fig:CER}
\end{figure*}

We ran the Targeted Search on the 79 GWs shown in Table \ref{tab:GWevents}.
\FermiGBM\ was in the SAA for 15 of those times, therefore the detectors were turned off and no data were obtained. Figure \ref{fig:CER} shows the cumulative rate above a given value of the marginalized log-likelihood ratio $\Lambda$ separated according to the best-fitting spectral template. The background distribution is determined by randomly selecting times during the O3 livetime without known GW triggers. This represents the FAR of the search and describes the frequency of expected false positives as a function of $\Lambda$.  No significant gamma-ray signals were found in coincidence with GWs.

\begin{figure*}
    \centering
    \includegraphics[width = 0.48\textwidth]{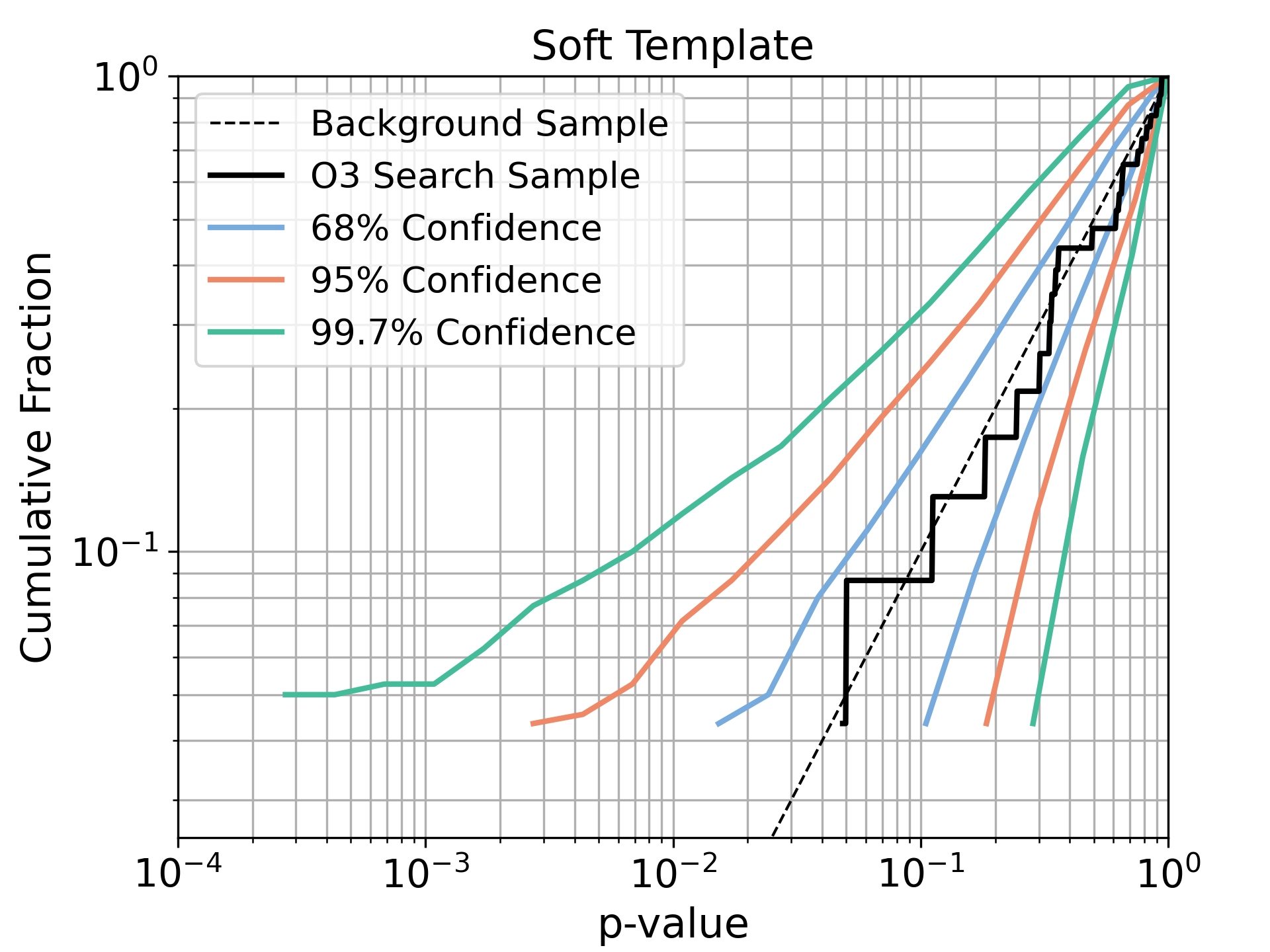}
    \includegraphics[width = 0.48\textwidth]{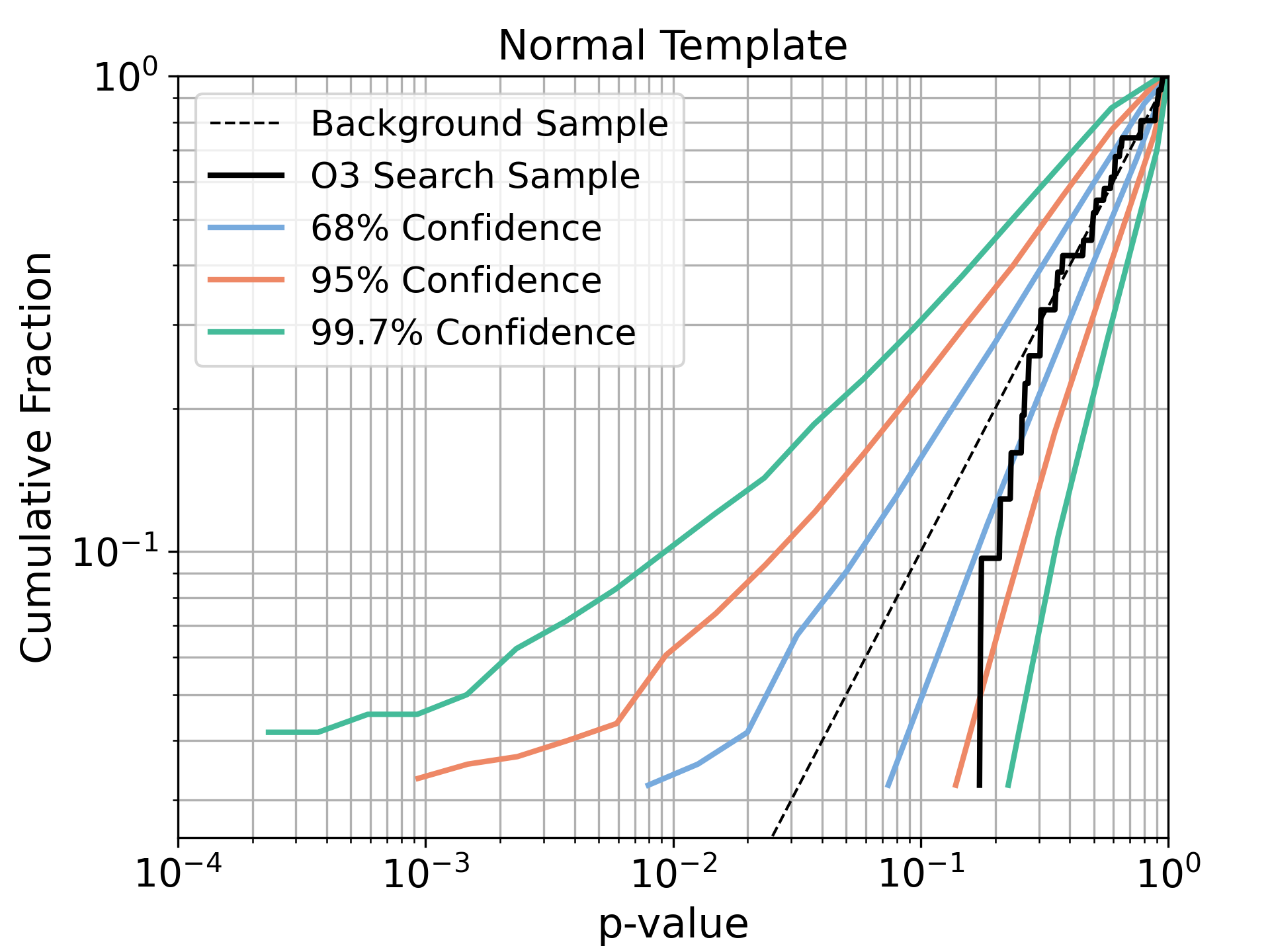}
    \includegraphics[width = 0.48\textwidth]{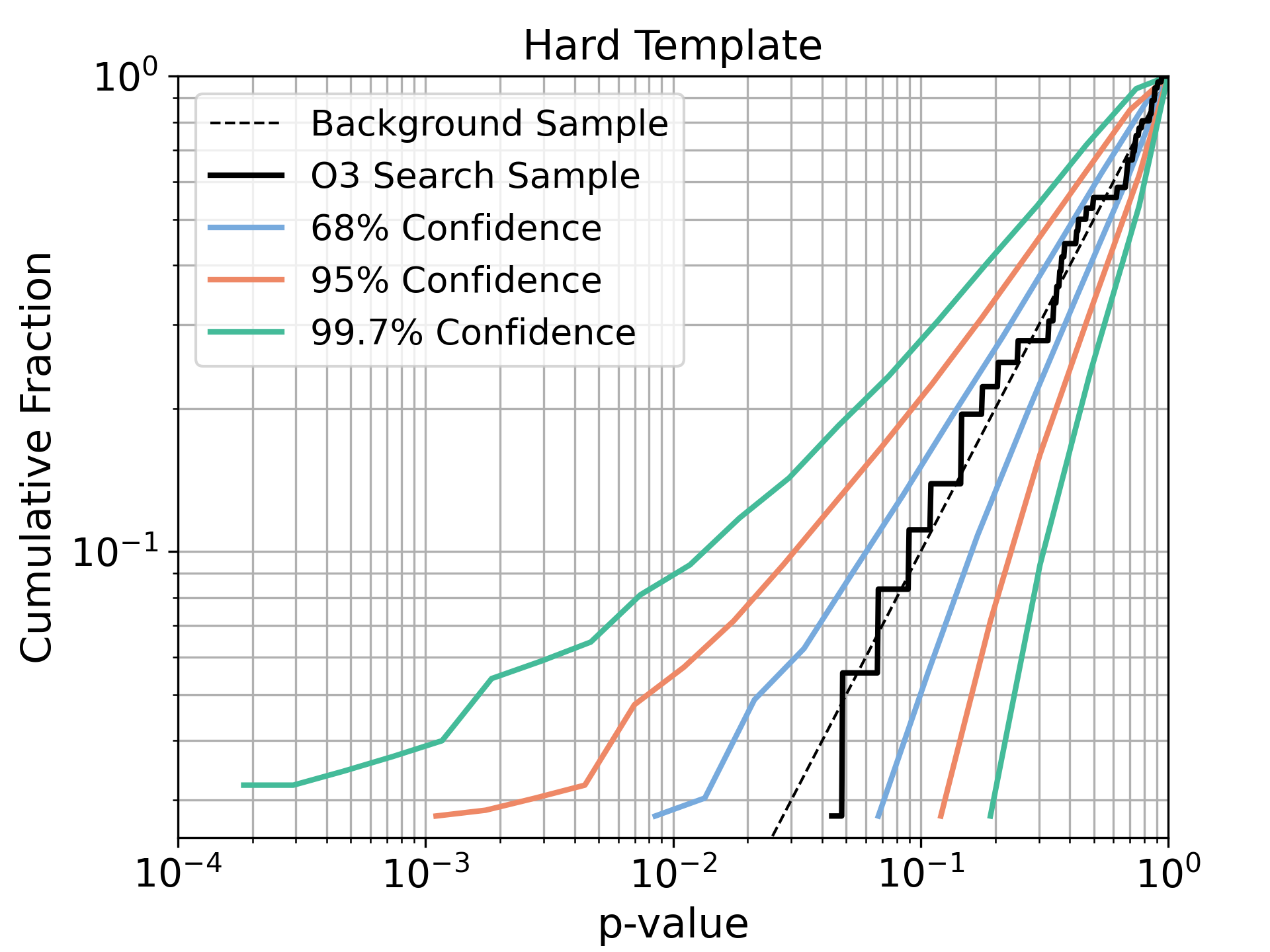}
    \caption{Cumulative fraction versus p-value of the updated ranking statistic $R$. The solid black line is the foreground distribution of GRB candidates found with the Targeted Search around the given GW merger time. The dashed black line is the randomly selected background sample and the blue, green and orange lines represent the 68\%, 95\% and 99.7\% confidence intervals for background, respectively.}
    \label{fig:ranking}
\end{figure*}

Figure \ref{fig:ranking} shows the ranking statistic $R$ from Section~\ref{ssec:ranking_statistic} mapped to a p-value, defined as the number of more highly ranked background candidates as compared to the total number of background candidates or $p_i = N(R > R_i)/N$, where $N$ is the number of background gamma-ray candidates and $i$ is the candidate index within the search sample. 
The plots show no significant deviations from background, yielding no evidence for a GRB counterpart to the GW signals.

\subsection{\Swift\ Results}\label{ssec:Swift_results}

We ran the pipeline described in Section~\ref{ssec:method_swift} on the 1~s binned light curves from \SwiftBAT\ for the 79 candidates listed in Table~\ref{tab:GWevents}. The goal of the pipeline is to examine whether any emission 
surpasses the 5$\sigma$ threshold above the observed background rate level. We visually inspect each light curve to ensure that no detector noise or malfunction affects the reported results. We identify detector noise as a fast duration peak seen only in some of the energy channels.
Once identified, the detector noise is subtracted down to the level of the average background rate. There are 14 GWs for which the data are either unavailable or background dominated because they occurred during either the \SwiftBAT\ SAA passage or a slew. Separately, 13 GWs were within the BAT FoV ($>$10\% partial coding) and had image survey data available at the time of interest.
We report no significant hard X-ray detection in the \SwiftBAT\ rate data coincident with the reported GW triggers at the $5\sigma$ level. We also ran the standard BAT analysis on the survey data (longer timescales, $\sim 300$~s) for the 13 inside-BAT-FoV candidates and also report no excess X-ray emission.\footnote{\url{https://heasarc.gsfc.nasa.gov/ftools/caldb/help/batsurvey.html}} 

\subsection{Results for Marginal GW Canidates}\label{ssec:marginal_results}
The \FermiGBM\ and \SwiftBAT\ searches were applied separately to the 6 marginal GW signals from Table~\ref{tab:margGWevents} in an effort to identify EM counterparts, that could prove an astrophysical origin. The closest time offset from the GBM Triggered and Untargeted Searches was observed for GW191118\_212859, which occurred 42 minutes before the on-board trigger of GRB 191118A and corresponds to a p-value of 0.1. The most significant Targeted Search candidate was found for GW200105\_162426 using the hard spectral template. It has a pre-trial p-value of 0.1 as estimated by the ranking statistic distribution for background described in Section~\ref{ssec:ranking_statistic}. Applying a trials factor of 3 to account for the 3 spectral templates used by the Targeted Search increases the p-value to 0.3. No 5$\sigma$ detections were found for any marginal candidates using the \SwiftBAT\ rates data search.
%
\section{Science Discussion}\label{sec:discussion}
Compact binary mergers containing a neutron star component are likely candidates for gamma-ray emission, particularly if the inspiral process results in tidal disruption of the neutron star \citep{Burns2020}.
In contrast, BBH mergers are not expected to produce gamma-rays outside of a few exotic scenarios (e.g., \citealt{Loeb_2016,Perna+16BBHmodel, Zhang16chargedBBH,Dai_2017}). Therefore, using the standard convention of $m_1 > m_2$, we divide our discussion into two sections based on the secondary component mass $m_2$:
\begin{enumerate}
    \item \textbf{Mergers with a possible neutron star:}
    $m_2 \le$ 3 \Msun\ (5\% credible level)
    
   \item \textbf{Probable BBH mergers:} $m_2 >$ 3 \Msun\ (95\% credible level)
\end{enumerate}

\begin{figure*}
\centering
\begin{minipage}[b]{.47\textwidth}
    \includegraphics[width=\textwidth]{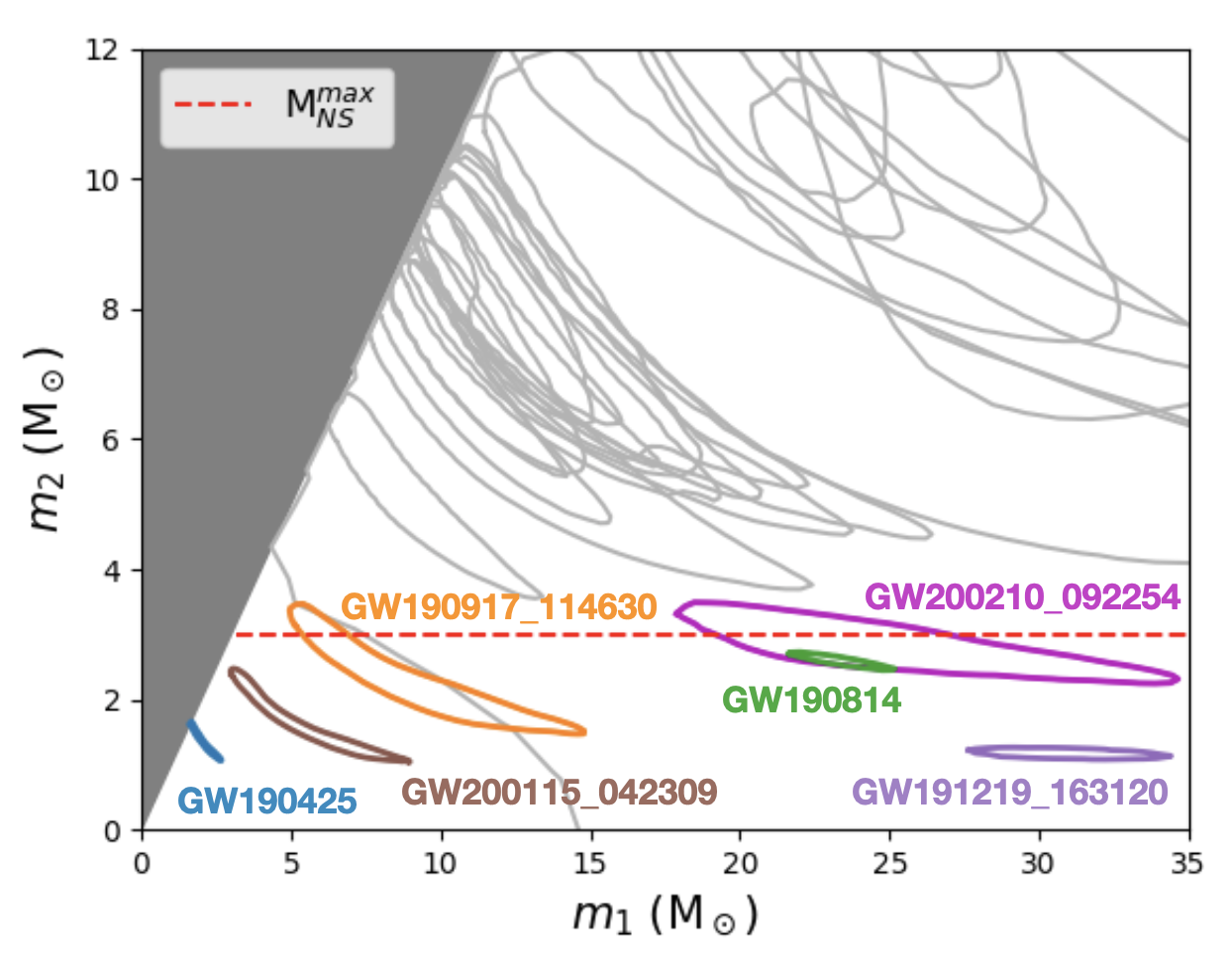}
    \caption{The inferred 90\% credible regions of the component masses for all GWs with \pastro\ $>$ 0.5 from O3 are shown in grey \citep{GWTC-2,GWTC2.1,GWTC-3}. The red dashed line marks the upper bound of $m_2$ allowed for our classification of systems with a possible neutron star component, which are marked by colored contours.}
    \label{fig:masscut}
\end{minipage}\qquad
\begin{minipage}[b]{.47\textwidth}
    \includegraphics[width=\textwidth]{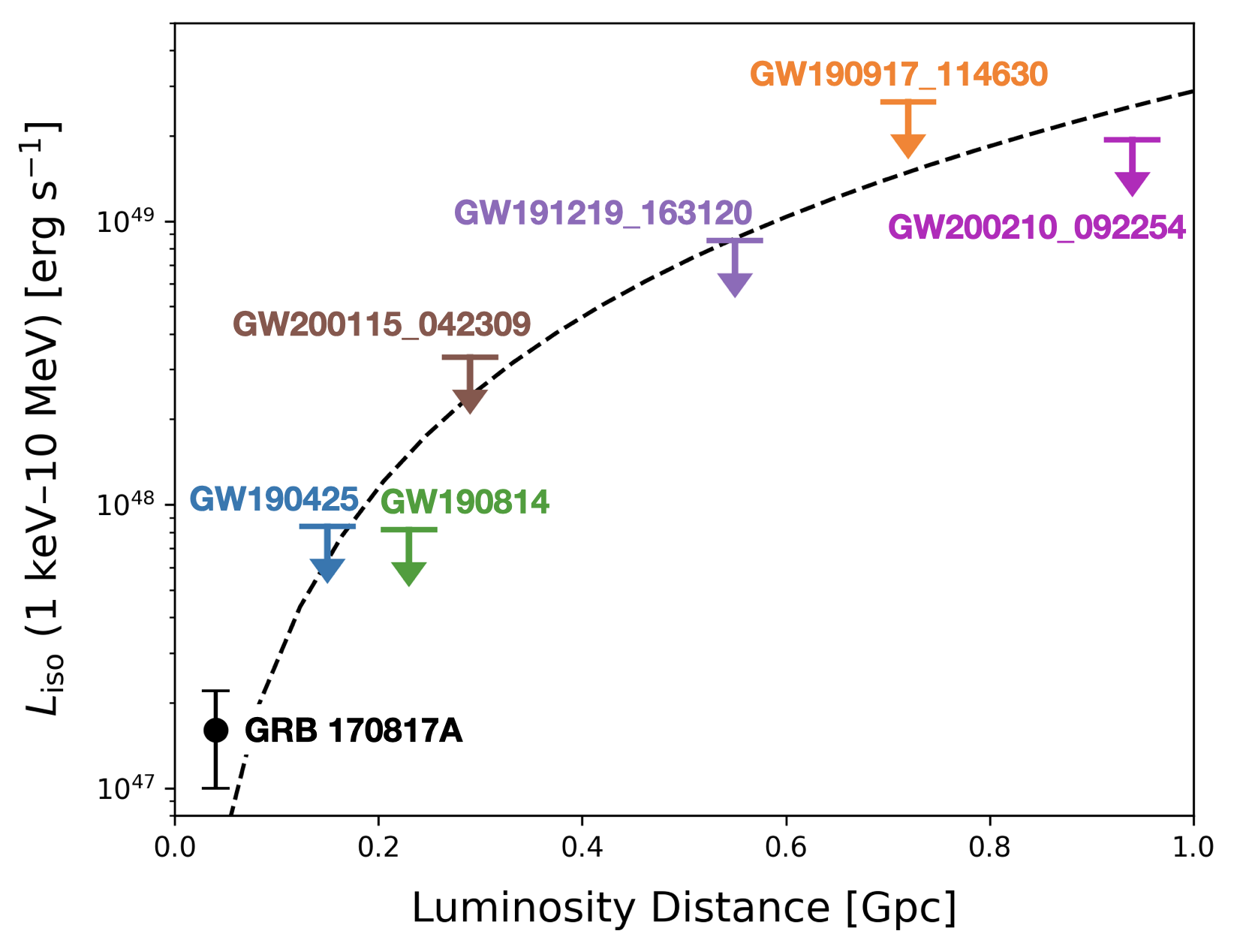}
    \caption{The 5$\sigma$ upper-limits on isotropic-equivalent luminosity $L_\mathrm{iso}$ for the 6 GWs in O3 identified with a possible neutron star component and \pastro\ $>$ 0.5. The black data point is the measured $L_\mathrm{iso}$ from GRB 170817A and the black dashed line is the approximate lower bound for $L_\mathrm{iso}$ of GRBs detected on-board \FermiGBM\ \citep{Abbott+17170817GWGRB}.}
    \label{fig:liso_limits}
\end{minipage}
\end{figure*}

The cut $m_2 \le$ 3 \Msun\ was chosen to include all systems with at least one neutron star component up to the maximal allowed neutron star mass of 2.16--3.0 \Msun\ \citep{MaxNS1996_1, MaxNS1996_2, 2017ApJ...850L..19M, Rezzolla_2018}. It may include a few ambiguous BBH mergers due to the uncertainty on the maximum allowed neutron star mass. We favor this approach over a stricter cut due to the limited number of systems in O3 with light secondary component masses. Additionally, the discussion of possible BBH mergers does not suffer from the loss of a few ambiguous candidates, 
particularly
given the large number of systems with $m_2 >$ 3 \Msun.

In Section \ref{sec:upperlimits_region1} we discuss the absence of GRB detections in coincidence with the 6 GWs classified with a possible neutron star component and present flux upper limits from \FermiGBM\ and \SwiftBAT. 
In Section \ref{sec:bbh_region3} we 
discuss the BBH mergers, providing flux upper limits and exploring
how these limits may rule out certain theoretical models.

\subsection{Possible Neutron Star in the System}\label{sec:upperlimits_region1}
There are 6 GWs (i.e., GW190425, GW190814, GW190917\_114630, GW191219\_163120, GW200115 \_042309, and GW200210\_092254) where $\ge$~5\% of posterior probability lies below the dashed red line in Figure \ref{fig:masscut}. GW190425 is the least massive system from O3 and the second BNS merger detected by LIGO-Virgo. It has a primary mass $m_1 =2.1^{+0.5}_{-0.4}$ \Msun\ and a secondary mass of  $m_2 =1.3^{+0.3}_{-0.2}$ \Msun\ \citep{GWTC2.1, Abbott_2020_190425}. GW190814 has a low mass secondary component, estimated at $m_2 = 2.6^{+0.1}_{-0.1}$ \Msun, while its primary component has an estimated mass of $m_1 = 23.3^{+1.4}_{-1.4}$ \Msun. It is unclear whether this source is a BBH or a NSBH merger, since the secondary component could either be a light black hole or a heavy neutron star \citep{GWTC2.1}.
GW190917\_114630 was identified as a BBH merger by the GstLAL pipeline, but its secondary component mass of $m_2 = 2.1^{+1.1}_{-0.4}$~\Msun\ is a strong indicator for a NSBH origin \citep{GWTC2.1}. GW191219\_163120 has a large primary component mass of $m_1 = 31.1^{+2.2}_{-2.8}$~\Msun\ and the lowest secondary component mass $m_2 =1.17^{+0.07}_{-0.06}$~\Msun\ of all the GWs with a possible neutron star \citep{GWTC-3}; it represents a potential NSBH merger. GW200115\_04230 has a primary mass of $m_1 =5.9^{+2.0}_{-2.5}$ \Msun\, suggesting a low-mass black hole, and a secondary mass of $m_2 =1.44^{+0.85}_{-0.29}$ \Msun\ which is consistent with a neutron star \citep{GWTC-3}. GW200210\_092254 possesses a primary component mass of $m_1 = 24.1^{+7.5}_{-4.6}$ \Msun\ and a secondary component mass of $m_2 = 2.83^{+0.47}_{-0.42}$ \Msun\ that could either be a light black hole or a heavy neutron star \citep{GWTC-3}. It is unclear if this source is a BBH or NSBH merger.

\begin{table*}
    \footnotesize 
    \centering
    \caption{
    Flux upper limits for the 6 GWs from O3 with \pastro~$>$~0.5 that are classified with a possible neutron star component. The 3$\sigma$ upper limits are computed for the 10--1000 keV energy range over the FoV of \FermiGBM. The 5$\sigma$ upper limits are computed for the combined coverage of \FermiGBM\ and \SwiftBAT\ with both instruments matched to the 15--350 keV energy range of \SwiftBAT. The columns labeled Min and Max correspond, respectively, to the minimum and maximum upper limits obtained for points within the 90\% credible region of the GW localization. The Marginal upper limit is computed by integrating the upper limits produced at individual locations over the full sky using the GW localization as a weighted prior, normalized to the visible portion of the sky.
    Also shown is the visible coverage percentage of the full GW localization for \FermiGBM\ alone, \SwiftBAT\ alone, and the combined FoV from both instruments.}
    \label{tab:upperlimits_bns/nsbh}
    \begin{tabular}{HHlHrrrcccccc}\hline
& & & \multicolumn{1}{c}{} & \multicolumn{3}{c}{} & \multicolumn{3}{c}{3$\sigma$ Flux U.L. [erg s$^{-1}$ cm$^{-2}$]} & \multicolumn{3}{c}{5$\sigma$ Flux U.L. [erg s$^{-1}$ cm$^{-2}$]} \\
& & & \multicolumn{1}{c}{} & \multicolumn{3}{c}{Coverage [\%]} & \multicolumn{3}{c}{10--1000 keV} & \multicolumn{3}{c}{15--350 keV} \\ \cline{5-7}
& & Event Name & \multicolumn{1}{H}{Date/Time} & GBM & BAT & \multicolumn{1}{r}{Combined} & Min & Max & \multicolumn{1}{c}{Marginal} & Min & Max & Marginal \\ \hline\hline  
possible NS with EM 7 &  7 & GW190425             & 2019-04-25T08:18:05.014133 &    56.70 &    10.81 &    57.81 & 1.37$\times 10^{-7}$ & 2.47$\times 10^{-7}$ & 1.66$\times 10^{-7}$ & 6.12$\times 10^{-8}$ & 2.31$\times 10^{-7}$ & 1.51$\times 10^{-7}$ \\
possible NS with EM 33 & 33 & GW190814             & 2019-08-14T21:10:38.990470 &   100.00 &   100.00 &   100.00 & 1.17$\times 10^{-7}$ & 1.26$\times 10^{-7}$ & 1.21$\times 10^{-7}$ & 4.71$\times 10^{-8}$ & 7.64$\times 10^{-8}$ & 6.26$\times 10^{-8}$ \\
possible NS with EM 39 & 39 & GW190917\_114630     & 2019-09-17T11:46:30.035623 &    88.92 &     6.56 &    95.07 & 1.48$\times 10^{-7}$ & 4.33$\times 10^{-7}$ & 2.33$\times 10^{-7}$ & 5.26$\times 10^{-8}$ & 3.83$\times 10^{-7}$ & 2.08$\times 10^{-7}$ \\
possible NS unlikely EM 56 & 12 & GW191219\_163120     & 2019-12-19T16:31:20.451572 &    61.06 &      N/A &    61.06 & 1.03$\times 10^{-7}$ & 2.36$\times 10^{-7}$ & 1.20$\times 10^{-7}$ & 1.00$\times 10^{-7}$ & 2.27$\times 10^{-7}$ & 1.15$\times 10^{-7}$ \\
possible NS with EM 60 & 16 & GW200115\_042309     & 2020-01-15T04:23:09.757607 &    96.26 &     4.80 &    96.26 & 1.31$\times 10^{-7}$ & 2.83$\times 10^{-7}$ & 1.78$\times 10^{-7}$ & 8.53$\times 10^{-8}$ & 2.56$\times 10^{-7}$ & 1.60$\times 10^{-7}$ \\
possible NS with EM 67 & 23 & GW200210\_092254     & 2020-02-10T09:22:54.982977 &    61.79 &    50.55 &    65.85 & 1.28$\times 10^{-7}$ & 3.13$\times 10^{-7}$ & 2.01$\times 10^{-7}$ & 4.41$\times 10^{-8}$ & 1.47$\times 10^{-7}$ & 9.00$\times 10^{-8}$ \\
\hline \hline
    \end{tabular}
\end{table*}

\begin{figure*}
    \centering
    \includegraphics[width=0.75\textwidth]{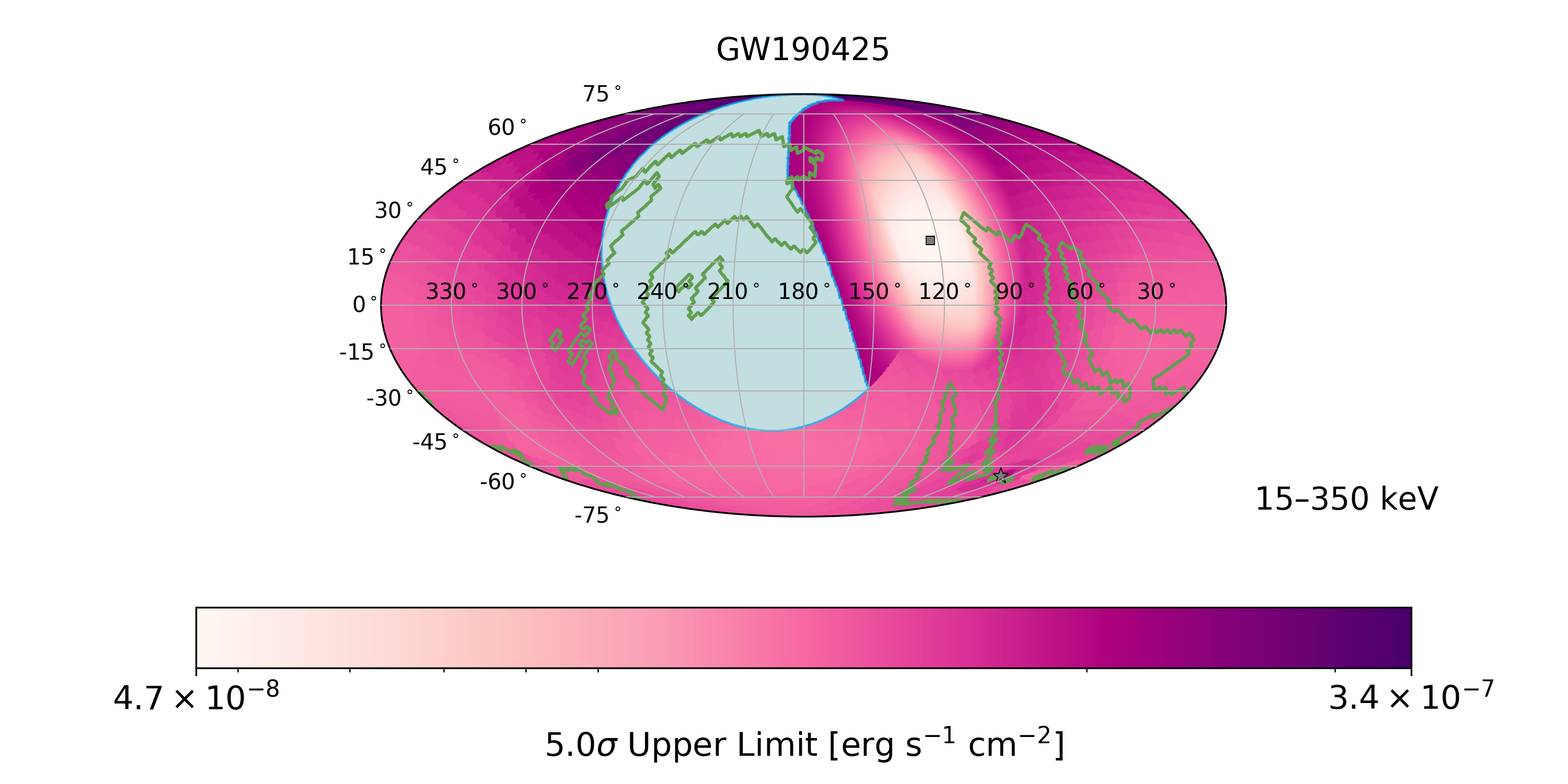}
    \includegraphics[width=0.75\textwidth]{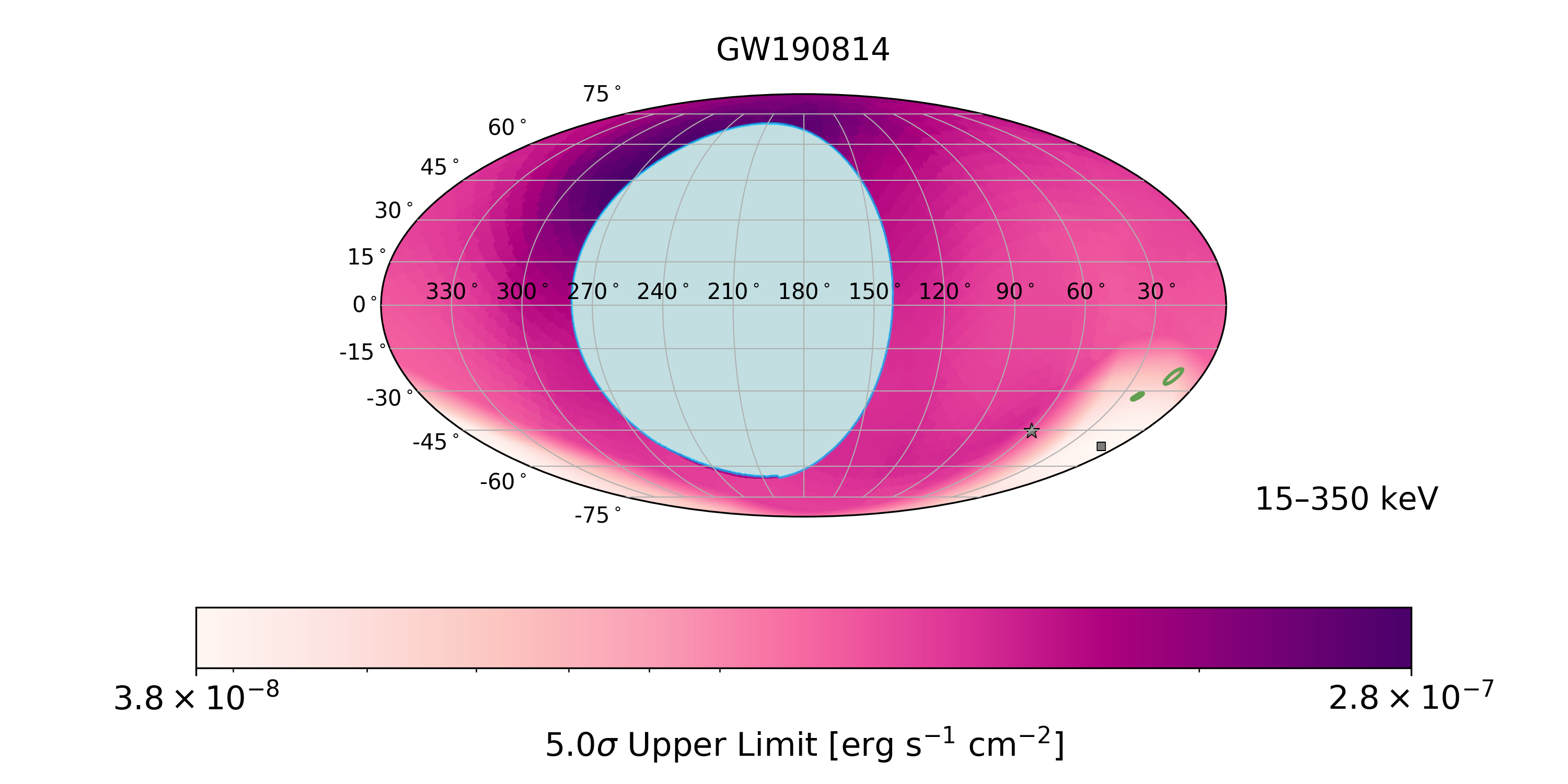}
    \includegraphics[width=0.75\textwidth]{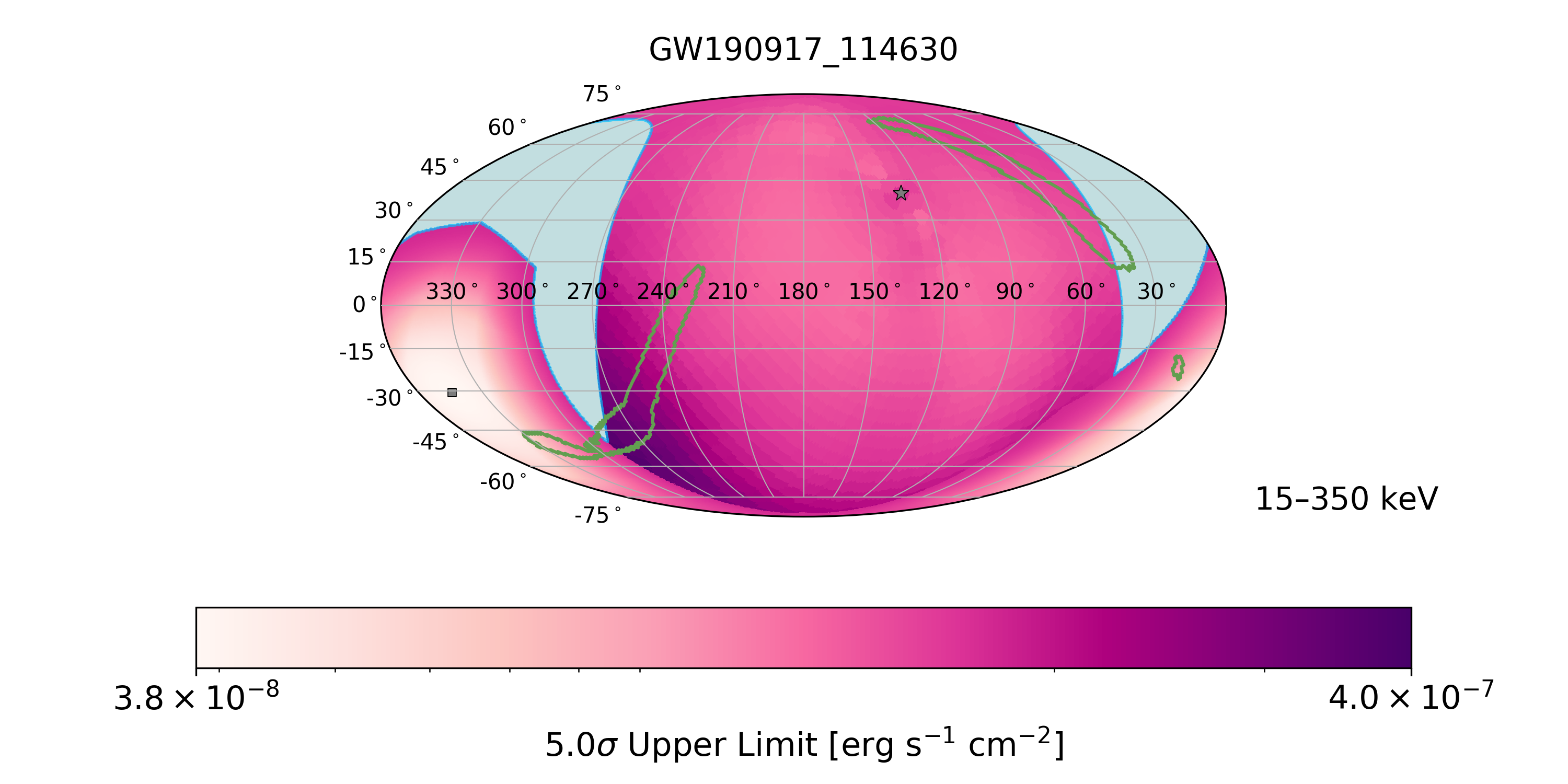}
    \caption{The 5$\sigma$ flux upper-limit as a function of sky position for the 6 GWs from O3 identified with a possible neutron star component and \pastro\ $>$ 0.5. The purple gradient represents the combined \FermiGBM\ and \SwiftBAT\ flux upper limits for source positions at each point on the sky. The star symbol represents the zenith direction of \FermiGBM, the square symbol represents the center of the \SwiftBAT\ FoV, and the green contour represents the 90\% credible area of the LVK localization. The blue region is the non-visible portion of the sky which is occulted by the Earth for \FermiGBM\ and outside the \SwiftBAT\ FoV. }
    \label{fig:possible_ns_skymaps}
\end{figure*}

All GWs except GW191219\_163120 were observable by both \FermiGBM\ and \SwiftBAT\ (Table~\ref{tab:upperlimits_bns/nsbh}). GW191219\_163120 was observed by \FermiGBM\ but no \SwiftBAT\ data are available due to a slewing behavior of the spacecraft at the GW time. Neither instrument detected an EM counterpart to these GWs. Therefore, we compute flux upper limits for each GW using the methods described in Section~\ref{sec:method}.
Table~\ref{tab:upperlimits_bns/nsbh} presents the minimum and maximum flux upper limits from \FermiGBM\ and \SwiftBAT\ over the 90\% credible regions of the GW localizations, as well as sky-marginalized flux upper limits. Incorporating the combined measurements from \FermiGBM\ and \SwiftBAT, we also use the sky-marginalized 5$\sigma$ flux limits to generate upper limits on the isotropic-equivalent luminosity (Figure~\ref{fig:liso_limits}).
The combined 5$\sigma$ flux upper limits over the 15--350 keV energy range can be seen in Figure \ref{fig:possible_ns_skymaps}.

The lack of a gamma-ray counterpart to BNS merger GW190425 has three plausible explanations. First, the combined coverage of $\sim$60\% of the total GW localization implies that the GW source may not have been visible to \FermiGBM\ and \SwiftBAT. Second, GW190425 has an estimated luminosity distance of $D_\mathrm{L} =0.15^{+0.08}_{-0.06}$~Gpc which is 4 times larger than that to GW170817 \citep{GWTC-1}. At this distance, the luminosity of GW170817 would fall well below the upper limit for GW190425\footnote{Note, the luminosity upper limit for GW1901425 is also consistent with that found in \citealt{Hosseinzadeh_2019}, which uses preliminary \FermiGBM\ flux upper limits reported during the O3 online analysis} \citep{S190425z_gbm_gcn}.
(Figure~\ref{fig:liso_limits}), indicating that a counterpart similar to the one for GW170817 would have been unobservable to \FermiGBM\ or \SwiftBAT. Finally, the inclination angle of this GW is poorly constrained, with the 90\% credible level extending to a viewing angle of 70$^\circ$ with respect to the jet axis. This encompasses scenarios where the observed off-axis flux would be below the detection limits, even if the central engine of GW190425 was powerful enough to be detected on-axis by \FermiGBM\ and \SwiftBAT\ at its measured distance.

\renewcommand\thefigure{5 continued}
\begin{figure*}
    \centering
    \includegraphics[width=0.75\textwidth]{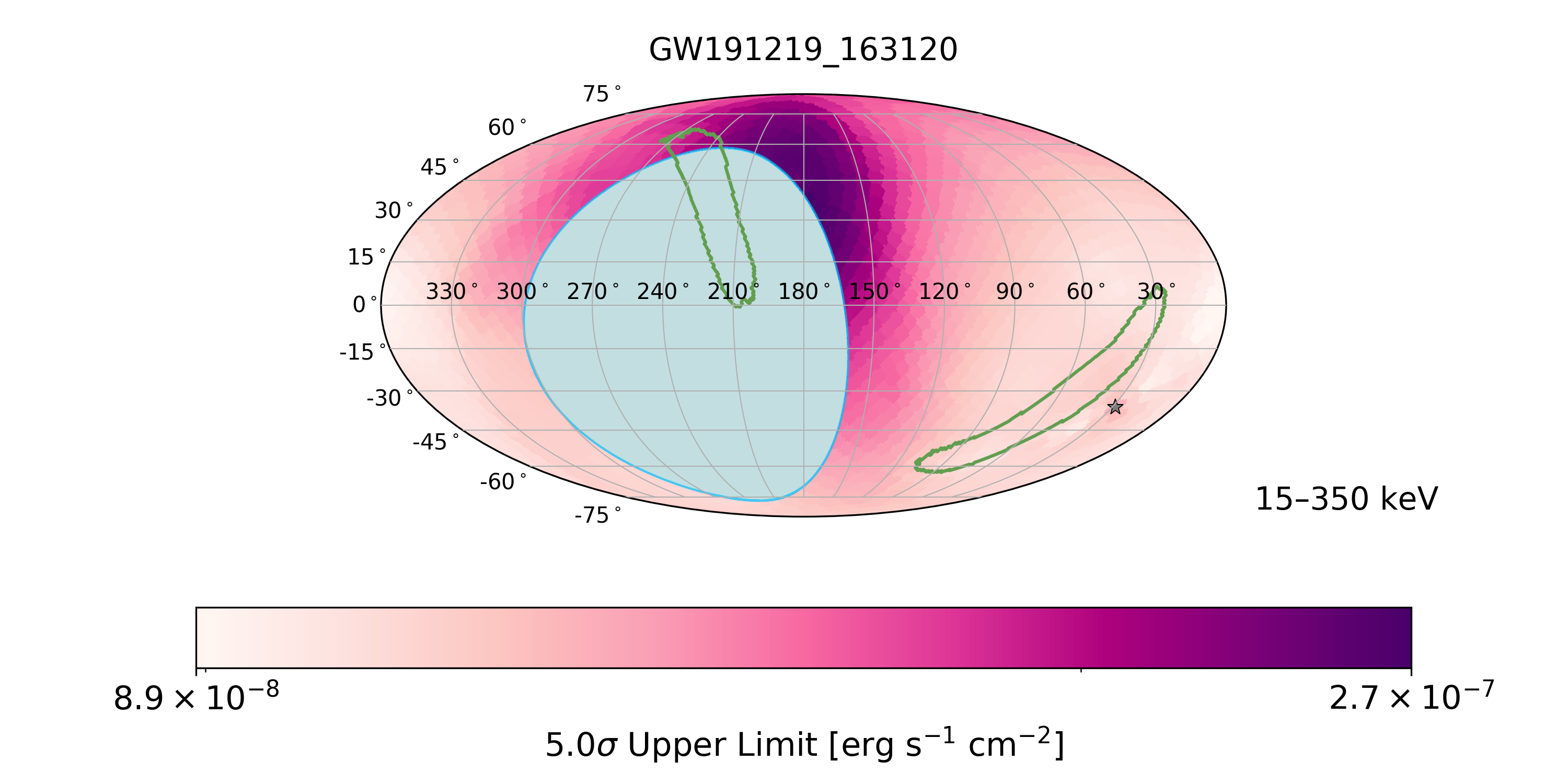}
    \includegraphics[width=0.75\textwidth]{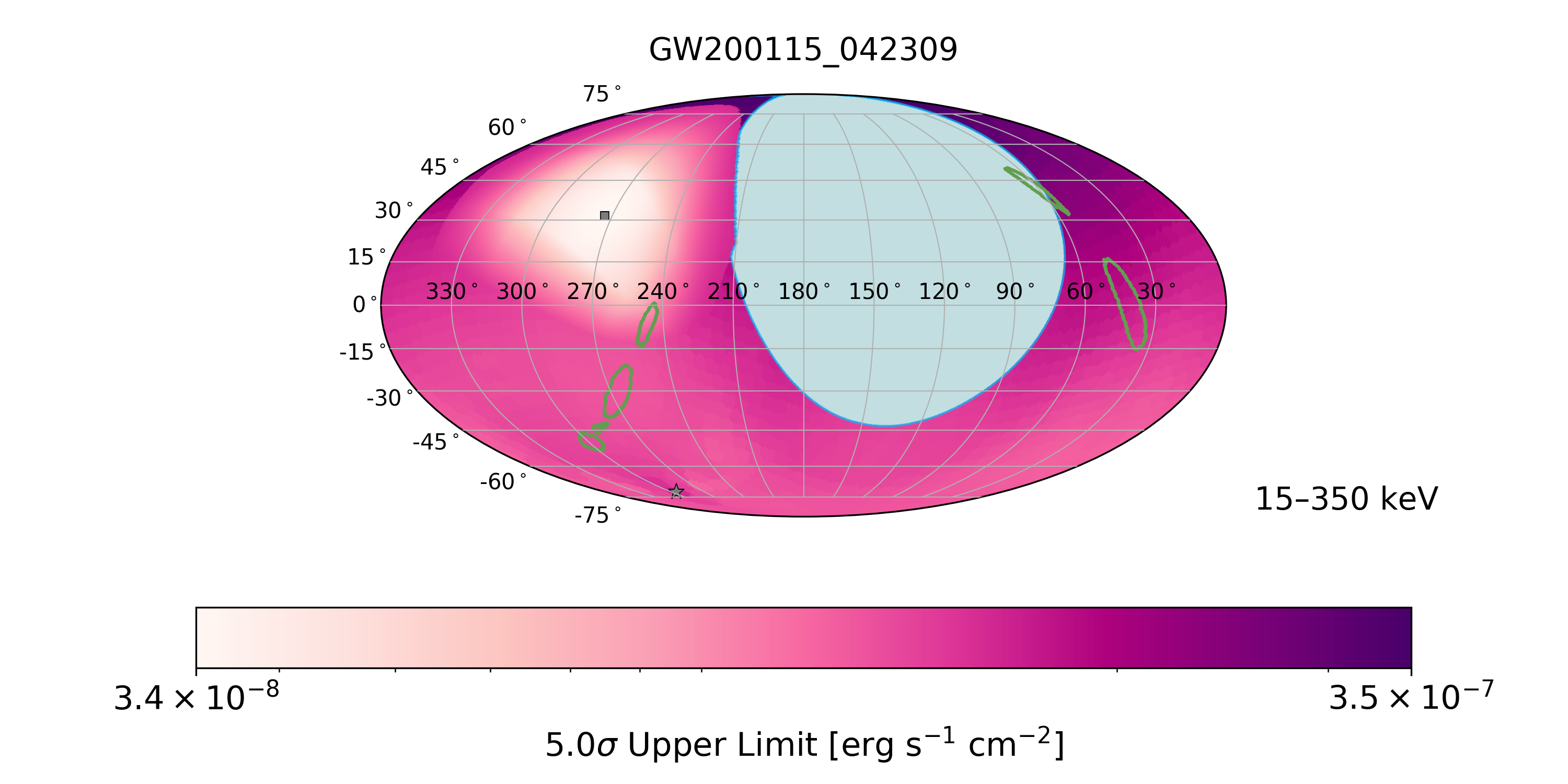}
    \includegraphics[width=0.75\textwidth]{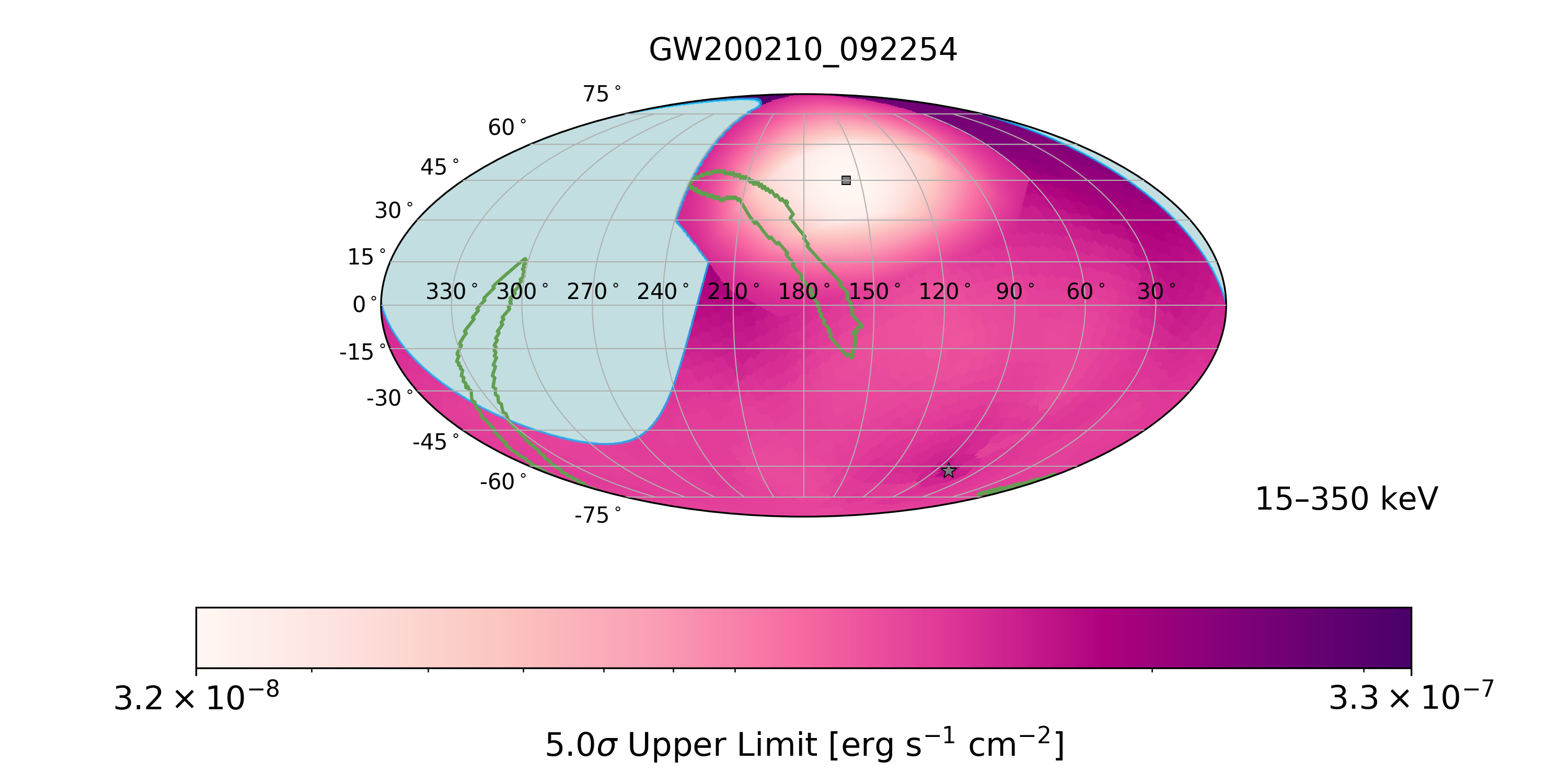}
    \caption{}
    \label{fig:possible_ns_skymaps2}
\end{figure*}
\renewcommand\thefigure{\arabic{figure}}
\setcounter{figure}{6} 

\subsection{Probable BBH Mergers}\label{sec:bbh_region3}

There are a total of 73 GWs with the criterion of $m_2 >$ 3 \Msun\ in $>$ 95\% of the posterior probability. All of these have estimated primary and secondary component masses much larger than the maximum expected neutron star mass of 3 \Msun. Therefore, they are most likely GW signals from BBH mergers.

Of these GWs, 10 occurred during SAA for \FermiGBM, but had data from \SwiftBAT. Likewise, \SwiftBAT\ does not have data for 9 GWs, either because \SwiftBAT\ was in the SAA or slewing, but data are available from \FermiGBM. Finally, there are 5 GWs that do not have data from either \FermiGBM\ or \SwiftBAT\, due to being in the SAA and/or slewing. Neither instrument identified an EM counterpart for the GWs with data coverage. As a result, we compute flux upper limits for each GW according to the methods described in Section~\ref{sec:method}. Table~\ref{tab:upperlimitsbbh} presents the minimum and maximum upper limits over the 90\% credible region of the GW localization as well as the marginalized flux upper limits. Joint flux upper limits as a function of sky position and the corresponding isotropic-equivalent luminosity limits for the GWs that have data coverage are provided in a separate data release~\citep{O3FollowupRelease}. For the GWs with \FermiGBM\ data, we look into constraining theoretical models of gamma-ray emission from BBHs using the 3$\sigma$ flux upper limits, as they provide broad spectral coverage over the  10--1000~keV energy range.


\subsubsection{Constraining gamma-ray emission models from BBH mergers}

EM radiation from BBH mergers is not expected due to the challenges associated with forming an accretion disk during the merger process. Nevertheless, \citet{Connaughton+16BBH} reported GW150914-GBM, a weak gamma-ray signal following the first LIGO--Virgo detection of BBH GW150914 \citep{Abbott+16GW150914}, and, more recently, the Zwicky Transient Facility identified a potential EM counterpart to GW190521 \citep{PhysRevLett.124.251102}. While the associations between these detections and the corresponding GWs remain nebulous \citep{Connaughton_2018, Ashton2021, Bustillo:2021tga, 2020MNRAS.499L..87D, 2021ApJ...914L..34P}, a wide spectrum of models have been developed to invoke EM emission from BBH-mergers, all with non-negligible difficulties (e.g., \citealt{Loeb16BBHmodel, Perna+16BBHmodel,Zhang16chargedBBH, perna_2018}).

To test the association between GW150914 and GW150914-GBM, \citet{Veres+19BBHEM} assumed some of these BBH emission models and derived a model-dependent BBH-to-GRB ratio which represents the expected number of BBH mergers to be detected by LVK before a gamma-ray counterpart might be observed by \FermiGBM. 
Since the number of LVK BBH merger detections has reached the BBH-to-GRB ratio for a few models reported by \citet{Veres+19BBHEM}, we attempt to constrain them by computing gamma-ray flux upper limits for each model and examining the implications with respect to individual BBH mergers.

We consider four models for relating potential gamma-ray emission to the energy present in BBH mergers: a neutrino--antineutrino annihilation powered jet mechanism ($\nu\bar{\nu}$; \citealt{Ruffert+98neutrino}), a charged BH (Q; \citealt{Zhang16chargedBBH}), the Blandford--Znajek mechanism (BZ; \citealt{BZ77}), and a model where the gamma-ray energy is proportional to the emitted GW energy ($\rm{E}_{GW}$).
A detailed summary of these models and their parameters can be found in \cite{Veres+19BBHEM}.
We note that all of the above scenarios suffer from non-trivial critiques but are used here to be widely-inclusive of the broad spectrum of proposed mechanisms for gamma-ray production.
The intrinsic properties of each model (e.g., magnetic field strength, charge of the black hole, etc.) are determined by setting them to values consistent with the observed luminosity of GW150914-GBM \citep{Connaughton+16BBH}.
 
For each GW, we use the posterior distributions of BBH parameters (e.g., final mass, distance, inclination, rotation parameter, etc.) from GWTC-3 and derive a distribution of gamma-ray fluxes for the different models. We then compare the distribution of fluxes to the 3$\sigma$ marginalized flux upper limit from \FermiGBM, shown in Table \ref{tab:upperlimitsbbh}. This is performed for three GRB jet geometries: an isotropic emitter (i.e., an opening angle of 90$^\circ$), an opening angle distributed uniformly between 10--40$^\circ$, and a fixed 20$^\circ$ opening angle. All jets are assumed to have a top-hat angular structure and are assigned pointing directions by sampling from the inclination posterior of each GW. Due to relativistic beaming, emission is strongly suppressed for GRBs with jet opening angles smaller than the viewing angle. In order to simplify the treatment of such cases, we assign zero flux to jets whose inclination is larger than the opening angle.

\begin{figure}
    \centering
        \includegraphics[width=0.8\columnwidth]{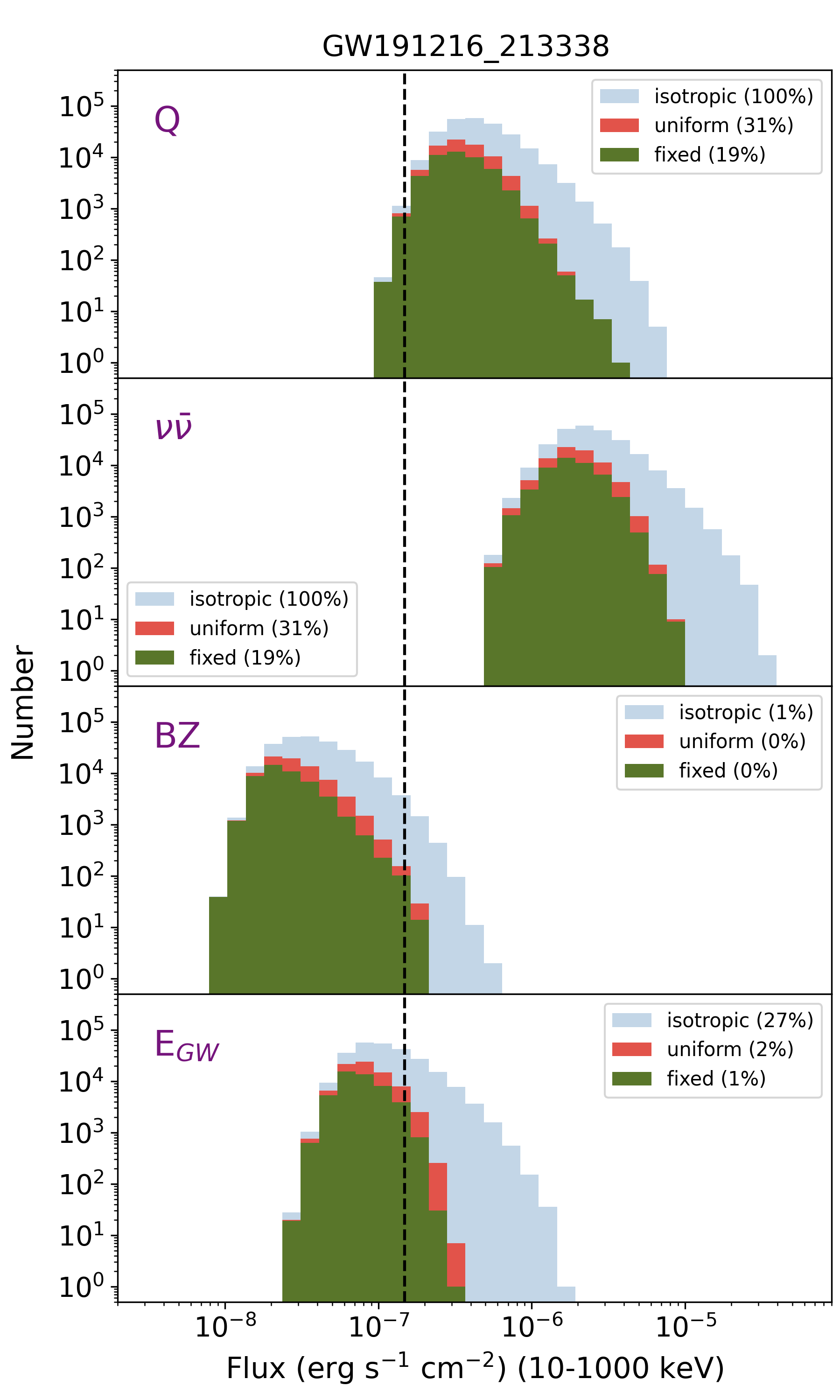}
    \caption{ Example of gamma-ray flux expected for the 4 different models: charged BH model (Q), neutrino--antineutrino annihilation powered jet model ($\nu\bar{\nu}$), Blandford--Znajek (BZ), and gamma-ray energy as a fraction of GW energy (E$_{\mathrm{GW}}$) in the case of GW191216\_213338. The 3$\sigma$ (10--1000 keV) marginalized flux upper limit is indicated by the vertical line. The legend contains the fraction of cases above the 3$\sigma$ upper limit; note that for the jetted emission models (uniform, fixed), cases with zero expected flux are not shown. Here the limit is violated in 100\% and $>$99\% of cases for the  $\nu\bar{\nu}$ and Q models, respectively (assuming isotropic emission). } 
    \label{fig:em1}
\end{figure}

Figure \ref{fig:em1} shows an example of the flux distribution for the four different models and the three jet geometries compared to the GBM upper limits for GW191216\_213338. There is a dearth of jetted cases (green, red) compared to isotropic emission (light blue) at higher fluxes. This is explained by the inclination angle-distance degeneracy of the GW parameter estimation. Point estimates with smaller distances and thus higher flux will preferentially have jets pointed away from our line of sight. When we impose a jet opening angle $\lesssim 40^{\circ}$ on such systems, they will not include the observer within their aperture in most of the cases. Conversely, the point estimates with largest distances point preferentially towards the observer, thus there will be no strong differences between the jetted and isotropic cases at low flux values.

We classify GW191216\_213338 as noteworthy because the predicted gamma-ray flux 
distribution from at least one model violates the GBM flux upper limit by more than 10\%. The 10\% limit in the rest of this section refers to the isotropic emission model. Of the 58 probable BBH mergers with \FermiGBM\ data coverage described in Section~\ref{sec:bbh_region3}, 18 are considered noteworthy according to this criterion. The remaining 40 did not yield the necessary number of cases above the GBM flux upper limit in any of the models.

Out of the four models considered here, the $\nu\bar{\nu}$ model violates the \FermiGBM\ flux upper limit in most of the cases. Of the noteworthy GWs (denoted by $*$ in Table \ref{tab:ULmodel}, Appendix \ref{appendix}), 15 exceed the GBM limit in more than 10\% of cases for this model. In particular, GW190924\_021846, GW191216\_213338 (Figure \ref{fig:em1}) and GW200202\_154313  exceed the GBM limit in \{100, 30, 20\}\%, \{100, 31, 19\}\% and \{100, 36, 24\}\% of the cases respectively (the 3 numbers represent the 3 different opening angle choices). Interestingly, for these three events, in the isotropic emission scenario the $\nu\bar{\nu}$ can be ruled out, and the non-detection in gamma-rays can constrain the jet geometry. This is due to GW191216\_213338 and GW200202\_154313 being the two closest BBH signals observed during O3 with luminosity distances of $0.34^{+0.12}_{-0.13}$~Gpc and $0.41^{+0.15}_{-0.16}$~Gpc, respectively \citep{GWTC-3}. In addition to being nearby (luminosity distance of $0.55\pm{0.22}$~Gpc), GW190924\_021846 has a relatively low final mass ($13.9^{+2.8}_{-0.9}$~\Msun), which leads to higher flux in the $\nu\bar{\nu}$ model.

For the Q model, the most constraining GWs are also GW191216\_213338 and GW200202\_154313. They violate the GBM flux upper limit for the different jet geometries in \{100, 31, 19\}\% and \{96, 33, 22\}\%  of cases, respectively.
In total, 12 of the probable BBH mergers have larger than 10\% of their flux estimates above the upper limit for this model.

GW191109\_010717 is the most constraining for the BZ scenario. It violates the gamma-ray upper limit in \{26, $<$0.1, $<$0.1\}\% of cases for the 3 different opening angle choices. It has the fourth-highest total mass $M = 112^{+20}_{-16}$ \Msun\ in O3 and is reasonably close at $D_\mathrm{L} = 1.29^{+1.13}_{-0.65}$ Gpc \citep{GWTC-3}. The only other GW with more than 10\% of the flux estimates above the upper limit for the BZ mechanism is GW190521\_074359.

For the $\rm{E}_{GW}$ scenario, GW191216\_213338 is again the most constrained. It violates the gamma-ray upper limit in \{27, 1.8, 0.7\}\% of the cases for the three jet opening angle choices. There are 3 GWs with 10\% of the flux estimates from this model above the GBM flux upper limit (Table~\ref{tab:ULmodel_B}, Appendix~\ref{appendix}).

In summary, we provide constraints on theoretical models of gamma-ray emission from BBH mergers using the flux upper limits from \Fermi-GBM. We find that for most BBH mergers the models considered here do not predict gamma-ray flux over the upper limit. Under our model assumptions, this can be understood as a consequence of the larger average distance of BBH mergers during O3 compared to that of GW150914.
We also find that the $\nu\bar{\nu}$ model is the most constrained. This model has the lowest BBH-to-GRB ratio in \cite{Veres+16BBHmodel}, and indeed, observations reveal that \jw{18} out of 58 cases for the $\nu\bar{\nu}$ model yield an appreciable flux above the upper limit. The expected flux in this model is inversely proportional with the final mass; the average BBH merger in O3 was less massive than GW150914, resulting in larger predicted gamma-ray flux.

\begin{table*}[ht!]
    \centering
    \caption{Flux upper limits for possible EM counterparts to probable BBH candidates detected during O3 with $\pastro > 0.5$. The 3$\sigma$ upper limits are computed for the 10--1000 keV energy range over the FoV of \FermiGBM. The 5$\sigma$ upper limits are computed for the combined coverage of the \FermiGBM~and \SwiftBAT\ with both instruments matched to the 15-350 keV energy range of \SwiftBAT. The columns labeled Min and Max correspond, respectively, to the minimum and maximum upper limit values obtained for points within the 90\% credible level of the GW localization. The Marginal upper limit is computed by integrating the upper limits produced at individual locations over the full sky using the GW localization as a weighted prior, normalized to the visible portion of the sky.}
    \label{tab:upperlimitsbbh}
    \begin{tabular}{HHlHrrrcccccc}\hline

& & & \multicolumn{1}{c}{} & \multicolumn{3}{c}{} & \multicolumn{3}{c}{3$\sigma$ Flux U.L. [erg s$^{-1}$ cm$^{-2}$]} & \multicolumn{3}{c}{5$\sigma$ Flux U.L. [erg s$^{-1}$ cm$^{-2}$]} \\
& & & \multicolumn{1}{c}{} & \multicolumn{3}{c}{Coverage [\%]} & \multicolumn{3}{c}{10--1000 keV} & \multicolumn{3}{c}{15--350 keV} \\ \cline{5-7}
& & Event Name & \multicolumn{1}{H}{Date/Time} & GBM & BAT & \multicolumn{1}{r}{Combined} & Min & Max & \multicolumn{1}{c}{Marginal} & Min & Max & Marginal \\ \hline\hline  
    
 1 &  1 & GW190403\_051519     & 2019-04-03T05:15:19.212116 &    76.61 &    24.76 &    82.91 & 1.07$\times 10^{-7}$ & 3.09$\times 10^{-7}$ & 1.78$\times 10^{-7}$ & 3.67$\times 10^{-8}$ & 2.80$\times 10^{-7}$ & 1.50$\times 10^{-7}$ \\
 2 &  2 & GW190408\_181802     & 2019-04-08T18:18:02.300251 &      SAA &     0.00 &     0.00 &        - &        - &        - &        - &        - & 3.50$\times 10^{-7}$ \\
 3 &  3 & GW190412             & 2019-04-12T05:30:44.177351 &    97.27 &     3.45 &    99.80 & 1.00$\times 10^{-7}$ & 1.22$\times 10^{-7}$ & 1.11$\times 10^{-7}$ & 9.81$\times 10^{-8}$ & 1.15$\times 10^{-7}$ & 1.07$\times 10^{-7}$ \\
 4 &  4 & GW190413\_052954     & 2019-04-13T05:29:54.489733 &    33.38 &     0.05 &    33.42 & 1.21$\times 10^{-7}$ & 1.97$\times 10^{-7}$ & 1.36$\times 10^{-7}$ & 1.15$\times 10^{-7}$ & 2.12$\times 10^{-7}$ & 1.28$\times 10^{-7}$ \\
 5 &  5 & GW190413\_134308     & 2019-04-13T13:43:08.711010 &      SAA &     6.94 &     6.94 &        - &        - &        - & 7.40$\times 10^{-8}$ & 2.10$\times 10^{-7}$ & 1.25$\times 10^{-7}$ \\
 6 &  6 & GW190421\_213856     & 2019-04-21T21:38:56.233623 &    65.97 &    40.81 &    99.97 & 1.38$\times 10^{-7}$ & 1.58$\times 10^{-7}$ & 1.44$\times 10^{-7}$ & 5.02$\times 10^{-8}$ & 2.04$\times 10^{-7}$ & 1.30$\times 10^{-7}$ \\
 8 &  8 & GW190426\_190642     & 2019-04-26T19:06:42.630194 &    88.70 &      SAA &    88.70 & 1.10$\times 10^{-7}$ & 2.48$\times 10^{-7}$ & 1.34$\times 10^{-7}$ & 1.09$\times 10^{-7}$ & 2.29$\times 10^{-7}$ & 1.31$\times 10^{-7}$ \\
 9 &  9 & GW190503\_185404     & 2019-05-03T18:54:04.278387 &    96.59 &      SAA &    96.59 & 1.32$\times 10^{-7}$ & 1.36$\times 10^{-7}$ & 1.33$\times 10^{-7}$ & 1.25$\times 10^{-7}$ & 1.28$\times 10^{-7}$ & 1.26$\times 10^{-7}$ \\
10 & 10 & GW190512\_180714     & 2019-05-12T18:07:14.411427 &    30.95 &     0.00 &    30.95 & 1.65$\times 10^{-7}$ & 1.79$\times 10^{-7}$ & 1.76$\times 10^{-7}$ & 1.54$\times 10^{-7}$ & 1.66$\times 10^{-7}$ & 1.64$\times 10^{-7}$ \\
11 & 11 & GW190513\_205428     & 2019-05-13T20:54:28.744647 &    84.97 &     0.00 &    84.97 & 1.07$\times 10^{-7}$ & 1.31$\times 10^{-7}$ & 1.13$\times 10^{-7}$ & 1.06$\times 10^{-7}$ & 1.19$\times 10^{-7}$ & 1.09$\times 10^{-7}$ \\
12 & 12 & GW190514\_065416     & 2019-05-14T06:54:16.868963 &    83.33 &    68.64 &    83.75 & 1.09$\times 10^{-7}$ & 3.03$\times 10^{-7}$ & 1.35$\times 10^{-7}$ & 3.89$\times 10^{-8}$ & 2.79$\times 10^{-7}$ & 8.77$\times 10^{-8}$ \\
13 & 13 & GW190517\_055101     & 2019-05-17T05:51:01.837191 &     6.81 &     4.07 &    10.57 & 1.32$\times 10^{-7}$ & 1.35$\times 10^{-7}$ & 1.34$\times 10^{-7}$ & 4.41$\times 10^{-8}$ & 1.26$\times 10^{-7}$ & 1.02$\times 10^{-7}$ \\
14 & 14 & GW190519\_153544     & 2019-05-19T15:35:44.391148 &    40.53 &     0.00 &    40.53 & 1.13$\times 10^{-7}$ & 1.61$\times 10^{-7}$ & 1.26$\times 10^{-7}$ & 1.06$\times 10^{-7}$ & 1.54$\times 10^{-7}$ & 1.18$\times 10^{-7}$ \\
15 & 15 & GW190521             & 2019-05-21T03:02:29.436318 &    58.61 &    61.27 &    99.98 & 1.64$\times 10^{-7}$ & 3.54$\times 10^{-7}$ & 2.19$\times 10^{-7}$ & 4.20$\times 10^{-8}$ & 3.20$\times 10^{-7}$ & 1.35$\times 10^{-7}$ \\
16 & 16 & GW190521\_074359     & 2019-05-21T07:43:59.468497 &   100.00 &     0.00 &   100.00 & 1.20$\times 10^{-7}$ & 1.61$\times 10^{-7}$ & 1.51$\times 10^{-7}$ & 1.16$\times 10^{-7}$ & 1.48$\times 10^{-7}$ & 1.40$\times 10^{-7}$ \\
17 & 17 & GW190527\_092055     & 2019-05-27T09:20:55.788773 &    72.51 &     0.05 &    72.51 & 1.14$\times 10^{-7}$ & 3.30$\times 10^{-7}$ & 1.91$\times 10^{-7}$ & 1.10$\times 10^{-7}$ & 2.96$\times 10^{-7}$ & 1.74$\times 10^{-7}$ \\
18 & 18 & GW190602\_175927     & 2019-06-02T17:59:27.090972 &    65.84 &      SAA &    65.84 & 1.53$\times 10^{-7}$ & 2.08$\times 10^{-7}$ & 1.89$\times 10^{-7}$ & 1.51$\times 10^{-7}$ & 1.92$\times 10^{-7}$ & 1.76$\times 10^{-7}$ \\
19 & 19 & GW190620\_030421     & 2019-06-20T03:04:21.320868 &      SAA &     4.10 &     4.10 &        - &        - &        - & 8.21$\times 10^{-8}$ & 1.79$\times 10^{-7}$ & 1.39$\times 10^{-7}$ \\
20 & 20 & GW190630\_185205     & 2019-06-30T18:52:05.174378 &    78.32 &      SAA &    78.32 & 1.17$\times 10^{-7}$ & 2.16$\times 10^{-7}$ & 1.30$\times 10^{-7}$ & 1.08$\times 10^{-7}$ & 1.97$\times 10^{-7}$ & 1.20$\times 10^{-7}$ \\
21 & 21 & GW190701\_203306     & 2019-07-01T20:33:06.571927 &   100.00 &    99.51 &   100.00 & 1.27$\times 10^{-7}$ & 1.33$\times 10^{-7}$ & 1.28$\times 10^{-7}$ & 1.01$\times 10^{-7}$ & 1.20$\times 10^{-7}$ & 1.15$\times 10^{-7}$ \\
22 & 22 & GW190706\_222641     & 2019-07-06T22:26:41.331680 &    66.90 &    12.80 &    73.80 & 1.03$\times 10^{-7}$ & 2.95$\times 10^{-7}$ & 1.63$\times 10^{-7}$ & 4.66$\times 10^{-8}$ & 2.79$\times 10^{-7}$ & 1.53$\times 10^{-7}$ \\
23 & 23 & GW190707\_093326     & 2019-07-07T09:33:26.173894 &    42.31 &      SAA &    42.31 & 1.38$\times 10^{-7}$ & 2.34$\times 10^{-7}$ & 1.60$\times 10^{-7}$ & 1.30$\times 10^{-7}$ & 2.12$\times 10^{-7}$ & 1.48$\times 10^{-7}$ \\
24 & 24 & GW190708\_232457     & 2019-07-08T23:24:57.387952 &    56.01 &      SAA &    56.01 & 1.28$\times 10^{-7}$ & 4.24$\times 10^{-7}$ & 1.93$\times 10^{-7}$ & 1.22$\times 10^{-7}$ & 3.77$\times 10^{-7}$ & 1.75$\times 10^{-7}$ \\
25 & 25 & GW190719\_215514     & 2019-07-19T21:55:14.920692 &    74.97 &    15.00 &    89.79 & 1.17$\times 10^{-7}$ & 3.74$\times 10^{-7}$ & 1.85$\times 10^{-7}$ & 3.65$\times 10^{-8}$ & 3.40$\times 10^{-7}$ & 1.60$\times 10^{-7}$ \\
26 & 26 & GW190720\_000836     & 2019-07-20T00:08:36.706197 &    87.90 &      SAA &    87.90 & 1.08$\times 10^{-7}$ & 2.64$\times 10^{-7}$ & 1.19$\times 10^{-7}$ & 1.01$\times 10^{-7}$ & 2.46$\times 10^{-7}$ & 1.12$\times 10^{-7}$ \\
27 & 27 & GW190725\_174728     & 2019-07-25T17:47:28.453920 &      SAA &      SAA &        - &        - &        - &        - &        - &        - &        - \\
28 & 28 & GW190727\_060333     & 2019-07-27T06:03:33.988299 &    61.20 &     0.01 &    61.20 & 1.61$\times 10^{-7}$ & 1.93$\times 10^{-7}$ & 1.74$\times 10^{-7}$ & 1.49$\times 10^{-7}$ & 1.72$\times 10^{-7}$ & 1.58$\times 10^{-7}$ \\
29 & 29 & GW190728\_064510     & 2019-07-28T06:45:10.546183 &    74.03 &    71.13 &    74.03 & 1.07$\times 10^{-7}$ & 2.36$\times 10^{-7}$ & 1.21$\times 10^{-7}$ & 6.33$\times 10^{-8}$ & 2.28$\times 10^{-7}$ & 9.27$\times 10^{-8}$ \\
30 & 30 & GW190731\_140936     & 2019-07-31T14:09:36.628868 &    61.08 &     3.17 &    61.08 & 1.20$\times 10^{-7}$ & 1.99$\times 10^{-7}$ & 1.40$\times 10^{-7}$ & 1.13$\times 10^{-7}$ & 1.88$\times 10^{-7}$ & 1.32$\times 10^{-7}$ \\
31 & 31 & GW190803\_022701     & 2019-08-03T02:27:01.873990 &      SAA &    47.43 &    47.43 &        - &        - &        - & 7.34$\times 10^{-8}$ & 1.47$\times 10^{-7}$ & 1.16$\times 10^{-7}$ \\
32 & 32 & GW190805\_211137     & 2019-08-05T21:11:37.335853 &    91.07 &     7.64 &    98.63 & 1.15$\times 10^{-7}$ & 1.60$\times 10^{-7}$ & 1.43$\times 10^{-7}$ & 4.04$\times 10^{-8}$ & 1.48$\times 10^{-7}$ & 1.27$\times 10^{-7}$ \\
34 & 34 & GW190828\_063405     & 2019-08-28T06:34:05.739895 &    90.53 &    24.93 &    90.53 & 1.34$\times 10^{-7}$ & 2.01$\times 10^{-7}$ & 1.81$\times 10^{-7}$ & 8.58$\times 10^{-8}$ & 1.83$\times 10^{-7}$ & 1.56$\times 10^{-7}$ \\
35 & 35 & GW190828\_065509     & 2019-08-28T06:55:09.868595 &    12.79 &     8.82 &    12.79 & 1.60$\times 10^{-7}$ & 2.98$\times 10^{-7}$ & 2.00$\times 10^{-7}$ & 5.19$\times 10^{-8}$ & 2.67$\times 10^{-7}$ & 1.32$\times 10^{-7}$ \\
36 & 36 & GW190910\_112807     & 2019-09-10T11:28:07.324587 &      SAA &    34.37 &    34.37 &        - &        - &        - & 3.85$\times 10^{-8}$ & 1.77$\times 10^{-7}$ & 8.11$\times 10^{-8}$ \\
37 & 37 & GW190915\_235702     & 2019-09-15T23:57:02.703981 &    94.82 &    76.04 &    94.87 & 1.59$\times 10^{-7}$ & 2.15$\times 10^{-7}$ & 1.86$\times 10^{-7}$ & 5.59$\times 10^{-8}$ & 1.74$\times 10^{-7}$ & 1.13$\times 10^{-7}$ \\
38 & 38 & GW190916\_200658     & 2019-09-16T20:06:58.899706 &    56.57 &     0.00 &    56.57 & 1.20$\times 10^{-7}$ & 1.61$\times 10^{-7}$ & 1.31$\times 10^{-7}$ & 1.12$\times 10^{-7}$ & 1.42$\times 10^{-7}$ & 1.21$\times 10^{-7}$ \\
40 & 40 & GW190924\_021846     & 2019-09-24T02:18:46.826574 &   100.00 &    92.40 &   100.00 & 1.39$\times 10^{-7}$ & 1.61$\times 10^{-7}$ & 1.47$\times 10^{-7}$ & 4.25$\times 10^{-8}$ & 1.43$\times 10^{-7}$ & 8.43$\times 10^{-8}$ \\
41 & 41 & GW190925\_232845     & 2019-09-25T23:28:45.115602 &      SAA &      SAA &        - &        - &        - &        - &        - &        - &        - \\
42 & 42 & GW190926\_050336     & 2019-09-26T05:03:36.066208 &    60.68 &     0.02 &    60.68 & 1.53$\times 10^{-7}$ & 2.84$\times 10^{-7}$ & 2.10$\times 10^{-7}$ & 1.43$\times 10^{-7}$ & 2.57$\times 10^{-7}$ & 1.88$\times 10^{-7}$ \\
43 & 43 & GW190929\_012149     & 2019-09-29T01:21:49.493417 &    73.05 &    34.84 &    73.05 & 1.35$\times 10^{-7}$ & 2.25$\times 10^{-7}$ & 1.75$\times 10^{-7}$ & 4.79$\times 10^{-8}$ & 2.14$\times 10^{-7}$ & 1.59$\times 10^{-7}$ \\
44 & 44 & GW190930\_133541     & 2019-09-30T13:35:41.243613 &    63.05 &     0.56 &    63.05 & 1.55$\times 10^{-7}$ & 2.87$\times 10^{-7}$ & 2.39$\times 10^{-7}$ & 1.46$\times 10^{-7}$ & 2.71$\times 10^{-7}$ & 2.25$\times 10^{-7}$ \\
\hline
\multicolumn{13}{c}{\multirow{2}{*}{continued on next page}} \\

    \end{tabular}
\end{table*}

\renewcommand\thetable{5 continued}
\begin{table*}[htb]
    \centering
    \caption{}
    \label{tab:upperlimitsbbh2}
    \begin{tabular}{HHlHrrrcccccc}\hline

& & & \multicolumn{1}{c}{} & \multicolumn{3}{c}{} & \multicolumn{3}{c}{3$\sigma$ Flux U.L. [erg s$^{-1}$ cm$^{-2}$]} & \multicolumn{3}{c}{5$\sigma$ Flux U.L. [erg s$^{-1}$ cm$^{-2}$]} \\
& & & \multicolumn{1}{c}{} & \multicolumn{3}{c}{Coverage [\%]} & \multicolumn{3}{c}{10--1000 keV} & \multicolumn{3}{c}{15--350 keV} \\ \cline{5-7}
& & Event Name & \multicolumn{1}{H}{Date/Time} & GBM & BAT & \multicolumn{1}{r}{Combined} & Min & Max & \multicolumn{1}{c}{Marginal} & Min & Max & Marginal \\ \hline\hline  

45 &  1 & GW191103\_012549     & 2019-11-03T01:25:49.534235 &    76.96 &    56.21 &    97.35 & 1.57$\times 10^{-7}$ & 3.60$\times 10^{-7}$ & 1.98$\times 10^{-7}$ & 1.25$\times 10^{-7}$ & 3.24$\times 10^{-7}$ & 1.65$\times 10^{-7}$ \\
46 &  2 & GW191105\_143521     & 2019-11-05T14:35:21.926697 &    77.45 &     8.05 &    80.39 & 1.35$\times 10^{-7}$ & 2.01$\times 10^{-7}$ & 1.69$\times 10^{-7}$ & 1.06$\times 10^{-7}$ & 1.91$\times 10^{-7}$ & 1.58$\times 10^{-7}$ \\
47 &  3 & GW191109\_010717     & 2019-11-09T01:07:17.208776 &    89.38 &    29.05 &    89.38 & 1.29$\times 10^{-7}$ & 2.22$\times 10^{-7}$ & 1.55$\times 10^{-7}$ & 1.25$\times 10^{-7}$ & 2.05$\times 10^{-7}$ & 1.47$\times 10^{-7}$ \\
48 &  4 & GW191113\_071753     & 2019-11-13T07:17:53.819268 &    72.98 &     2.27 &    73.00 & 1.53$\times 10^{-7}$ & 2.27$\times 10^{-7}$ & 1.72$\times 10^{-7}$ & 1.45$\times 10^{-7}$ & 2.08$\times 10^{-7}$ & 1.61$\times 10^{-7}$ \\
49 &  5 & GW191126\_115259     & 2019-11-26T11:52:59.632015 &    59.81 &     7.88 &    59.81 & 1.12$\times 10^{-7}$ & 2.09$\times 10^{-7}$ & 1.36$\times 10^{-7}$ & 7.09$\times 10^{-8}$ & 1.96$\times 10^{-7}$ & 1.25$\times 10^{-7}$ \\
50 &  6 & GW191127\_050227     & 2019-11-27T05:02:27.545395 &    89.03 &    77.16 &    89.04 & 1.12$\times 10^{-7}$ & 2.42$\times 10^{-7}$ & 1.40$\times 10^{-7}$ & 3.94$\times 10^{-8}$ & 2.23$\times 10^{-7}$ & 8.96$\times 10^{-8}$ \\
51 &  7 & GW191129\_134029     & 2019-11-29T13:40:29.194979 &      SAA &      SAA &        - &        - &        - &        - &        - &        - &        - \\
52 &  8 & GW191204\_110529     & 2019-12-04T11:05:29.542637 &    48.10 &    25.20 &    66.23 & 1.09$\times 10^{-7}$ & 2.77$\times 10^{-7}$ & 1.42$\times 10^{-7}$ & 3.10$\times 10^{-8}$ & 1.43$\times 10^{-7}$ & 1.07$\times 10^{-7}$ \\
53 &  9 & GW191204\_171526     & 2019-12-04T17:15:26.076726 &      SAA &    87.15 &    87.15 &        - &        - &        - & 2.02$\times 10^{-7}$ & 9.13$\times 10^{-7}$ & 3.97$\times 10^{-7}$ \\
54 & 10 & GW191215\_223052     & 2019-12-15T22:30:52.344737 &    51.87 &    21.09 &    51.87 & 1.41$\times 10^{-7}$ & 1.69$\times 10^{-7}$ & 1.53$\times 10^{-7}$ & 4.40$\times 10^{-8}$ & 1.55$\times 10^{-7}$ & 1.29$\times 10^{-7}$ \\
55 & 11 & GW191216\_213338     & 2019-12-16T21:33:38.487513 &    94.70 &     1.76 &    94.76 & 1.40$\times 10^{-7}$ & 1.64$\times 10^{-7}$ & 1.48$\times 10^{-7}$ & 1.26$\times 10^{-7}$ & 1.50$\times 10^{-7}$ & 1.33$\times 10^{-7}$ \\
57 & 13 & GW191222\_033537     & 2019-12-22T03:35:37.111327 &      SAA &     0.89 &     0.89 &        - &        - &        - & 1.84$\times 10^{-7}$ & 1.92$\times 10^{-7}$ & 1.89$\times 10^{-7}$ \\
58 & 14 & GW191230\_180458     & 2019-12-30T18:04:58.398136 &    40.80 &     0.00 &    40.80 & 1.09$\times 10^{-7}$ & 1.72$\times 10^{-7}$ & 1.39$\times 10^{-7}$ & 1.09$\times 10^{-7}$ & 1.60$\times 10^{-7}$ & 1.33$\times 10^{-7}$ \\
59 & 15 & GW200112\_155838     & 2020-01-12T15:58:38.099159 &      SAA &      SAA &        - &        - &        - &        - &        - &        - &        - \\
61 & 17 & GW200128\_022011     & 2020-01-28T02:20:11.897212 &    45.58 &    23.04 &    45.58 & 1.22$\times 10^{-7}$ & 3.08$\times 10^{-7}$ & 1.42$\times 10^{-7}$ & 3.29$\times 10^{-8}$ & 2.83$\times 10^{-7}$ & 1.07$\times 10^{-7}$ \\
62 & 18 & GW200129\_065458     & 2020-01-29T06:54:58.422370 &     1.36 &     1.16 &     1.36 &        - &        - & 1.42$\times 10^{-7}$ &        - &        - & 6.49$\times 10^{-8}$ \\
63 & 19 & GW200202\_154313     & 2020-02-02T15:43:13.581732 &    99.99 &      SAA &    99.99 & 1.18$\times 10^{-7}$ & 1.26$\times 10^{-7}$ & 1.21$\times 10^{-7}$ & 1.10$\times 10^{-7}$ & 1.18$\times 10^{-7}$ & 1.14$\times 10^{-7}$ \\
64 & 20 & GW200208\_130117     & 2020-02-08T13:01:17.942877 &    99.70 &     0.00 &    99.70 & 1.33$\times 10^{-7}$ & 1.36$\times 10^{-7}$ & 1.36$\times 10^{-7}$ & 1.22$\times 10^{-7}$ & 1.25$\times 10^{-7}$ & 1.25$\times 10^{-7}$ \\
65 & 21 & GW200208\_222617     & 2020-02-08T22:26:17.970136 &      SAA &     5.85 &     5.85 &        - &        - &        - & 6.09$\times 10^{-8}$ & 1.77$\times 10^{-7}$ & 1.06$\times 10^{-7}$ \\
66 & 22 & GW200209\_085452     & 2020-02-09T08:54:52.182059 &    61.47 &     7.05 &    61.63 & 1.27$\times 10^{-7}$ & 1.69$\times 10^{-7}$ & 1.35$\times 10^{-7}$ & 3.71$\times 10^{-8}$ & 1.28$\times 10^{-7}$ & 1.17$\times 10^{-7}$ \\
68 & 24 & GW200216\_220804     & 2020-02-16T22:08:04.896274 &      SAA &    38.42 &    38.42 &        - &        - &        - & 8.98$\times 10^{-8}$ & 1.71$\times 10^{-7}$ & 1.47$\times 10^{-7}$ \\
69 & 25 & GW200219\_094415     & 2020-02-19T09:44:15.187768 &    20.36 &      SAA &    20.36 & 1.37$\times 10^{-7}$ & 1.73$\times 10^{-7}$ & 1.52$\times 10^{-7}$ & 1.29$\times 10^{-7}$ & 1.58$\times 10^{-7}$ & 1.41$\times 10^{-7}$ \\
70 & 26 & GW200220\_061928     & 2020-02-20T06:19:28.669171 &    99.63 &     0.05 &    99.65 & 1.33$\times 10^{-7}$ & 2.12$\times 10^{-7}$ & 1.65$\times 10^{-7}$ & 1.22$\times 10^{-7}$ & 1.87$\times 10^{-7}$ & 1.48$\times 10^{-7}$ \\
71 & 27 & GW200220\_124850     & 2020-02-20T12:48:50.151796 &    63.37 &    21.46 &    80.25 & 1.10$\times 10^{-7}$ & 2.34$\times 10^{-7}$ & 1.28$\times 10^{-7}$ & 3.07$\times 10^{-8}$ & 2.18$\times 10^{-7}$ & 1.10$\times 10^{-7}$ \\
72 & 28 & GW200224\_222234     & 2020-02-24T22:22:34.393320 &      SAA &    98.76 &    98.76 &        - &        - &        - & 1.13$\times 10^{-7}$ & 1.40$\times 10^{-7}$ & 1.27$\times 10^{-7}$ \\
73 & 29 & GW200225\_060421     & 2020-02-25T06:04:21.404204 &    87.61 &     1.30 &    87.61 & 1.37$\times 10^{-7}$ & 3.32$\times 10^{-7}$ & 2.36$\times 10^{-7}$ & 1.28$\times 10^{-7}$ & 3.10$\times 10^{-7}$ & 2.23$\times 10^{-7}$ \\
74 & 30 & GW200302\_015811     & 2020-03-02T01:58:11.520688 &    67.41 &    23.35 &    67.92 & 1.04$\times 10^{-7}$ & 3.40$\times 10^{-7}$ & 1.62$\times 10^{-7}$ & 3.27$\times 10^{-8}$ & 1.77$\times 10^{-7}$ & 1.19$\times 10^{-7}$ \\
75 & 31 & GW200306\_093714     & 2020-03-06T09:37:14.127646 &    72.37 &    30.46 &    90.36 & 1.18$\times 10^{-7}$ & 2.54$\times 10^{-7}$ & 1.46$\times 10^{-7}$ & 5.38$\times 10^{-8}$ & 2.40$\times 10^{-7}$ & 1.33$\times 10^{-7}$ \\
76 & 32 & GW200308\_173609     & 2020-03-08T17:36:09.687320 &    70.53 &     4.40 &    70.89 & 1.19$\times 10^{-7}$ & 3.39$\times 10^{-7}$ & 1.92$\times 10^{-7}$ & 8.13$\times 10^{-8}$ & 3.20$\times 10^{-7}$ & 1.77$\times 10^{-7}$ \\
77 & 33 & GW200311\_115853     & 2020-03-11T11:58:53.376063 &      SAA &      N/A &        - &        - &        - &        - &        - &        - &        - \\
78 & 34 & GW200316\_215756     & 2020-03-16T21:57:56.166711 &    15.69 &    13.49 &    15.69 & 1.14$\times 10^{-7}$ & 1.51$\times 10^{-7}$ & 1.34$\times 10^{-7}$ & 1.09$\times 10^{-7}$ & 1.36$\times 10^{-7}$ & 1.22$\times 10^{-7}$ \\
79 & 35 & GW200322\_091133     & 2020-03-22T09:11:33.283116 &    75.18 &    13.82 &    89.00 & 1.06$\times 10^{-7}$ & 3.70$\times 10^{-7}$ & 1.54$\times 10^{-7}$ & 4.49$\times 10^{-8}$ & 3.38$\times 10^{-7}$ & 1.39$\times 10^{-7}$ \\
\hline \hline
    \end{tabular}
\end{table*}

\subsection{Marginal GWs}\label{ssec:marginal_discussion}
Although all 6 marginal GWs from Table~\ref{tab:margGWevents} have \pastro~$<$~0.5 they are of interest for EM follow-up. This is because GW200311\_103121 may have a possible BNS origin and the remaining 5 candidates have possible NSBH origins \citep{GWTC2.1,GWTC-3}. In particular, the possible NSBH merger GW200105\_162426 was noted as a clear outlier from experimental backgrounds despite not satisfying the \pastro~$>$~0.5 criteria used to identify GW signals with a likely astrophysical origin. It also has the highest observed $\pastro$ of all the marginal GWs.

\begin{figure*}
    \centering
    \includegraphics[width=0.75\textwidth]{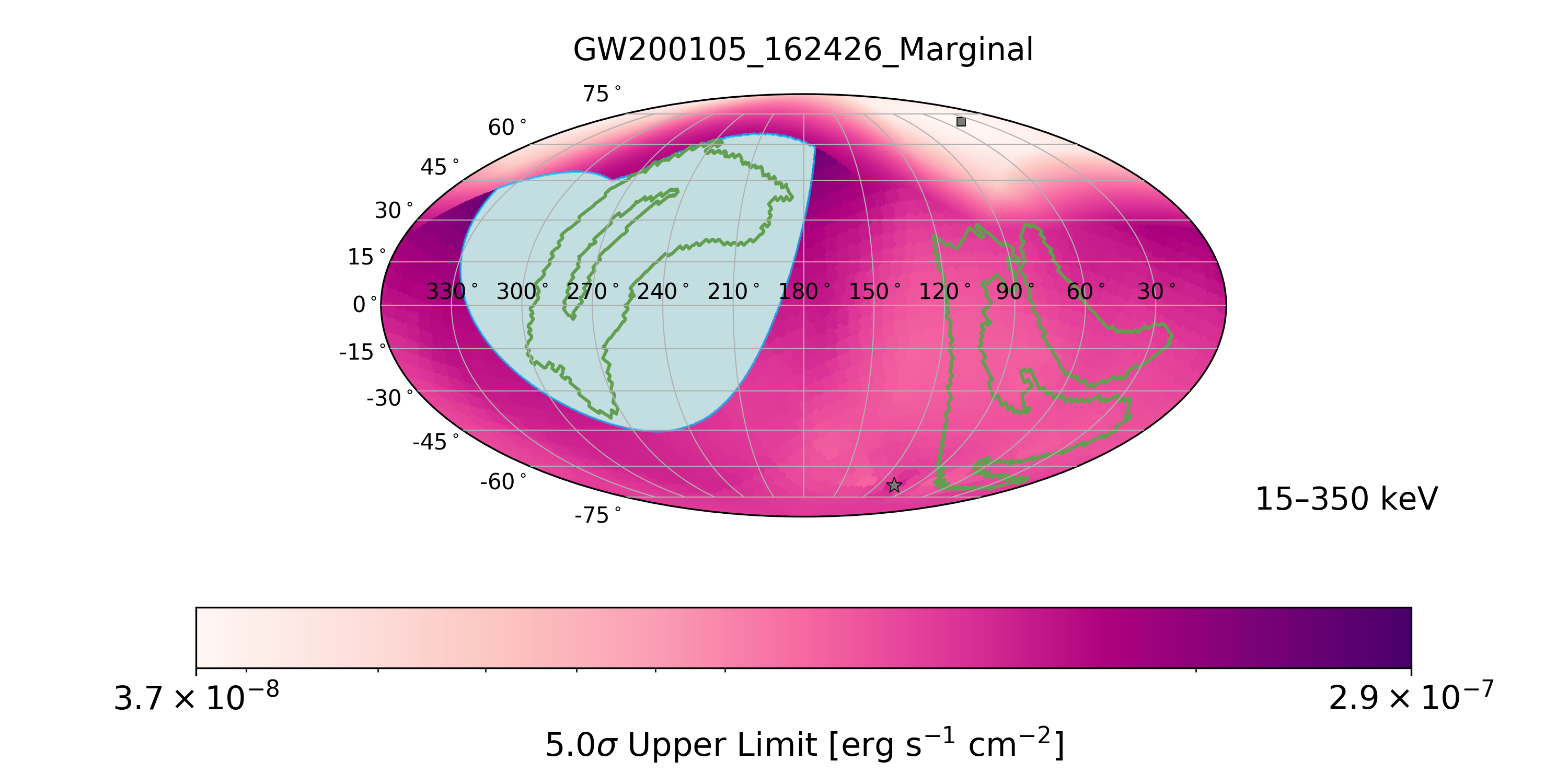}
    \caption{The 5$\sigma$ upper-limits as a function of sky position for GW200105\_162426, the marginal GW with the highest \pastro. The purple gradient represents the combined \FermiGBM\ and \SwiftBAT\ flux upper limits for source positions at each point on the sky. The star symbol represents the zenith direction of \FermiGBM, the square symbol represents the center of the \SwiftBAT\ field-of-view, and the green contour represents the 90\% credible area of the LVK localization. The blue region is the non-visible portion of the sky which is occulted by the Earth for \FermiGBM\ and outside the \SwiftBAT\ FoV.}
    \label{fig:GW200105_ul_skymap}
\end{figure*}

The 5 marginal GWs with possible NSBH origins were visible to \FermiGBM\ while the remaining one, GW200311\_103121, occurred when \FermiGBM\ was in the SAA. None of the marginal GWs have appreciable coverage in \SwiftBAT. No significant counterparts were found. As with GW190425, this may be due to unfavorable viewing angles with respect to the jet axis, larger observational distances such as the $0.27_{-0.11}^{+0.12}$ Gpc distance to GW200105\_162426 \citep{GWTC-3}, and partial sky coverage for candidates other than GW190426\_152155. It therefore remains ambiguous as to whether these signals are real compact binary coalescences. Nevertheless, we provide in Table~\ref{tab:marginal_upperlimits} the flux upper limits for each marginal GW calculated according to the same methods described in Section~\ref{sec:upperlimits_region1} since they may provide emission model constraints if future analyses can identify an astrophysical progenitor with a favorable viewing angle with respect to the jet axis. Figure~\ref{fig:GW200105_ul_skymap} displays the 5$\sigma$ confidence level flux upper limit map for GW200105\_162426 since it is the marginal GW with the highest probability of having an astrophysical origin. The marginalized 5$\sigma$ flux upper limit of GW200105\_162426 yields an isotropic-equivalent luminosity upper limit of $L_\mathrm{iso} = 2.1 \times 10^{48}$~erg~s$^{-1}$ when combined with its 0.27 Gpc distance. The data release~\citep{O3FollowupRelease} associated with this work provides flux upper-limits as a function of sky position for the remaining marginal GWs.

\renewcommand\thetable{6}
\begin{table*}
    \centering
    \caption{Flux upper limits for possible EM counterparts to marginal GW candidates with (FAR $<$ 2 yr$^{-1}$, \pastro~$<$~0.5). The 3$\sigma$ upper limits are computed for the 10--1000 keV energy range over the FoV of   \FermiGBM. The 5$\sigma$ upper limits are computed for the combined coverage of \FermiGBM~and \SwiftBAT\ with both instruments matched to the 15--350 keV energy range of \SwiftBAT. The columns labeled Min and Max correspond, respectively, to the minimum and maximum upper limits obtained for points within the 90\% credible level of the GW candidate localization. The Marginal upper limit is computed by integrating the upper limits produced at individual locations over the full sky using the GW localization as a weighted prior, normalized to the visible portion of the sky.
    Also shown is the visible coverage percentage of the full GW localization for \FermiGBM\ alone, \SwiftBAT\ alone, and the combined FoV from both instruments.}
    \label{tab:marginal_upperlimits}

    \begin{tabular}{HHlHrrrcccccc}\hline
& & & \multicolumn{1}{c}{} & \multicolumn{3}{c}{} & \multicolumn{3}{c}{3$\sigma$ Flux U.L. [erg s$^{-1}$ cm$^{-2}$]} & \multicolumn{3}{c}{5$\sigma$ Flux U.L. [erg s$^{-1}$ cm$^{-2}$]} \\
& & & \multicolumn{1}{c}{} & \multicolumn{3}{c}{Coverage [\%]} & \multicolumn{3}{c}{10--1000 keV} & \multicolumn{3}{c}{15--350 keV} \\ \cline{5-7}
& & Event Name & \multicolumn{1}{H}{Date/Time} & GBM & BAT & \multicolumn{1}{r}{Combined} & Min & Max & \multicolumn{1}{c}{Marginal} & Min & Max & Marginal \\ \hline\hline  
 1 &  1 & GW190426\_152155 & 2019-04-26T15:21:55.360838 &   100.00 &      SAA &   100.00 & 1.03$\times 10^{-7}$ & 1.65$\times 10^{-7}$ & 1.30$\times 10^{-7}$ & 1.00$\times 10^{-7}$ & 1.53$\times 10^{-7}$ & 1.21$\times 10^{-7}$ \\
 2 &  2 & GW190531\_023648 & 2019-05-31T02:36:48.327961 &    86.90 &     0.03 &    86.91 & 1.35$\times 10^{-7}$ & 2.63$\times 10^{-7}$ & 1.74$\times 10^{-7}$ & 1.26$\times 10^{-7}$ & 2.42$\times 10^{-7}$ & 1.60$\times 10^{-7}$ \\
 3 &  1 & GW191118\_212859 & 2019-11-18T21:28:59.431714 &    93.63 &      SAA &    93.63 & 1.07$\times 10^{-7}$ & 2.93$\times 10^{-7}$ & 1.19$\times 10^{-7}$ & 1.05$\times 10^{-7}$ & 2.71$\times 10^{-7}$ & 1.13$\times 10^{-7}$ \\
 4 &  2 & GW200105\_162426 & 2020-01-05T16:24:26.046855 &    53.57 &     3.01 &    54.36 & 1.13$\times 10^{-7}$ & 1.53$\times 10^{-7}$ & 1.26$\times 10^{-7}$ & 1.06$\times 10^{-7}$ & 1.69$\times 10^{-7}$ & 1.18$\times 10^{-7}$ \\
 5 &  3 & GW200201\_203549 & 2020-02-01T20:35:49.033606 &    86.01 &      SAA &    86.01 & 1.17$\times 10^{-7}$ & 1.80$\times 10^{-7}$ & 1.31$\times 10^{-7}$ & 1.09$\times 10^{-7}$ & 1.64$\times 10^{-7}$ & 1.22$\times 10^{-7}$ \\
 6 &  4 & GW200311\_103121 & 2020-03-11T10:31:21.731916 &      SAA &     0.01 &     0.01 &        - &        - &        - &        - &        - & 1.09$\times 10^{-7}$ \\
\hline \hline
    \end{tabular}
\end{table*}

\section{Summary and Future Directions}\label{sec:conclusion}
Using the 79 GW candidates with \pastro\ $>$ 0.5 from O3 that were reported in GWTC-3, we searched for coincident EM counterparts with \FermiGBM\ and \SwiftBAT. This represents the most comprehensive follow-up to date of the O3 run in the hard X-ray and gamma-ray regime. We found no significant counterparts in either instrument. For the one BNS merger, GW190425, with \pastro\ $>$ 0.5 there are several possible reasons for the non-detection of a counterpart: 
\begin{itemize} 
\item The combined \FermiGBM\ and \SwiftBAT\ coverage of the GW localization area was $\sim$60\%, meaning that the GW source may have been outside the FoV for both instruments.
\item The distance to GW190425 was four times larger than the estimated distance for GW170817, causing the observed flux to be below the detection threshold in both instruments if it had the same intrinsic luminosity and viewing angle as GW170817. 
\item The viewing angle may have been too far away from the jet axis to detect emission under scenarios with a clean or structured jet.
\end{itemize}
GW190425 is therefore unconstrained by our observations.

In contrast to GW190425, the large number of BBH detections in this sample allowed us to begin placing constraints on certain models of gamma-ray emission from BBH mergers, despite the larger average distance for this class compared to the BNS mergers. The most constrained model was the $\nu\bar{\nu}$ model where two GWs, GW191216\_213338 and GW200202\_154313, were predicted to produce an observable flux in \FermiGBM. With the number of GWs from BBH mergers to increase in the fourth LVK observing run (O4), we expect this model to become more constrained and ruled out as a potential explanation of EM emission from BBH mergers.

With O4 having begun on May 24, 2023, we expect an increase in GW detections by a factor of 5 \citep{2018LivingReviewLVK, Petrov2022} and a bountiful regime of EM follow-up data.  This will greatly increase the need for further instantaneous, wide FoV gamma-ray/X-ray observations in order to detect the EM counterparts to these GWs and localize them, especially given the absence of a counterpart detection during O3. Towards this end, the \FermiGBM\ Targeted Search updates presented in this paper will be used in future LVK observing runs. In the absence of detections, flux upper limits, both marginalized and as a function of sky position, will be provided to the community. Additionally, \SwiftBAT\ GUANO and NITRATES will be used during the next observing run.

\section*{Acknowledgements}
\label{sec:acknowledgements}
The USRA co-authors gratefully acknowledge NASA funding through contract 80MSFC17M0022. R.H. acknowledges funding from the European Union’s Horizon 2020 research and innovation programme under the Marie Skłodowska-Curie grant agreement No 945298-ParisRegionFP. The UAH co-authors gratefully acknowledge NASA funding from co-operative agreement 80MSFC22M0004. The NASA authors gratefully acknowledge NASA funding through the Fermi GBM project.  This research also made use of Astropy, a community developed core Python package for Astronomy \citep{astropy:2022}; NumPy \citep{harris2020array}; SciPy \citep{2020SciPy-NMeth} and matplotlib, a
Python library for publication quality graphics \citep{Hunter:2007}.

The Swift authors acknowledge the use of public data from the \textit{Swift} data archive. MC acknowledges support from NASA under award number 80GSFC21M0002 and from Vetenskapsr\r{a}det through project number 31004019. JD acknowledges support by the NSF under award numbers PHY-1913607 and PHY-2209445.


This material is based upon work supported by NSF’s LIGO Laboratory which is a major facility
fully funded by the National Science Foundation.
The authors also gratefully acknowledge the support of
the Science and Technology Facilities Council (STFC) of the
United Kingdom, the Max-Planck-Society (MPS), and the State of
Niedersachsen/Germany for support of the construction of Advanced LIGO 
and construction and operation of the GEO\,600 detector. 
Additional support for Advanced LIGO was provided by the Australian Research Council.
The authors gratefully acknowledge the Italian Istituto Nazionale di Fisica Nucleare (INFN),  
the French Centre National de la Recherche Scientifique (CNRS) and
the Netherlands Organization for Scientific Research (NWO), 
for the construction and operation of the Virgo detector
and the creation and support  of the EGO consortium. 
The authors also gratefully acknowledge research support from these agencies as well as by 
the Council of Scientific and Industrial Research of India, 
the Department of Science and Technology, India,
the Science \& Engineering Research Board (SERB), India,
the Ministry of Human Resource Development, India,
the Spanish Agencia Estatal de Investigaci\'on (AEI),
the Spanish Ministerio de Ciencia e Innovaci\'on and Ministerio de Universidades,
the Conselleria de Fons Europeus, Universitat i Cultura and the Direcci\'o General de Pol\'{\i}tica Universitaria i Recerca del Govern de les Illes Balears,
the Conselleria d'Innovaci\'o, Universitats, Ci\`encia i Societat Digital de la Generalitat Valenciana and
the CERCA Programme Generalitat de Catalunya, Spain,
the National Science Centre of Poland and the European Union – European Regional Development Fund; Foundation for Polish Science (FNP),
the Swiss National Science Foundation (SNSF),
the Russian Foundation for Basic Research, 
the Russian Science Foundation,
the European Commission,
the European Social Funds (ESF),
the European Regional Development Funds (ERDF),
the Royal Society, 
the Scottish Funding Council, 
the Scottish Universities Physics Alliance, 
the Hungarian Scientific Research Fund (OTKA),
the French Lyon Institute of Origins (LIO),
the Belgian Fonds de la Recherche Scientifique (FRS-FNRS), 
Actions de Recherche Concertées (ARC) and
Fonds Wetenschappelijk Onderzoek – Vlaanderen (FWO), Belgium,
the Paris \^{I}le-de-France Region, 
the National Research, Development and Innovation Office Hungary (NKFIH), 
the National Research Foundation of Korea,
the Natural Science and Engineering Research Council Canada,
Canadian Foundation for Innovation (CFI),
the Brazilian Ministry of Science, Technology, and Innovations,
the International Center for Theoretical Physics South American Institute for Fundamental Research (ICTP-SAIFR), 
the Research Grants Council of Hong Kong,
the National Natural Science Foundation of China (NSFC),
the Leverhulme Trust, 
the Research Corporation,
the National Science and Technology Council (NSTC), Taiwan,
the United States Department of Energy,
and
the Kavli Foundation.
The authors gratefully acknowledge the support of the NSF, STFC, INFN and CNRS for provision of computational resources.

This work was supported by MEXT, JSPS Leading-edge Research Infrastructure Program, JSPS Grant-in-Aid for Specially Promoted Research 26000005, JSPS Grant-inAid for Scientific Research on Innovative Areas 2905: JP17H06358, JP17H06361 and JP17H06364, JSPS Core-to-Core Program A. Advanced Research Networks, JSPS Grantin-Aid for Scientific Research (S) 17H06133 and 20H05639 , JSPS Grant-in-Aid for Transformative Research Areas (A) 20A203: JP20H05854, the joint research program of the Institute for Cosmic Ray Research, University of Tokyo, National Research Foundation (NRF), Computing Infrastructure Project of Global Science experimental Data hub Center (GSDC) at KISTI, Korea Astronomy and Space Science Institute (KASI), and Ministry of Science and ICT (MSIT) in Korea, Academia Sinica (AS), AS Grid Center (ASGC) and the National Science and Technology Council (NSTC) in Taiwan under grants including the Rising Star Program and Science Vanguard Research Program, Advanced Technology Center (ATC) of NAOJ, and Mechanical Engineering Center of KEK.

Additional LSC-Virgo-KAGRA acknowledgements for support of individual authors may be found in the following document: \\
\texttt{https://dcc.ligo.org/LIGO-M2300033/public}.
%
\bibliographystyle{aasjournal}
\bibliography{main}


\clearpage
\appendix
\section{Flux Upper Limits for Probable BBH Mergers}\label{appendix}
Here we present the sky-marginalized 3$\sigma$ flux upper limits for the probable BBH mergers described in Section \ref{sec:bbh_region3} as well as the 0.95 percentile fluxes for different models of BBH emission. The upper limits are constructed over a 10--1000 keV energy range according to the method in Section~\ref{sec:method}. They assume the spectral shape of potential emission follows the normal spectral template from Table~\ref{tab:targeted_search_templates}
with a $1\,\mathrm{s}$ emission duration.

\clearpage
\renewcommand\thetable{7}
\begin{table}
    \centering
\begin{tabular}{ccccccc}
Name & Waveform (visible frac.) & F$_{Q}$ &  F$_{\nu\bar{\nu}}$ &  F$_{\rm BZ}$ & F$_{\rm GW}$&  UL\\
\hline
GW190403\_051519     & IMRPhenomXPHM (49, 36) & ${74, 70, 67}$ & ${7.3, 6.7, 6.4}$ & ${7.3, 5.8, 5.6}$ & ${2.5, 1.8, 1.7}$ &   178 \\
GW190403\_051519     & SEOBNRv4PHM (39, 30) & ${44, 35, 35}$ & ${5.0, 4.0, 3.8}$ & ${6.5, 4.6, 4.5}$ & ${3.4, 1.9, 1.7}$ &   178 \\
GW190412             & IMRPhenomXPHM (6, 0) & *${137, 83, 95}$ & *${299, 165, 173}$ & ${59, 30, 25}$ & ${64, 28, 22}$ &   111 \\
GW190412             & SEOBNRv4PHM (12, 4) & *${148, 85, 82}$ & *${356, 178, 172}$ & ${48, 25, 23}$ & ${70, 29, 25}$ &   111 \\
GW190413\_052954     & IMRPhenomXPHM (30, 19) & ${11, 5.8, 5.2}$ & ${10, 5.7, 5.2}$ & ${10, 4.4, 4.0}$ & ${10, 4.2, 3.6}$ &   136 \\
GW190421\_213856     & IMRPhenomXPHM (23, 14) & ${21, 8.1, 7.5}$ & ${14, 6.1, 5.8}$ & ${36, 13, 12}$ & ${27, 10, 9.6}$ &   144 \\
GW190426\_190642     & IMRPhenomXPHM (22, 14) & ${32, 31, 29}$ & ${2.7, 2.3, 2.1}$ & ${75, 24, 22}$ & ${26, 8.0, 7.1}$ &   134 \\
GW190503\_185404     & IMRPhenomXPHM (30, 20) & ${50, 20, 18}$ & ${38, 16, 15}$ & ${68, 29, 28}$ & ${54, 20, 18}$ &   133 \\
GW190503\_185404     & SEOBNRv4PHM (25, 16) & ${42, 22, 21}$ & ${32, 17, 16}$ & ${53, 25, 24}$ & ${44, 19, 18}$ &   133 \\
GW190512\_180714     & IMRPhenomXPHM (27, 17) & ${49, 24, 24}$ & ${107, 55, 54}$ & ${20, 7.2, 6.6}$ & ${28, 10, 9.2}$ &   176 \\
GW190512\_180714     & SEOBNRv4PHM (22, 13) & ${45, 18, 17}$ & ${104, 46, 43}$ & ${18, 6.1, 5.2}$ & ${28, 9.5, 8.2}$ &   176 \\
GW190513\_205428     & IMRPhenomXPHM (31, 19) & ${42, 24, 23}$ & ${35, 21, 20}$ & ${16, 9.2, 8.4}$ & ${17, 8.2, 7.5}$ &   113 \\
GW190513\_205428     & SEOBNRv4PHM (30, 19) & ${52, 26, 24}$ & ${36, 20, 19}$ & ${15, 8.3, 7.9}$ & ${17, 8.0, 7.0}$ &   113 \\
GW190514\_065416     & IMRPhenomXPHM (24, 15) & ${12, 5.1, 4.8}$ & ${7.8, 3.8, 3.7}$ & ${21, 8.7, 7.3}$ & ${14, 6.0, 5.4}$ &   135 \\
GW190514\_065416     & SEOBNRv4PHM (21, 13) & ${12, 6.9, 6.1}$ & ${7.5, 4.4, 4.2}$ & ${17, 8.0, 7.8}$ & ${12, 6.0, 5.5}$ &   135 \\
GW190517\_055101     & IMRPhenomXPHM (10, 5) & *${485, 198, 173}$ & *${202, 85, 74}$ & ${48, 10, 7.9}$ & ${60, 18, 14}$ &   134 \\
GW190517\_055101     & SEOBNRv4PHM (14, 8) & *${360, 171, 145}$ & *${142, 70, 62}$ & ${29, 9.3, 7.6}$ & ${53, 16, 14}$ &   134 \\
GW190519\_153544     & IMRPhenomXPHM (6, 3) & ${49, 19, 19}$ & ${14, 6.1, 5.8}$ & ${47, 12, 9.7}$ & ${32, 7.1, 5.5}$ &   126 \\
GW190519\_153544     & SEOBNRv4PHM (5, 2) & ${49, 21, 17}$ & ${12, 6.3, 4.9}$ & ${38, 17, 14}$ & ${26, 9.6, 7.7}$ &   126 \\
GW190521\_074359     & IMRPhenomXPHM (7, 3) & *${222, 58, 49}$ & ${142, 44, 38}$ & *${259, 76, 69}$ & *${228, 65, 57}$ &   151 \\
GW190521\_074359     & SEOBNRv4PHM (25, 14) & *${99, 51, 47}$ & ${62, 31, 29}$ & *${94, 37, 35}$ & *${87, 35, 31}$ &   151 \\
GW190521             & IMRPhenomXPHM (19, 11) & ${15, 3.4, 3.2}$ & ${2.9, 0.77, 0.72}$ & ${121, 21, 18}$ & ${38, 6.9, 5.9}$ &   219 \\
GW190527\_092055     & IMRPhenomXPHM (21, 13) & ${44, 16, 15}$ & ${31, 14, 13}$ & ${27, 13, 12}$ & ${23, 9.5, 8.7}$ &   191 \\
GW190527\_092055     & SEOBNRv4PHM (22, 14) & ${34, 16, 15}$ & ${29, 15, 14}$ & ${23, 10, 9.7}$ & ${24, 10, 10}$ &   191 \\
GW190602\_175927     & IMRPhenomXPHM (31, 20) & ${24, 14, 13}$ & ${6.8, 4.2, 3.9}$ & ${61, 33, 32}$ & ${29, 13, 12}$ &   189 \\
GW190630\_185205     & IMRPhenomXPHM (21, 14) & *${225, 75, 70}$ & *${215, 78, 73}$ & ${145, 56, 51}$ & ${167, 56, 51}$ &   130 \\
GW190630\_185205     & SEOBNRv4PHM (26, 18) & *${195, 93, 88}$ & *${185, 89, 83}$ & ${131, 56, 53}$ & ${145, 60, 57}$ &   130 \\
GW190701\_203306     & IMRPhenomXPHM (32, 20) & ${20, 11, 10}$ & ${10, 6.0, 5.5}$ & ${51, 26, 24}$ & ${31, 15, 13}$ &   128 \\
GW190701\_203306     & SEOBNRv4PHM (35, 23) & ${21, 14, 14}$ & ${9.7, 6.9, 6.7}$ & ${48, 29, 26}$ & ${29, 16, 15}$ &   128 \\
GW190706\_222641     & IMRPhenomXPHM (20, 12) & ${53, 37, 34}$ & ${11, 8.4, 8.2}$ & ${63, 43, 40}$ & ${31, 19, 18}$ &   163 \\
GW190706\_222641     & SEOBNRv4PHM (19, 10) & ${41, 27, 29}$ & ${9.1, 6.9, 7.5}$ & ${62, 44, 46}$ & ${33, 22, 23}$ &   163 \\
GW190707\_093326     & IMRPhenomXPHM (37, 26) & ${174, 114, 109}$ & *${961, 632, 602}$ & ${21, 13, 12}$ & ${63, 38, 35}$ &   160 \\
GW190707\_093326     & SEOBNRv4PHM (31, 20) & ${151, 76, 70}$ & *${821, 429, 389}$ & ${20, 9.8, 8.9}$ & ${57, 27, 24}$ &   160 \\
GW190708\_232457     & IMRPhenomXPHM (39, 26) & ${152, 96, 91}$ & *${415, 260, 248}$ & ${35, 20, 18}$ & ${70, 39, 36}$ &   193 \\
GW190708\_232457     & SEOBNRv4PHM (36, 25) & ${125, 74, 72}$ & *${338, 203, 197}$ & ${37, 19, 19}$ & ${64, 34, 34}$ &   193 \\
GW190719\_215514     & IMRPhenomXPHM (25, 16) & ${51, 29, 28}$ & ${25, 13, 13}$ & ${16, 10, 9.2}$ & ${14, 7.2, 6.4}$ &   185 \\
GW190720\_000836     & IMRPhenomXPHM (44, 32) & *${261, 234, 266}$ & *${1054, 759, 785}$ & ${19, 12, 12}$ & ${54, 28, 26}$ &   119 \\
GW190720\_000836     & SEOBNRv4PHM (35, 24) & *${194, 126, 119}$ & *${852, 555, 515}$ & ${19, 11, 10}$ & ${51, 28, 26}$ &   119 \\
GW190727\_060333     & IMRPhenomXPHM (30, 19) & ${23, 11, 10}$ & ${14, 7.8, 7.3}$ & ${16, 6.8, 6.1}$ & ${15, 6.1, 5.4}$ &   174 \\
GW190727\_060333     & SEOBNRv4PHM (25, 16) & ${21, 12, 11}$ & ${12, 7.8, 7.3}$ & ${14, 6.8, 6.3}$ & ${14, 6.1, 5.6}$ &   174 \\
GW190728\_064510     & IMRPhenomXPHM (32, 21) & *${253, 125, 125}$ & *${1130, 462, 459}$ & ${21, 8.4, 7.6}$ & ${58, 17, 15}$ &   121 \\
GW190728\_064510     & SEOBNRv4PHM (30, 19) & *${208, 79, 75}$ & *${1015, 408, 388}$ & ${17, 6.0, 5.6}$ & ${54, 18, 16}$ &   121 \\
GW190731\_140936     & IMRPhenomXPHM (28, 18) & ${20, 11, 10}$ & ${12, 7.4, 6.8}$ & ${24, 11, 10}$ & ${19, 9.3, 8.7}$ &   140 \\
GW190731\_140936     & SEOBNRv4PHM (27, 18) & ${26, 17, 16}$ & ${13, 9.4, 9.1}$ & ${24, 12, 10}$ & ${20, 11, 9.9}$ &   140 \\
GW190805\_211137     & IMRPhenomXPHM (22, 14) & ${28, 13, 12}$ & ${8.7, 4.8, 4.6}$ & ${7.4, 2.2, 1.9}$ & ${7.3, 2.2, 1.9}$ &   143 \\
GW190805\_211137     & SEOBNRv4PHM (18, 11) & ${21, 13, 13}$ & ${6.6, 4.7, 4.8}$ & ${7.9, 2.6, 2.2}$ & ${6.9, 2.4, 2.3}$ &   143 \\
\hline
\end{tabular} 
\caption{Table showing the 0.95 percentile fluxes from different models of BBH emission. The two numbers after the waveform names indicate the percentage of the cases where the jet is pointing towards Earth in the uniform 10-40 degree opening angle and in the fixed 20 degrees opening angle case respectively. Flux units are $10^{-9}$ erg cm$^{-2}$ s$^{-1}$. The three numbers in each cell represent the isotropic emission, the uniform-distributed jet opening angle  and the fixed jet opening angle. The upper limits (UL) are the 3$\sigma$, 10--1000 keV range values from Table \ref{tab:upperlimitsbbh}. Stars mark instances where the isotropic emission exceeds the UL in more than 10\% of the cases.}
    \label{tab:ULmodel}
\end{table}

\clearpage
\renewcommand\thetable{7 continued}
\begin{table}
    \centering
\begin{tabular}{ccccccc}
Name & Waveform (visible frac.)  & F$_{Q}$ &  F$_{\nu\bar{\nu}}$ &  F$_{\rm BZ}$ & F$_{\rm GW}$&  UL\\
\hline
GW190828\_063405     & IMRPhenomXPHM (31, 21) & ${58, 25, 23}$ & ${48, 22, 21}$ & ${27, 7.7, 6.8}$ & ${33, 9.9, 8.5}$ &   181 \\
GW190828\_065509     & IMRPhenomXPHM (22, 12) & ${43, 19, 18}$ & ${103, 48, 44}$ & ${16, 6.4, 5.4}$ & ${23, 7.6, 6.3}$ &   200 \\
GW190915\_235702     & IMRPhenomXPHM (18, 10) & ${39, 17, 14}$ & ${37, 18, 16}$ & ${28, 11, 10}$ & ${29, 11, 10}$ &   186 \\
GW190915\_235702     & SEOBNRv4PHM (20, 12) & ${42, 25, 24}$ & ${39, 23, 22}$ & ${31, 12, 10}$ & ${32, 13, 11}$ &   186 \\
GW190916\_200658     & IMRPhenomXPHM (28, 18) & ${13, 6.6, 6.2}$ & ${6.8, 3.5, 3.2}$ & ${8.4, 4.0, 3.6}$ & ${6.7, 2.3, 2.0}$ &   131 \\
GW190916\_200658     & SEOBNRv4PHM (25, 16) & ${12, 6.1, 6.0}$ & ${6.1, 3.4, 3.4}$ & ${8.4, 4.9, 4.2}$ & ${5.9, 2.4, 2.2}$ &   131 \\
GW190924\_021846     & IMRPhenomXPHM (32, 22) & *${351, 258, 281}$ & *${3217, 1887, 1884}$ & ${21, 12, 12}$ & ${78, 31, 28}$ &   147 \\
GW190924\_021846     & SEOBNRv4PHM (25, 16) & *${317, 131, 120}$ & *${3158, 1364, 1253}$ & ${17, 6.4, 5.8}$ & ${75, 28, 25}$ &   147 \\
GW190926\_050336     & IMRPhenomXPHM (10, 6) & ${14, 2.9, 2.6}$ & ${11, 3.3, 3.0}$ & ${21, 3.9, 3.5}$ & ${13, 2.5, 2.2}$ &   210 \\
GW190929\_012149     & IMRPhenomXPHM (9, 5) & ${12, 2.8, 2.6}$ & ${4.9, 1.5, 1.4}$ & ${34, 14, 12}$ & ${14, 4.3, 3.7}$ &   175 \\
GW190930\_133541     & IMRPhenomXPHM (34, 23) & *${363, 236, 252}$ & *${1483, 790, 776}$ & ${23, 12, 12}$ & ${69, 26, 23}$ &   239 \\
GW190930\_133541     & SEOBNRv4PHM (32, 21) & *${264, 141, 136}$ & *${1086, 566, 534}$ & ${24, 11, 11}$ & ${60, 26, 24}$ &   239 \\
GW191103\_012549     & IMRPhenomXPHM (34, 23) & *${332, 234, 230}$ & *${1349, 831, 801}$ & ${15, 9.3, 8.6}$ & ${53, 27, 24}$ &   198 \\
GW191103\_012549     & SEOBNRv4PHM (27, 18) & *${303, 164, 158}$ & *${1355, 764, 710}$ & ${16, 8.6, 8.3}$ & ${58, 28, 26}$ &   198 \\
GW191105\_143521     & IMRPhenomXPHM (35, 24) & ${86, 43, 42}$ & *${509, 262, 247}$ & ${8.5, 3.5, 3.3}$ & ${26, 10, 9.5}$ &   169 \\
GW191105\_143521     & SEOBNRv4PHM (27, 17) & ${84, 35, 31}$ & *${509, 229, 208}$ & ${8.3, 3.0, 2.7}$ & ${26, 9.8, 8.6}$ &   169 \\
GW191109\_010717     & IMRPhenomXPHM (5, 2) & ${169, 11, 10}$ & ${43, 5.5, 5.0}$ & *${437, 54, 43}$ & ${217, 27, 22}$ &   155 \\
GW191109\_010717     & SEOBNRv4PHM (16, 10) & ${47, 24, 22}$ & ${16, 10.0, 9.5}$ & *${198, 113, 102}$ & ${89, 50, 43}$ &   155 \\
GW191113\_071753     & IMRPhenomXPHM (15, 8) & ${35, 20, 20}$ & ${86, 53, 51}$ & ${25, 8.3, 7.3}$ & ${16, 5.9, 5.3}$ &   172 \\
GW191113\_071753     & SEOBNRv4PHM (15, 9) & ${74, 42, 35}$ & ${118, 72, 62}$ & ${17, 8.6, 8.0}$ & ${14, 6.3, 5.7}$ &   172 \\
GW191126\_115259     & IMRPhenomXPHM (36, 25) & ${103, 68, 68}$ & *${394, 252, 244}$ & ${5.6, 2.9, 2.7}$ & ${18, 9.1, 8.4}$ &   136 \\
GW191126\_115259     & SEOBNRv4PHM (28, 18) & ${104, 58, 52}$ & *${419, 234, 214}$ & ${6.5, 3.1, 2.7}$ & ${21, 9.8, 8.8}$ &   136 \\
GW191127\_050227     & IMRPhenomXPHM (29, 19) & ${59, 58, 61}$ & ${15, 11, 11}$ & ${48, 44, 41}$ & ${11, 8.8, 8.8}$ &   140 \\
GW191127\_050227     & SEOBNRv4PHM (17, 10) & ${32, 15, 16}$ & ${16, 7.4, 7.1}$ & ${20, 10, 11}$ & ${13, 4.8, 4.5}$ &   140 \\
GW191204\_110529     & IMRPhenomXPHM (22, 13) & ${107, 56, 51}$ & ${118, 68, 64}$ & ${46, 26, 25}$ & ${57, 32, 30}$ &   142 \\
GW191204\_110529     & SEOBNRv4PHM (15, 9) & ${119, 78, 70}$ & ${121, 84, 82}$ & ${42, 29, 31}$ & ${55, 35, 36}$ &   142 \\
GW191215\_223052     & IMRPhenomXPHM (15, 8) & ${34, 14, 11}$ & ${54, 22, 18}$ & ${19, 5.2, 4.1}$ & ${25, 6.5, 5.4}$ &   153 \\
GW191215\_223052     & SEOBNRv4PHM (14, 8) & ${31, 13, 12}$ & ${49, 21, 19}$ & ${17, 4.9, 4.3}$ & ${22, 6.7, 5.8}$ &   153 \\
GW191216\_213338     & IMRPhenomXPHM (31, 19) & *${1097, 696, 684}$ & *${5893, 3599, 3466}$ & ${96, 60, 54}$ & *${287, 152, 140}$ &   148 \\
GW191216\_213338     & SEOBNRv4PHM (25, 15) & *${1180, 586, 545}$ & *${6516, 3304, 3024}$ & ${93, 45, 40}$ & *${311, 150, 135}$ &   148 \\
GW191230\_180458     & IMRPhenomXPHM (21, 13) & ${10, 3.2, 2.9}$ & ${4.6, 1.8, 1.7}$ & ${18, 5.5, 4.9}$ & ${11, 3.5, 3.1}$ &   139 \\
GW191230\_180458     & SEOBNRv4PHM (23, 14) & ${8.4, 4.2, 4.1}$ & ${3.8, 2.1, 2.1}$ & ${14, 5.1, 4.5}$ & ${9.3, 3.5, 3.0}$ &   139 \\
GW200128\_022011     & IMRPhenomXPHM (20, 12) & ${35, 13, 11}$ & ${18, 8.1, 7.0}$ & ${32, 11, 9.8}$ & ${27, 9.5, 8.3}$ &   142 \\
GW200128\_022011     & SEOBNRv4PHM (26, 15) & ${36, 29, 21}$ & ${15, 13, 9.6}$ & ${26, 18, 11}$ & ${23, 17, 10}$ &   142 \\
GW200129\_065458     & IMRPhenomXPHM (33, 18) & *${220, 149, 145}$ & ${170, 108, 105}$ & ${113, 64, 59}$ & ${128, 62, 57}$ &   142 \\
GW200129\_065458     & SEOBNRv4PHM (18, 9) & *${235, 108, 103}$ & ${211, 96, 88}$ & ${165, 60, 59}$ & ${188, 75, 71}$ &   142 \\
GW200202\_154313     & IMRPhenomXPHM (37, 25) & *${666, 364, 350}$ & *${4383, 2342, 2214}$ & ${50, 23, 21}$ & *${174, 78, 71}$ &   121 \\
GW200202\_154313     & SEOBNRv4PHM (32, 21) & *${703, 330, 300}$ & *${4700, 2248, 2051}$ & ${50, 22, 20}$ & *${185, 80, 72}$ &   121 \\
GW200208\_130117     & IMRPhenomXPHM (31, 20) & ${18, 9.3, 8.4}$ & ${16, 8.7, 7.9}$ & ${25, 11, 9.6}$ & ${21, 9.3, 7.8}$ &   136 \\
GW200208\_130117     & SEOBNRv4PHM (31, 20) & ${17, 11, 11}$ & ${14, 9.8, 9.2}$ & ${22, 11, 10}$ & ${19, 10, 9.0}$ &   136 \\
GW200209\_085452     & IMRPhenomXPHM (22, 13) & ${21, 4.4, 4.0}$ & ${16, 4.6, 4.3}$ & ${25, 4.8, 4.2}$ & ${21, 4.4, 4.0}$ &   135 \\
GW200209\_085452     & SEOBNRv4PHM (22, 14) & ${10, 5.7, 5.6}$ & ${8.4, 4.9, 4.8}$ & ${12, 4.5, 3.9}$ & ${10, 4.0, 3.7}$ &   135 \\
GW200219\_094415     & IMRPhenomXPHM (18, 11) & ${11, 3.1, 2.7}$ & ${8.9, 3.1, 2.8}$ & ${16, 3.6, 3.2}$ & ${13, 3.2, 2.7}$ &   152 \\
GW200219\_094415     & SEOBNRv4PHM (22, 14) & ${10, 4.9, 4.6}$ & ${7.7, 3.9, 3.7}$ & ${11, 4.1, 3.8}$ & ${9.7, 3.4, 3.2}$ &   152 \\
GW200220\_061928     & IMRPhenomXPHM (22, 14) & ${8.3, 3.3, 3.1}$ & ${1.3, 0.63, 0.61}$ & ${52, 11, 10}$ & ${12, 4.1, 3.4}$ &   165 \\
GW200220\_061928     & SEOBNRv4PHM (22, 14) & ${10, 6.2, 5.3}$ & ${1.4, 0.88, 0.8}$ & ${34, 12, 10}$ & ${12, 4.6, 4.1}$ &   165 \\
GW200220\_124850     & IMRPhenomXPHM (20, 13) & ${13, 4.4, 4.3}$ & ${8.8, 3.6, 3.5}$ & ${16, 6.3, 6.0}$ & ${12, 5.0, 4.8}$ &   128 \\
GW200220\_124850     & SEOBNRv4PHM (17, 12) & ${11, 4.8, 5.3}$ & ${8.0, 3.8, 3.9}$ & ${18, 6.2, 7.5}$ & ${13, 4.9, 5.5}$ &   128 \\
\hline
\end{tabular}    
\caption{}
    \label{tab:ULmodel_B}
\end{table}
\clearpage

\clearpage
\renewcommand\thetable{7 continued}
\begin{table}
    \centering
\begin{tabular}{ccccccc}
Name & Waveform (visible frac.) & F$_{Q}$ &  F$_{\nu\bar{\nu}}$ &  F$_{\rm BZ}$ & F$_{\rm GW}$&  UL\\
\hline
GW200225\_060421     & IMRPhenomXPHM (16, 8) & ${97, 45, 40}$ & ${224, 114, 107}$ & ${33, 16, 14}$ & ${53, 26, 24}$ &   236 \\
GW200225\_060421     & SEOBNRv4PHM (22, 13) & ${71, 43, 47}$ & ${174, 97, 102}$ & ${25, 12, 12}$ & ${43, 18, 20}$ &   236 \\
GW200302\_015811     & IMRPhenomXPHM (15, 8) & ${62, 27, 26}$ & ${60, 29, 29}$ & ${62, 27, 26}$ & ${53, 23, 22}$ &   162 \\
GW200302\_015811     & SEOBNRv4PHM (20, 12) & ${66, 35, 35}$ & ${58, 30, 30}$ & ${54, 22, 21}$ & ${48, 20, 18}$ &   162 \\
GW200306\_093714     & IMRPhenomXPHM (28, 18) & ${133, 73, 67}$ & ${113, 60, 56}$ & ${19, 10, 9.7}$ & ${24, 9.9, 9.1}$ &   146 \\
GW200306\_093714     & SEOBNRv4PHM (25, 16) & ${132, 86, 82}$ & ${102, 63, 61}$ & ${20, 9.3, 8.1}$ & ${23, 10, 9.5}$ &   146 \\
GW200308\_173609     & IMRPhenomXPHM (18, 11) & ${51, 55, 51}$ & ${24, 24, 22}$ & ${74, 19, 12}$ & ${7.3, 2.2, 2.2}$ &   192 \\
GW200308\_173609     & SEOBNRv4PHM (14, 8) & ${71, 66, 67}$ & ${32, 26, 25}$ & ${13, 2.8, 3.3}$ & ${4.7, 2.1, 2.1}$ &   192 \\
GW200316\_215756     & IMRPhenomXPHM (21, 11) & ${112, 68, 62}$ & *${482, 258, 237}$ & ${13, 6.3, 5.3}$ & ${29, 10, 9.2}$ &   134 \\
GW200316\_215756     & SEOBNRv4PHM (24, 15) & ${103, 45, 41}$ & *${484, 223, 205}$ & ${10, 3.6, 3.2}$ & ${28, 10, 9.1}$ &   134 \\
GW200322\_091133     & IMRPhenomXPHM (14, 9) & ${147, 476, 499}$ & ${72, 58, 64}$ & ${42, 26, 27}$ & ${8.7, 3.3, 3.7}$ &   154 \\
GW200322\_091133     & SEOBNRv4PHM (4, 4) & ${3.7, 3.3, 0.22}$ & ${2.0, 0.45, 0.17}$ & ${4.8, 0.48, 0.37}$ & ${0.77, 0.27, 0.26}$ &   154 \\
\hline
\end{tabular}    
\caption{}
    \label{tab:ULmodel_C}
\end{table}
\renewcommand\thetable{\arabic{table}}
\setcounter{table}{7}
\clearpage

\end{document}